%% file: main.tex
\documentclass[11pt]{article}

\usepackage[margin=1.0in]{geometry}               
\usepackage{graphicx}
\usepackage{xcolor}
\usepackage{amsmath,amssymb,amsfonts,amsthm,bm}
\usepackage{authblk}   
\usepackage{hyperref} 
\usepackage{bbm} 
\usepackage{upgreek} 
\usepackage{tikz-cd}    
\usepackage{mathtools}
\usepackage{multirow}
\usepackage{booktabs}
\usepackage{subcaption}

\title{A Hamiltonian structure-preserving discretization of Maxwell's equations in nonlinear media}
\author[1,3]{William Barham}
\author[4]{Yaman G{\"u}{\c c}l{\"u}}
\author[2,3]{Philip J. Morrison}
\author[4,5]{Eric Sonnendr{\"u}cker}
\affil[1]{Oden Institute for Computational Engineering and Sciences, The University of Texas at Austin}
\affil[2]{Department of Physics, The University of Texas at Austin}
\affil[3]{Institute for Fusion Studies, The University of Texas at Austin}
\affil[4]{Max-Planck-Institut f{\"u}r Plasmaphysik, Numerische Methoden der Plasmaphysik}
\affil[5]{Technische Universit{\"a}t M{\"u}nchen, Zentrum Mathematik}
\date{\today}                     
\newcommand{\bmsf}[1]{\boldsymbol{\mathsf{#1}}}        
\newcommand{\msf}[1]{\mathsf{#1}}      

\hypersetup{%
  colorlinks=true,
  linkcolor=blue,
  linkbordercolor=red,
  citecolor=red,
}

\newtheorem{definition}{Definition}
\newtheorem{theorem}{Theorem}
\newtheorem{lemma}{Lemma}
\newtheorem{prop}{Proposition}

\begin{document}

\maketitle

\section*{Abstract}
A simple Hamiltonian modeling framework for general models in nonlinear optics is given. This framework is specialized to describe the Hamiltonian structure of electromagnetic phenomena in cubicly nonlinear optical media. The model has a simple Poisson bracket structure with the Hamiltonian encoding all of the nonlinear coupling of the fields. The field-independence of the Poisson bracket facilitates a straightforward Hamiltonian structure-preserving discretization using finite element exterior calculus. The generality and relative simplicity of this Hamiltonian framework makes it amenable for simulating a broad class of time-domain nonlinear optical problems. The main contribution of this work is a finite element discretization of Maxwell's equations in cubicly nonlinear media which is energy-stable and exactly conserves Gauss's laws. Moreover, this approach may be readily adapted to consider more general nonlinear media in subsequent work. 

\tableofcontents

\section{Introduction}


A large number of interactions between matter and the electromagnetic field may be adequately captured by approximate models which reduce the interaction to a phenomenological nonlinear polarization of the medium. In many ways, this concept is the heart of the field of nonlinear optics \cite{BoydNLO, New_2011, ShenNLO}, but also bears relevance in plasma physics in which asymptotic approximations often induce field dependent polarization and magnetization of the electromagnetic fields: see e.g. \cite{morrison_gauge_free_lifting} for nonlinear electromagnetic media arising from lifting particle-orbit asymptotics to kinetic theories, and \cite{brizard_and_hahm_2007, BURBY20152073, brizard_2021} for nonlinear media arising from the gyrokinetic approximation. The connection to plasma physics is mentioned because this paper leverages the Hamiltonian perspective used ubiquitously in plasma physics \cite{PhysRevLett.45.790, MORRISON1980383, PB-fluids-and-plasmas} as a novel modeling perspective for nonlinear optics. This paper describes a Hamiltonian modeling paradigm for Maxwell's equations in general nonlinear media, and then derives a general approach to its Hamiltonian structure-preserving discretization which might be applied to broad class of such models. This general approach is then applied to study the particular case of cubicly nonlinear media using a B-spline finite element exterior calculus (FEEC) method \cite{BUFFA20101143, BUFFA2011, güçlü2022broken}. By considering the Hamiltonian structure of Maxwell's equations in general nonlinear media in designing a structure-preserving discretization, it becomes straightforward to derive a semi-discrete model which conserves Gauss's laws and a discrete energy. This may then be integrated in time with a symplectic integrator \cite{geometric_numerical_integration}. Moreover, one can incorporate dissipative dynamics as a simple gradient flow generated by the Hamiltonian ensuring that the spatially discrete equations dissipate energy at the correct rate. The principle contribution of this paper is to elucidate how an infinite dimensional Hamiltonian formulation of Maxwell's equations in nonlinear media might be leveraged to derive discrete models which preserve the conservation laws of the continuous model. 


By far the most prevalent approach to the numerical study of nonlinear optical phenomena are those based on asymptotic assumptions, e.g.\ methods solving the nonlinear Schr{\"o}dinger equation, and the beam propagation method \cite{BoydNLO, New_2011, ShenNLO}. These are attractive methods because they are relatively inexpensive compared with full time-domain Maxwell solvers \cite{Taflove-FDFT}, however their validity depends on the validity of the asymptotic assumptions, e.g.\ the slowly varying amplitude and paraxial approximations \cite{BoydNLO}. For this reason, there is growing interest in developing efficient and accurate time-domain methods for nonlinear optics. A time-domain solver for the same nonlinear optical model studied in this work was previously studied using a discontinuous Galerkin approach was given in \cite{BOKIL2017420, lyu_et_al_2021, lyu_et_al_2022} and using an FDFT approach in \cite{bokil_et_al_2018}. These methods yielded energy conserving methods which are second order accurate in time due to their use of the Leapfrog method for time-stepping. This work pursues a markedly different discretization approach, which exploits the Hamiltonian structure of the model, generalizing a method used to derive a structure-preserving discretization of a model of the nonlinear ponderomotive force \cite{10.1063/5.0178935}. The method studied in this work likewise conserves energy (or rather shadows it in the manner of a symplectic integrator \cite{geometric_numerical_integration}), allows for the use of broken elements (so that computation might be localized to each element), and allows for arbitrary discretization order in both space and time. However, because the method preserves the Hamiltonian structure, it further preserves Gauss's laws. While this work only considers a second-order method for the temporal discretization, one may, in principle, use higher-order splitting methods to obtain higher-order methods in time, although the number of sub-stages needed for higher-order splitting methods rapidly increases with order. This structure-preserving discretization of Maxwell's equations in nonlinear media makes use of the geometrically inspired finite element exterior calculus (FEEC) method \cite{ArnoldDouglasN2010Feec}. This approach leverages the manner in which Maxwell's equations may be written in terms of differential forms. A general Hamiltonian framework for the Maxwell-Vlasov system in electromagnetic media with field dependent polarization and magnetization was given in \cite{morrison_gauge_free_lifting}. This modeling framework is the basis for the numerical methods presented in this work. For work describing the geometric interpretation of this Hamiltonian modeling framework, see \cite{BARHAM2023100016}.  


The geometric interpretation of Maxwell's equations in terms of differential forms provides benefits not only from a model building perspective, but also helps elucidate and motivate successful numerical methods for electromagnetism. Substantial research in the theory of finite element methods has been conducted in this vein in the past half century with many of the successful finite element methods for electromagnetic simulations being in some manner based on designing a sequence of finite element spaces which preserve the structure of the de Rham complex, see \cite{bossavit_1988a, bossavit_1988b, bossavit_1988c} for early work outlining the connection between mixed finite element methods and Whitney forms \cite{geometric_intergration_theory}. These various finite element methods are united by a common interpretation in terms of exterior calculus, frequently referred to as FEEC, which synthesizes the key elements of this subject in an elegant and perspicuous mathematical theory \cite{ArnoldDouglasN2010Feec}. More recently, there have been efforts to design finite element methods which preserve the de Rham complex which are built from broken elements, i.e\ discontinuous shape functions which are local to each element \cite{campospinto:hal-00969326, conga_pic_1, conga_pic_2, campos-pinto2016:CONGA, campospinto2024brokenfeec, güçlü2022broken}. These are referred to variously as CONGA or broken-FEEC methods. While this work uses a conforming B-spline FEEC method to spatially discretize the fields, virtually any discretization method which preserves the de Rham cohomology (e.g. FEEC methods of any kind \cite{ArnoldDouglasN2010Feec, BUFFA20101143, kreeft2011mimetic}, mimetic finite difference methods \cite{Marsden2003DiscreteEC, desbrun2005discrete}, and spectral methods \cite{pinto2021geometric}) may be used to discretize the Hamiltonian structure following the approach taken in this work to obtain numerical methods which likewise preserve the continuous system's conservation laws. A general theory of the properties common to many structure-preserving discretizations of exterior calculus is given in \cite{10.1007/0-387-38034-5_5}. 


The evolution equations for Maxwell's equations in cubicly nonlinear media considered in this work are generated by a simple Poisson bracket: the bracket is a direct sum of the Poisson brackets for Maxwell's equations and canonical brackets for the additional degrees of freedom mediating the nonlinear polarization.  Because these brackets are field-free (i.e., the Poisson bivector does not depend on the dynamical fields), a structure-preserving discretization is easily accomplished using a FEEC method: the domain of the functionals in the Hamiltonian formulation are simply restricted to act on the finite-element vector spaces. In general, self-consistent models of the interaction of a plasma with the electromagnetic field come from field-dependent Poisson brackets making structure-preserving discretizations rare. For example, Hamiltonian structure-preserving discretizations of the Vlasov-Maxwell system and 2D vorticity dynamics generally rely on particle based discretizations: see \cite{GEMPIC, 10.1063/1.4976849} for a Hamiltonian structure-preserving discretization of the Vlasov-Maxwell system, and see \cite{kirchhoff_point_vortex} for an early reference on the Hamiltonian mechanics of point vorticies. The simple Poisson bracket of the nonlinear Maxwell equations (and indeed of many Hamiltonian nonlinear wave equations), with all nonlinear coupling of the fields occurring in the Hamiltonian, facilitates the approach taken herein. 


This paper is organized as follows. In section \ref{section:general_nonlinear_maxwell}, the Hamiltonian structure of Maxwell's equations in general nonlinear media and the special case of cubicly nonlinear media is described. In section \ref{sec:general_feec_overview}, a brief overview of FEEC methods and the discretization of functional derivatives is given. In section \ref{sec:general_disc_theory}, a Hamiltonian structure-preserving discretization of Maxwell's equations in general nonlinear media is given. In section \ref{sec:disc_cubic_maxwell}, this general approach is applied to obtain a discretization of Maxwell's equations in cubicly nonlinear media. Finally, in section \ref{sec:numerical_results}, the numerical results are presented. Some material was moved to the appendices to aid in the readability of the paper. Appendix \ref{appendix:variational-NL-maxwell} briefly explains how the Hamiltonian structure of Maxwell's equations in nonlinear media arises from the Lagrangian variational framework. Appendix \ref{appendix:integral_formulas} provides a result which is needed to prove the conservation of several Casimir invariants of the semi-discretized models. Appendix \ref{appendix:lipschitz-bound} provides a technical result needed to prove the stability of the time-stepper of the solver derived in section \ref{sec:disc_cubic_maxwell}. Finally, appendix \ref{appendix:kink_shaped_waves} provides some background on a special class of exact traveling wave solutions to the cubicly nonlinear Maxwell equations in one spatial dimension. 


\input{continuous_model}

\input{feec_overview}

\input{general_discrete_theory}

\input{discrete_model}

\input{numerical_results}

\section{Conclusion}


This paper derived a general approach to discretize the Hamiltonian structure of Maxwell's equations in nonlinear media so as to obtain a semi-discretized system which was itself Hamiltonian and possessed discrete analogs of the continuous conservation laws, i.e.\ conserved semi-discrete Gauss's laws. The model was then specialized to consider isotropic cubicly nonlinear media. The spatial discretization was accomplished using a B-spline FEEC method on a uniform mesh. The preservation of the de Rham cohomology was essential for deriving a semi-discrete model which is Gauss-preserving. An explicit time-discretization was then derived using Hamiltonian splitting. This may in principle be extended to higher-order temporal methods, but this would result in a proliferation of sub-steps. Hence, this paper only considered second-order Strang splitting for which a CFL condition was derived. Finally, dissipation was included in the model as a gradient flow generated by the Hamiltonian. 


This work only considered isotropic, cubicly nonlinear media, but it is possible to derive Hamiltonian structure-preserving discretizations for nearly arbitrary models from nonlinear optics by adapting the approach taken in this work. In particular, anisotropic cubicly nonlinear media, and the Maxwell-Duffing system \cite{8932393} would be interesting to study in a future work using the Hamiltonian structure-preserving discretization framework. Moreover, there is nothing to prevent one from applying this paper's approach to Hamiltonian structure-preserving discretization to more general nonlinear vector and scalar wave equations known as nonlinear Hodge wave models. Such models, including nonlinear acoustic models, were discussed in general in \cite{Barham_dissertation}, and Westervelt's equation, a particular nonlinear acoustic wave model, was recently studied numerically with a mimetic finite difference method in \cite{barham2024mimeticdiscretizationwesterveltsequation}. 


Although Hamiltonian splitting yielded an efficient explicit method, implicit methods such as implicit Runge-Kutta \cite{geometric_numerical_integration} (which are symplectic) or energy conserving discrete gradient methods \cite{discrete_gradient} are worth investigating as alternative time-stepping schemes. An implicit method would lead to a more expensive nonlinear solve each time-step, but, in compensation, would overcome the relatively severe scaling of the CFL condition with spatial resolution. 

The method implemented in this work utilizes a conforming-FEEC method on unmapped grids. By considering unmapped grids, it was possible to exploit the Kronecker structure of the finite element mass matrices for fast matrix inversion. However, the algorithm readily generalizes to use broken-FEEC methods \cite{campospinto:hal-00969326, conga_pic_1, conga_pic_2, campos-pinto2016:CONGA, campospinto2024brokenfeec, güçlü2022broken} on mapped, multi-patch domains. The current implementation utilizes a library which already supports these features \cite{Gl2022PSYDACAH}. This approach has not been considered in the current work, but presents several distinct advantages. A mapped, multi-patch domain allows one to consider general, complex geometries. Additionally, in a broken-FEEC implementation, the nonlinear electric solve would fully decouple across each element and the mass matrices would be block diagonal. Therefore, broken-FEEC would yield a solver with significant potential for parallelism. This would facilitate higher-resolution two- and three-dimensional simulations than considered in this present work. Another involution-preserving discontinuous Galerkin approximation of the curl-curl operator is discussed in \cite{doi:10.1137/23M1555235}: such a spatial discretization would likewise yield a method based on broken spaces which nonetheless enjoys the structure-preserving properties of the method derived herein. 

As cubicly nonlinear media can exhibit wave-steepening in certain parameter regimes, it is possible for shocks to form. For the method to more effectively handle traveling-wave problems, it would be beneficial to investigate the feasibility of broken-FEEC methods with conforming projection operators which stabilize such discontinuous solutions. The framework developed here intentionally utilized a method which might easily accommodate a broken-FEEC method on mapped elements in a future work. Whether there exist high-order nonlinear flux functions which are compatible with the broken-FEEC framework appears to be an open question. Subsequent work will expand the approach derived in this work to seek a time-explicit, energy-stable, Gauss-preserving, high-order, shock-capturing, and scalable broken-FEEC method for the time-domain solution of Maxwell's equations in nonlinear media. 

\section*{Acknowledgments}
The authors gratefully acknowledge the support of U.S. Dept. of Energy Contract \# DE-FG02-04ER54742, NSF Graduate Research Fellowship \# DGE-1610403, and the Humboldt foundation. 

\bibliographystyle{plain}
\bibliography{references} 

\input{appendix}

\end{document}

%% file: continuous_model.tex
\section{Maxwell's equations in nonlinear media} \label{section:general_nonlinear_maxwell}

Many models in nonlinear electrodynamics take a generic form in which a nonlinear polarization or magnetization is specified by some self-consistent functional of the fields. It turns out that this class of models may be formulated as a Hamiltonian field theory \cite{morrison_gauge_free_lifting}. In addition to being a powerful model building tool, helping one to self-consistently understand complex models and the manner in which components of a model couple with each other, the Hamiltonian formulation also may be used to derive structure-preserving numerical schemes which better capture the physical content of the continuous model \cite{GEMPIC, morrison_structure_preserving_algorithms}. To begin, the Hamiltonian structure of the macroscopic Maxwell system is described in general before specializing to discuss cubicly nonlinear media. In what follows, the models are described in the context of the full three-dimensional Maxwell wave equations, but, mutatis mutandis, one may proceed in a like manner for its one- and two-dimensional analogs. 

\subsection{A general Hamiltonian model of Maxwell's equations in nonlinear media} 

Maxwell's equations in general macroscopic media may be written
\begin{equation}
	\begin{aligned}[c]
		\partial_t \bm{D} &= \nabla \times \bm{H} - \bm{J}_f, \\
		\partial_t \bm{B} &= - \nabla \times \bm{E}, 
	\end{aligned}
	\quad \text{and} \quad
	\begin{aligned}[c]
		\nabla \cdot \bm{D} &= \rho_f, \\
		\nabla \cdot \bm{B} &= 0 \,,
	\end{aligned}
\end{equation}
where the free charge and current distributions, $\bm{\rho}_f$ and $\bm{J}_f$ respectively, may depend on additional degrees of freedom, $\{ \chi_k \}$, and the displacement field and auxiliary magnetic field are defined by the constitutive relations
\begin{multline}
    \bm{D} 
    = \epsilon_0 \bm{E} + \bm{P}[\bm{E}, \bm{B}, \{ \chi_k \}, \nabla \bm{E}, \nabla \bm{B}, \{ \nabla \chi_k \}, \hdots] \\
	\text{and} \quad
    \bm{H}
    = \mu_0^{-1} \bm{B} + \bm{M}[\bm{E}, \bm{B}, \{ \chi_k \}, \nabla \bm{E}, \nabla \bm{B}, \{ \nabla \chi_k \}, \hdots] \,,
\end{multline}
where the polarization, $\bm{P}$, and magnetization, $\bm{M}$, are general functionals of the fields and additional degrees of freedom. These may be quite general allowing even derivatives of the fields in the constitutive relations. 

This system may be written formally as a Hamiltonian field theory with two caveats: (1) the conservative character of this system is dependent on the choice of boundary conditions, (2) the additional degrees of freedom, $\{\chi_k\}$, are treated as a black box and are assumed to be in Hamiltonian form. The following adapts ideas ideas found in \cite{morrison_gauge_free_lifting} to write a Hamiltonian formulation of general nonlinear optical systems. Appendix \ref{appendix:variational-NL-maxwell} builds some intuition for the relationship between nonlinear polarization and the electromagnetic energy. Consider a general matter model prescribed by
\begin{equation}
    K[ \bm{E}, \bm{B}, \{ \chi_k \}] 
    = \int_\Omega \mathcal{K}(\bm{x}, \bm{E}, \bm{B}, \{ \chi_k \}, \nabla \bm{E}, \nabla \bm{B}, \{ \nabla \chi_k \}, ... ) \mathsf{d}^3 \bm{x} \,,
\end{equation}
so that the polarization and magnetization are defined by
\begin{equation}
	\bm{P} = - \frac{\delta K}{\delta \bm{E}} \quad \text{and} \quad \bm{M} = \frac{\delta K}{\delta \bm{B}}.
\end{equation}
The Hamiltonian of the macroscopic Maxwell equations system is given by
\begin{equation}
    H[\bm{E},\bm{B}, \{ \chi_k \}] 
    = K - \int_\Omega \bm{E} \cdot \frac{\delta K}{\delta \bm{E}} \mathsf{d}^3 \bm{x} 
    + \frac{1}{2} \int_M ( \epsilon_0 \bm{E} \cdot \bm{E} + \mu_0^{-1} \bm{B} \cdot \bm{B} ) \mathsf{d}^3 \bm{x} \,.
\end{equation}
The Poisson bracket is defined to be
\begin{equation}
    \{ F, G \} 
    = \int_\Omega \bigg[ \frac{\delta F}{\delta \bm{D}} \cdot \nabla \times \frac{\delta G}{ \delta \bm{B}} 
    - \frac{\delta G}{\delta \bm{D}} \cdot \nabla \times \frac{\delta F}{\delta \bm{B}} \bigg] \mathsf{d}^3 \bm{x}
    + \{F, G\}_{\chi} \,,
\end{equation}
where $\{,\}_\chi$ is the bracket generating the flow for the additional degrees of freedom. One may show, see e.g.\ \cite{morrison_gauge_free_lifting, BARHAM2023100016}, that if $\overline{H}[\bm{D}, \bm{B}, \{ \chi_k \}] = H[\bm{E}, \bm{B}, \{ \chi_k \}]$, then
\begin{equation}
    \frac{\delta \overline{H}}{\delta \bm{D}} = \bm{E} 
    \quad \text{and} \quad 
    \frac{\delta \overline{H}}{\delta \bm{B}} = \bm{H} \,.
\end{equation}
This calculation will be shown in detail subsequently in the specialized context of cubicly nonlinear media. Hence, one obtains the equations of motion:
\begin{equation}
    \begin{cases}
    \partial_t \bm{D} = \{ \bm{D}, \overline{H} \} 
    = \nabla \times \bm{H} + \{\bm{D}, \overline{H} \}_\chi \\
    \partial_t \bm{B} = \{ \bm{B}, \overline{H} \} 
    = - \nabla \times \bm{E} + \{\bm{B}, \overline{H} \}_\chi \,.
   \end{cases}
\end{equation}
The free current is defined to be $-\bm{J}_f = \{\bm{D}, \overline{H}\}_\chi$, and one requires that $\{\bm{B}, \overline{H}\}_\chi = 0$. The precise characteristics of the model for $\{\chi_k\}$ cannot be further specified without choosing a particular system to model. 

\subsection{A Hamiltonian model of Maxwell's equations in cubicly nonlinear media} \label{sec:cts_cubic_maxwell}

Consider the following cubicly nonlinear optical model (in SI units):
\begin{equation} \label{eq:cubic_nl_maxwell_sys}
	\begin{aligned}
		\partial_t \bm{B} &= - \nabla \times \bm{E}, \quad 
		\partial_t \bm{D} = \mu_0^{-1} \nabla \times \bm{B} \\
		\bm{D} &= \epsilon_0 \left( \epsilon_\infty \bm{E} + \bm{P} + a(1 - \theta) | \bm{E} |^2 \bm{E} + a \theta Q \bm{E} \right) \\
		\partial_t \bm{P} &= \bm{J}, \quad 
		\partial_t \bm{J} = \omega_p^2 \bm{E} - \omega_0^2 \bm{P} \\
		\partial_t Q &= \sigma, \quad 
		\partial_t \sigma = \omega_v^2 | \bm{E} |^2 - \omega_v^2 Q.
	\end{aligned}
\end{equation}
The notation has been aligned with prior numerical studies of the same model \cite{bokil_et_al_2018, BOKIL2017420, lyu_et_al_2021, lyu_et_al_2022} for greater convenience in comparing these methods with those developed herein. This model includes fairly general features of cubicly nonlinear media although the medium simply assumed to be isotropic. It would be an interesting generalization to study anisotropic media as well. 

One can see that this is a macroscopic formulation of Maxwell's equations with a self-consistent polarization prescribed by both instantaneous effects and time delayed effects facilitated through auxiliary variables, $(\bm{P}, \bm{J}, Q, \sigma)$. The macroscopic field $\bm{D}$ may be divided into four parts: the instantaneous linear response, $\epsilon_0 \epsilon_\infty \bm{E}$; the dispersive (in time) linear response, $\epsilon_0 \bm{P}$; the instantaneous cubic nonlinearity, the optical Kerr effect, $\epsilon_0 a(1 - \theta) | \bm{E} |^2 \bm{E}$; and the dispersive (in time) cubic nonlinearity, the Raman molecular vibrational response, $\epsilon_0  a \theta Q \bm{E}$. The parameter $\theta \in [0,1]$ controls the relative importance of the two cubically nonlinear terms while $a > 0$ controls the strength of the cubic terms as a whole. As will be discussed at length in the subsequent discussion, this model possesses a conserved energy:
\begin{multline} \label{eq:cubic_hamiltonian}
    H[\bm{E}, \bm{B}, \bm{P}, \bm{J}, Q, \sigma] 
	= 
	\frac{1}{2} \int_\Omega \bigg[ \epsilon_0 \bigg( \epsilon_\infty | \bm{E} |^2 
	+ \frac{3 a ( 1 - \theta) | \bm{E} |^4}{2} + a \theta Q | \bm{E} |^2 \\
	+ \frac{\omega_0^2}{\omega_p^2} | \bm{P} |^2 + \frac{1}{\omega_p^2}  | \bm{J} |^2
	+ \frac{a \theta}{2} Q^2 + \frac{a \theta}{2 \omega_v^2} \sigma^2 \bigg) 
	+ \mu_0^{-1} | \bm{B} |^2 \bigg] \mathsf{d}^3 \bm{x}.
\end{multline}
Positivity of the energy may be ensured by restricting the range of the parameter $\theta$ \cite{lyu_et_al_2022}:
\begin{equation}
    \frac{3 a ( 1 - \theta) | \bm{E} |^4}{2} 
    + a \theta Q | \bm{E} |^2 + \frac{a \theta}{2} Q^2
    =
    \frac{a \theta}{4} ( | \bm{E} |^2 + Q )^2
    + \frac{a(3 - 4 \theta)}{4} | \bm{E} |^4 \geq 0
\end{equation}
if one requires that $\theta \in [0, 3/4]$. The parameters $\omega_0$ and $\omega_p$ are the resonance and plasma frequencies respectively. One may add additional resonant frequencies to the system, but this work considers only a single resonant frequency in the dispersion relation. Finally, $\omega_v$ is the resonance frequency for the Raman molecular vibrational response. 

One can see that the polarization vector $\bm{P}$ and Raman term may be described by second order equations:
\begin{equation}
	\begin{aligned}
		\partial_{tt} \bm{P} + \omega_0^2 \bm{P} &= \omega_p^2 \bm{E} 
		\implies \mathcal{F}[\bm{P}] = \frac{\omega_p^2}{\omega_0^2 - \omega^2} \mathcal{F}[\bm{E}] \\
		\partial_{tt} Q + \omega_v^2 Q &= \omega_v^2 | \bm{E} |^2
		\implies \mathcal{F}[Q] = \frac{\omega_v^2 }{\omega_v^2 - \omega^2} \mathcal{F}[ | \bm{E} |^2 ]
	\end{aligned}
\end{equation}
where $\mathcal{F}$ denotes the Fourier transform in time. The Fourier convolution theorem then implies
\begin{equation}
    \mathcal{F}[\bm{D}] = \epsilon_0 \left[ 
    \left( \epsilon_\infty + \frac{\omega_p^2}{\omega_0^2 - \omega^2} \right) \mathcal{F}[\bm{E}]
    + a \left( (1 - \theta) + \frac{\theta \omega_v^2}{\omega_v^2 - \omega^2} \right) \mathcal{F}[ | \bm{E} |^2 \bm{E}]
    \right] \,.
\end{equation}
Hence, one can see that $\bm{P}$ may be thought of as a time delayed component of the linear polarization and $Q$ as a time delayed component of the cubicly nonlinear polarization. Moreover, one can see that the cubic term leads to phase mixing and hence nonlinear optical effects such as higher harmonic generation \cite{BoydNLO, ShenNLO, New_2011}. Finally, it is often physically relevant to add linear damping terms to the equations for $\bm{P}$ and $Q$ in order to model absorption as was done in \cite{bokil_et_al_2018, BOKIL2017420, lyu_et_al_2021, lyu_et_al_2022}. The non-dissipative model conserves the energy in equation \eqref{eq:cubic_hamiltonian}. The remainder of this section specifies the Hamiltonian structure of these equations using a modeling framework developed in \cite{morrison_gauge_free_lifting}. 

The Poisson bracket of the cubicly nonlinear Maxwell system is a simple extension of the usual Poisson bracket Maxwell's equations:
\begin{multline} \label{eq:cubic_3d_pb}
	\{F, G\} = 
		\int_\Omega \left[ - \left( \frac{\delta F}{\delta \bm{B}} \cdot \nabla \times \frac{\delta G}{\delta \bm{D}} - 
				   \frac{\delta G}{\delta \bm{B}} \cdot \nabla \times \frac{\delta F}{\delta \bm{D}} \right) \right. \\
			\left. + \frac{\omega_p^2}{\epsilon_0} \left( \frac{\delta F}{\delta \bm{P}} \cdot \frac{\delta G}{\delta \bm{J}} - 
				  \frac{\delta G}{\delta \bm{P}} \cdot \frac{\delta F}{\delta \bm{J}} \right)
			+  \frac{2 \omega_v^2}{\epsilon_0 a \theta} \left( \frac{\delta F}{\delta Q} \frac{\delta G}{\delta \sigma} - 
				  \frac{\delta G}{\delta Q} \frac{\delta F}{\delta \sigma} \right) \right] \mathsf{d}^3 \bm{x} \,.
\end{multline}
The bracket is merely a direct sum of the Born-Infeld bracket for $(\bm{D}, \bm{B})$ with (nearly) canonical brackets for the canonically conjugate (modulo scaling) variables $(\bm{P}, \bm{J})$ and $(Q, \sigma)$ governing the time-dispersive effects. The nonlinear coupling of these three systems comes from the Hamiltonian. 

As in \cite{morrison_gauge_free_lifting}, the key to specifying the Hamiltonian structure of the macroscopic Maxwell system is the specification of an an energy functional $K$ such that $\bm{D} = \epsilon_0 \bm{E} - \delta K/\delta \bm{E}$. This is accomplished via defining the functional
\begin{multline} \label{eq:cubic_3d_matter_model}
	K[\bm{E}, \bm{B}, \bm{P}, \bm{J}, Q, \sigma] = 
		- \frac{\epsilon_0}{2} \int_\Omega \bigg[ (\epsilon_\infty - 1) | \bm{E} |^2 + 2 \bm{P} \cdot \bm{E} + \frac{a(1 - \theta) | \bm{E} |^4}{2} + a \theta Q | \bm{E} |^2 \\
		\left. - \left( \frac{\omega_0^2}{\omega_p^2} | \bm{P} |^2 + \frac{1}{\omega_p^2}  | \bm{J} |^2
		+ \frac{a \theta}{2} Q^2 + \frac{a \theta}{2 \omega_v^2} \sigma^2 \right) \right] \mathsf{d}^3 \bm{x}
\end{multline}
so that
\begin{equation}
	\frac{\delta K}{\delta \bm{E}} = - \epsilon_0 \left[ (\epsilon_\infty - 1) \bm{E} + \bm{P} + a(1 - \theta) | \bm{E} |^2 \bm{E} + a \theta Q \bm{E} \right].
\end{equation}
The additional terms, which are independent of $\bm{E}$, are necessary to specify the auxiliary equations for $\bm{P}$ and $Q$ and their canonically conjugate variables $\bm{J}$ and $\sigma$. The Hamiltonian is then defined to be
\begin{equation} \label{eq:cubic_3d_ham}
	\begin{aligned}
		H[\bm{E}, \bm{B}, \bm{P}, \bm{J}, Q, \sigma] 
			&= K - \int_\Omega \frac{\delta K}{\delta \bm{E}} \cdot \bm{E} \mathsf{d}^3 \bm{x} 
			+ \frac{1}{2} \int_\Omega \left( \epsilon_0 | \bm{E} |^2 + \mu_0^{-1} | \bm{B} |^2 \right) \mathsf{d}^3 \bm{x} \\
			&\hspace{-7em} = 
			\frac{1}{2} \int_\Omega \bigg[ \epsilon_0 \bigg( \epsilon_\infty | \bm{E} |^2 
			+ \frac{3 a ( 1 - \theta) | \bm{E} |^4}{2} + a \theta Q | \bm{E} |^2 \\
			+& \frac{\omega_0^2}{\omega_p^2} | \bm{P} |^2 + \frac{1}{\omega_p^2}  | \bm{J} |^2
			+ \frac{a \theta}{2} Q^2 + \frac{a \theta}{2 \omega_v^2} \sigma^2 \bigg) 
			+ \mu_0^{-1} | \bm{B} |^2 \bigg] \mathsf{d}^3 \bm{x}.
	\end{aligned}
\end{equation}
This corresponds with the conserved energy given in \cite{bokil_et_al_2018, BOKIL2017420, lyu_et_al_2021, lyu_et_al_2022} and is identical to that in equation \eqref{eq:cubic_hamiltonian}. As the equations of motion are actually in the variables $(\bm{D}, \bm{B}, \bm{P}, \bm{J}, Q, \sigma)$, one must change variables in order to recover the equations of motion from the Hamiltonian and Poisson bracket. That is, one needs to take derivatives of 
\begin{equation} \label{eq:cubic_3d_ham_2}
    \overline{H}[\bm{D}, \bm{B}, \bm{P}, \bm{J}, Q, \sigma]
    \coloneq H[\bm{E}, \bm{B}, \bm{P}, \bm{J}, Q, \sigma]
\end{equation}
where $\bm{E}$ is thought of as an implicitly defined function. 

\begin{theorem}
Let $\overline{H}[\bm{D}, \bm{B}, \bm{P}, \bm{J}, Q, \sigma] = H[\bm{E}, \bm{B}, \bm{P}, \bm{J}, Q, \sigma]$ where $H$ is defined as in equation \ref{eq:cubic_3d_ham}. Then the derivatives are given by
\begin{multline}
    \frac{\delta \overline{H}}{\delta \bm{D}} = \bm{E} \,, 
    \quad
    \frac{\delta \overline{H}}{\delta \bm{B}} = \mu_0^{-1} \bm{B} \,,
    \quad
    \frac{\delta \overline{H}}{\delta \bm{P}} = \epsilon_0 \left( \frac{ \omega_0^2}{\omega_p^2} \bm{P} - \bm{E} \right), \\
    \quad
    \frac{\delta \overline{H}}{\delta \bm{J}} = \frac{\epsilon_0}{\omega_p^2} \bm{J} \,,
    \quad
    \frac{\delta \overline{H}}{\delta Q} = \frac{\epsilon_0 a \theta}{2} \left( Q - | \bm{E} |^2 \right) \,,
    \quad \text{and} \quad
    \frac{\delta \overline{H}}{\delta \sigma} = \frac{\epsilon_0 a \theta}{2 \omega_v^2} \sigma \,.
\end{multline}
\end{theorem}
\noindent \textit{Proof:} The derivatives of the Hamiltonian $H$ are given by
\begin{multline}
    \frac{\delta H}{\delta \bm{E}} = \epsilon_0 \left( \epsilon_\infty \bm{E} 
    + 3 a (1 - \theta) | \bm{E} |^2 \bm{E} + a \theta Q \bm{E} \right) \,,
    \quad
    \frac{\delta H}{\delta \bm{B}} = \mu_0^{-1} \bm{B} \,, \\
    \quad
    \frac{\delta H}{\delta \bm{P}} = \frac{\epsilon_0 \omega_0^2}{\omega_p^2} \bm{P} \,,
    \quad
    \frac{\delta H}{\delta \bm{J}} = \frac{\epsilon_0}{\omega_p^2} \bm{J} \,,
    \quad
    \frac{\delta H}{\delta Q} = \frac{\epsilon_0 a \theta}{2} \left( | \bm{E} |^2 + Q \right) \,,
    \quad \text{and} \quad
    \frac{\delta H}{\delta \sigma} = \frac{\epsilon_0 a \theta}{2 \omega_v^2} \sigma \,.
\end{multline}
Note that
\begin{equation}
    \frac{\delta \overline{H}}{\delta \bm{D}} = \left( \frac{\delta \bm{E}}{\delta \bm{D}} \right)^\dagger \frac{\delta H}{\delta \bm{E}}
    \quad \text{and} \quad
    \frac{\delta \overline{H}}{\delta \Psi} = \frac{\delta H}{\delta \Psi} + \left( \frac{\delta \bm{E}}{\delta \Psi} \right)^\dagger \frac{\delta H}{\delta \bm{E}} \,,
\end{equation}
where $\Psi \in \{ \bm{B}, \bm{P}, \bm{J}, Q, \sigma \}$. 

Because $\bm{D} = \bm{E} - \delta K/\delta \bm{E}$, and
\begin{equation}
	H = K - \int_\Omega \frac{\delta K}{\delta \bm{E}} \cdot \bm{E} \mathsf{d}^3 \bm{x} + \frac{1}{2} \int_\Omega ( \epsilon_0 | \bm{E} |^2 + \mu_0^{-1} | \bm{B} |^2 ) \mathsf{d}^3 \bm{x}
\end{equation}
it follows that
\begin{equation}
	\frac{\delta H}{\delta \bm{E}} = \left( \epsilon_0 I - \frac{\delta^2 K}{\delta \bm{E} \delta \bm{E}} \right) \bm{E}
	\quad \text{and} \quad
	\frac{\delta \bm{D}}{\delta \bm{E}} = \epsilon_0 I - \frac{\delta^2 K}{\delta \bm{E} \delta \bm{E}}. 
\end{equation}
For convenience, define $\mathcal{E}[\bm{E}] = \epsilon_0 I - \frac{\delta^2 K}{\delta \bm{E} \delta \bm{E}}$. Consider the mapping 
\begin{equation}
    \Phi[\bm{E}, \bm{B}, \bm{P}, \bm{J}, Q, \sigma] = (\bm{D}, \bm{B}, \bm{P}, \bm{J}, Q, \sigma) \,.
\end{equation}
Then it follows that
\begin{equation}
	D \Phi[\bm{E}, \bm{B}, \bm{P}, \bm{J}, Q, \sigma] = 
	\begin{pmatrix}
		\mathcal{E}[\bm{E}] & 0 & \epsilon_0 I & 0 & \epsilon_0 a \theta (\bm{E} \cdot) & 0 \\
		& I & \\
		& & I \\
		& & & I \\
		& & & & I \\
		& & & & & I
	\end{pmatrix}
\end{equation}
where $(\bm{E} \cdot)$ denotes the operator which applies the dot product of $\bm{E}$ to a given vector. Hence
\begin{equation}
	D \Phi^{-1} = 
	\begin{pmatrix}
		\mathcal{E}[\bm{E}]^{-1} & 0 & - \epsilon_0 \mathcal{E}[\bm{E}]^{-1} & 0 & - \epsilon_0 a \theta \mathcal{E}[\bm{E}]^{-1} (\bm{E} \cdot) & 0 \\
		& I & \\
		& & I \\
		& & & I \\
		& & & & I \\
		& & & & & I
	\end{pmatrix} \,.
\end{equation}
Hence, since $\delta H/\delta \bm{E} = \mathcal{E} \bm{E}$, it follows that $\delta \overline{H}/\delta \bm{D} = \bm{E}$, 
\begin{equation}
    \frac{\delta \overline{H}}{\delta \bm{P}} 
    = \epsilon_0 \left( \frac{ \omega_0^2}{\omega_p^2} \bm{P} 
    - \mathcal{E}^{-1} \mathcal{E} \bm{E} \right) 
    = \epsilon_0 \left( \frac{ \omega_0^2}{\omega_p^2} \bm{P} - \bm{E} \right) \,,
\end{equation}
and
\begin{equation}
    \frac{\delta \overline{H}}{\delta Q} 
    = \frac{\epsilon_0 a \theta}{2} \left( | \bm{E} |^2 + Q \right) 
    - \epsilon_0 a \theta ( \bm{E} \cdot ) \mathcal{E}^{-1} \mathcal{E} \bm{E} 
    = \frac{\epsilon_0 a \theta}{2} \left( Q - | \bm{E} |^2 \right) \,.
\end{equation}
All the other functional derivatives remain unchanged. \qed

\begin{theorem}
Using the Poisson bracket given in equation \ref{eq:cubic_3d_pb} and the Hamiltonian given in equation \ref{eq:cubic_3d_ham_2}, we obtain the equations of motion
\begin{equation}
\begin{split}
    \partial_t \bm{D} &= \{ \bm{D}, \overline{H} \} = \mu_0^{-1} (\nabla \times)^* \bm{B} \\
    \partial_t \bm{B} &= \{ \bm{B}, \overline{H} \} = - \nabla \times \bm{E} \\
    \partial_t \bm{P} &= \{ \bm{P}, \overline{H} \} = \bm{J}
\end{split}
\qquad
\begin{split}
    \partial_t \bm{J} &= \{ \bm{J}, \overline{H} \} = \omega_p^2 \bm{E} - \omega_0^2 \bm{P} \\
    \partial_t Q &= \{Q, \overline{H} \} = \sigma \\
    \partial_t \sigma &= \{\sigma, \overline{H} \} = \omega_v^2 | \bm{E} |^2 - \omega_v^2 Q \,.
\end{split}
\end{equation}
where $(\nabla \times)^*$ denotes the adjoint curl operator. 
\end{theorem}
\noindent \textit{Proof:} This is a consequence of the fact that for any observable, $F$, its evolution is prescribed by $\dot{F} = \{F, \overline{H}\}$. \qed

In the case of homogeneous boundary conditions, the curl is self-adjoint. Hence, these are precisely the equations of motion from the original model, and this Hamiltonian and Poisson bracket generate the dynamics. If one wishes to accommodate non-homogeneous boundary conditions (such as absorbing or free boundary conditions), one must add boundary terms to the Hamiltonian in order to account for these loss mechanisms. 

\subsubsection{Dissipative dynamics}

It is typical to include absorption in the model by turning the Lorentz oscillators (the $(\bm{P}, \bm{J})$ and $(Q,\sigma)$ subsystems) into damped oscillators: 
\begin{equation}
\begin{split}
    \partial_t \bm{P} &= \bm{J} \,, \\
    \partial_t \bm{J} + \lambda_0 \bm{J} & = \omega_p^2 \bm{E} - \omega_0^2 \bm{P} \,,
\end{split}
\qquad
\begin{split}
    \partial_t Q &= \sigma \,, \\
    \partial_t \sigma + \lambda_v \sigma &= \omega_v^2 | \bm{E} |^2 - \omega_v^2 Q \,.
\end{split}
\end{equation}
With this change, energy is no longer conserved by the system. Rather, one can express the dissipative dynamics as a gradient flow generated by the Hamiltonian with the following symmetric bracket:
\begin{equation}
( F, G ) = - \frac{1}{\epsilon_0} \int_\Omega \left( 
    \omega_p^2 \lambda_0 \frac{\delta F}{\delta \bm{J}} \cdot \frac{\delta G}{\delta \bm{J}} 
    + \frac{2 \omega_v^2 \lambda_v}{a \theta} \frac{\delta F}{\delta \sigma} \cdot \frac{\delta G}{\delta \sigma} \right) \mathsf{d}^3 \bm{x} \,.
\end{equation}
For a general functional $F$, one then finds that its evolution is given by $\dot{F} = \{F, H\} + (F, H)$. Therefore, one finds that
\begin{equation} \label{eq:energy_dissipation_rate}
    \dot{H} = (H, H) = - \epsilon_0 \int \left( \frac{\lambda_0}{\omega_p^2} | \bm{J} |^2 + \frac{a \theta \lambda_v}{2 \omega_v^2} \sigma^2 \right) \mathsf{d}^3 \bm{x} \,.
\end{equation}
It will be found that it is straightforward to include this dissipation in the structure-preserving discretization framework. What is sought is a discretization such that the rate at which energy is dissipated in the spatially discrete model matches the known rate of the continuous model and which is Hamiltonian when one sets $\lambda_0 = \lambda_v = 0$.

\subsubsection{One- and two-dimensional models}

The model of Maxwell's equations in cubicly nonlinear media was formulated in three spatial dimensions, but for the purposes of this numerical study one- and two-dimensional simplifications of this model are discretized. These lower-dimensional models are obtained by assuming symmetries in the system. The full details of the discretization procedure will be presented when discussing the two-dimensional model. However, to avoid redundancy in exposition, only a sketch is provided in the discretization of the one-dimensional model which is analogous. 

A one-dimensional model is obtained by assuming $\bm{E} = E \hat{x}$ and $\bm{B} = B \hat{y}$ and that all fields be functions of $z$ only. From these assumptions, one obtains the Poisson bracket
\begin{multline} \label{eq:cont_1d_cubic_pb}
    \{F, G\} = 
    \int_\Omega \left[ - \left( \frac{\delta F}{\delta B} \partial_z \frac{\delta G}{\delta D} 
    - \frac{\delta G}{\delta B} \partial_z \frac{\delta F}{\delta D} \right) \right. \\
    \left. + \frac{\omega_p^2}{\epsilon_0} \left( \frac{\delta F}{\delta P} \frac{\delta G}{\delta J} - 
    \frac{\delta G}{\delta P} \frac{\delta F}{\delta J} \right)
    +  \frac{2 \omega_v^2}{\epsilon_0 a \theta} \left( \frac{\delta F}{\delta Q} \frac{\delta G}{\delta \sigma} - 
    \frac{\delta G}{\delta Q} \frac{\delta F}{\delta \sigma} \right) \right] \mathsf{d} z \,,
\end{multline}
and the Hamiltonian
\begin{equation} \label{eq:cont_1d_cubic_ham}
    \begin{aligned}
    H[E, B, P, J, Q, \sigma] 
    = 
    \frac{1}{2} \int_\Omega \bigg[ \epsilon_0 \bigg( \epsilon_\infty E^2 
    + \frac{3 a ( 1 - \theta) E^4}{2} + a \theta Q E^2 \\
    + \frac{\omega_0^2}{\omega_p^2} P^2 + \frac{1}{\omega_p^2}  J^2
    + \frac{a \theta}{2} Q^2 + \frac{a \theta}{2 \omega_v^2} \sigma^2 \bigg) 
    + \mu_0^{-1} B^2 \bigg] \mathsf{d} z \,.
    \end{aligned}
\end{equation}
These give rise to the equations of motion
\begin{equation}
\begin{split}
    \partial_t D &= \{ D, \overline{H} \} = \mu_0^{-1} \partial_z ^*B \\
    \partial_t B &= \{ B, \overline{H} \} = - \partial_z E \\
    \partial_t P &= \{P, \overline{H} \} = J
\end{split}
\qquad
\begin{split}
    \partial_t J &= \{ J, \overline{H} \} = \omega_p^2 E - \omega_0^2 P \\
    \partial_t Q &= \{Q, \overline{H} \} = \sigma \\
    \partial_t \sigma &= \{\sigma, \overline{H} \} = \omega_v^2 E^2 - \omega_v^2 Q \,,
\end{split}
\end{equation}
where $D = \epsilon_0 \left( \epsilon_\infty E + P + a(1 - \theta) E^3 + a \theta Q E \right)$ and, in the case of homogeneous boundary conditions, $\partial_z^* = - \partial_z$. 

This bracket has only two Casimir invariants (assuming homogeneous boundary conditions):
\begin{equation}
	C_D = \int_\Omega D \, \mathsf{d} z
	\quad \text{and} \quad
	C_B = \int_\Omega B \, \mathsf{d} z.
\end{equation}
That these Casimir invariants are conserved by the numerical implementation will be verified in section \ref{sec:numerical_results}. 

The two-dimensional model is obtained by assuming $\bm{B} = B \hat{z}$, $\bm{E} \cdot \hat{z} = 0$, and that all fields are functions of $(x,y)$ only. Note, one could have alternatively let $\bm{B} \cdot \hat{z} = 0$ and $\bm{E} = E \hat{z}$. From these assumptions, one obtains the Poisson bracket
\begin{multline}
    \{F, G\} = 
    \int_\Omega \left[ \left( \frac{\delta F}{\delta B} \msf{curl} \frac{\delta G}{\delta \bm{D} } - 
    \frac{\delta G}{\delta B} \msf{curl} \frac{\delta F}{\delta \bm{D} } \right) \right. \\
    \left. + \frac{\omega_p^2}{\epsilon_0} \left( \frac{\delta F}{\delta \bm{P}} \frac{\delta G}{\delta \bm{J}} - 
    \frac{\delta G}{\delta \bm{P}} \frac{\delta F}{\delta \bm{J}} \right)
    +  \frac{2 \omega_v^2}{\epsilon_0 a \theta} \left( \frac{\delta F}{\delta Q} \frac{\delta G}{\delta \sigma} - 
    \frac{\delta G}{\delta Q} \frac{\delta F}{\delta \sigma} \right) \right] \mathsf{d}^2 \bm{x} \,.
\end{multline}
where the two-dimensional scalar curl operator and its formal adjoint are given by
\begin{equation}
    \msf{curl} \bm{A} = \partial_x A_y - \partial_y A_x
    \quad \text{and} \quad
    \msf{curl}^*f = \bmsf{curl} f = 
    \begin{pmatrix}
    \partial_y f \\
    - \partial_x f
    \end{pmatrix} \,.
\end{equation}
The Hamiltonian is given by
\begin{equation}
    \begin{aligned}
    H[\bm{E}, \bm{B}, \bm{P}, \bm{J}, Q, \sigma] 
    = 
    \frac{1}{2} \int_\Omega \bigg[ \epsilon_0 \bigg( \epsilon_\infty | \bm{E} |^2 
    + \frac{3 a ( 1 - \theta) | \bm{E} |^4}{2} + a \theta Q | \bm{E} |^2 \\
    + \frac{\omega_0^2}{\omega_p^2} | \bm{P} |^2 + \frac{1}{\omega_p^2}  | \bm{J} |^2
    + \frac{a \theta}{2} Q^2 + \frac{a \theta}{2 \omega_v^2} \sigma^2 \bigg) 
    + \mu_0^{-1} B^2 \bigg] \mathsf{d}^2 \bm{x} \,.
    \end{aligned}
\end{equation}
As in the three-dimensional case, one defines a functional $K$ such that
\begin{equation}
    H = K - \int_\Omega \frac{\delta K}{\delta \bm{E}} \cdot \bm{E} \mathsf{d}^2 \bm{x} 
    + \frac{1}{2} \int_\Omega \left( \epsilon_0 | \bm{E} |^2 + \mu_0^{-1} B^2 \right) \mathsf{d}^2 \bm{x} \,.
\end{equation}
The $K$ functional is more or less identical to that in the 3D case, see equation \ref{eq:cubic_3d_matter_model}, and one finds that
\begin{equation}
    \bm{D} = \bm{E} - \frac{\delta K}{\delta \bm{E}} = \epsilon_0 \left( \epsilon_\infty \bm{E} + \bm{P} + a(1 - \theta) | \bm{E} |^2 \bm{E} + a \theta Q \bm{E} \right) \,.
\end{equation}

The Poisson bracket and Hamiltonian give rise to the equations of motion
\begin{equation}
\begin{split}
    \partial_t \bm{D} &= \{ \bm{D}, \overline{H} \} = \mu_0^{-1} \msf{curl}^* B \\
    \partial_t B &= \{ B, \overline{H} \} = - \msf{curl} \bm{E} \\
    \partial_t \bm{P} &= \{ \bm{P}, \overline{H} \} = \bm{J}
\end{split}
\qquad
\begin{split}
    \partial_t \bm{J} &= \{ \bm{J}, \overline{H} \} = \omega_p^2 \bm{E} - \omega_0^2 \bm{P} \\
    \partial_t Q &= \{Q, \overline{H} \} = \sigma \\
    \partial_t \sigma &= \{\sigma, \overline{H} \} = \omega_v^2 | \bm{E} |^2 - \omega_v^2 Q \,,
\end{split}
\end{equation}
where, in the case of homogeneous boundary conditions, $\bmsf{curl}^* = \msf{curl}$. 

Under the assumption of homogeneous boundary conditions, this model possesses several conservation laws. Recall, $\msf{curl} \circ \bmsf{grad} = 0$. Hence, one finds that $\partial_t \msf{grad}^* \bm{D} = \mu_0^{-1} \msf{grad}^* \msf{curl}^* B = 0$. Therefore, the bracket possesses a Casimir invariant corresponding to a weak version of Gauss's law and a conservation law for the magnetic field:
\begin{equation}
    C_D = \left( \bmsf{grad} \phi,  \bm{D} \right)_{L^2(\Omega)} \quad \forall \phi \in H^1(\Omega),
    \quad \text{and} \quad
    C_B = \int_\Omega B \mathsf{d}^2 \bm{x} \,.
\end{equation}
While $C_D$ clearly comes from the 3D Gauss law, there is no Gauss law for the magnetic field in this 2D setting. Rather, $C_B$ arises from the fact that the magnetic field is a scalar conservation law.

\subsubsection{Model in dimensionless units}
It is convenient for the purposes of simulation to non-dimensionalize the system. One begins by defining a characteristic length scale $L$, and letting space and time be non-dimensionalized as
\begin{equation}
    t \mapsto \frac{c t}{L}
    \quad \text{and} \quad
    \bm{x} \mapsto \frac{\bm{x}}{L} \,,
\end{equation}
where $c = (\mu_0 \epsilon_0)^{-1/2}$ is the speed of light in vacuum. Hence, light travels at unit speed in dimensionless units. Let $A$ be a characteristic field strength. The dimensionless fields are given by
\begin{equation}
    \bm{E} \mapsto \frac{\bm{E}}{A} \,,
    \quad 
    \bm{B} \mapsto \frac{\mu_0 \bm{B}}{A} \,,
    \quad \text{and} \quad
    \bm{D} \mapsto \frac{\bm{D}}{\epsilon_0 A} \,.
\end{equation}
Let $(\bm{P}, \bm{J}, Q, \sigma)$ scale as
\begin{equation}
    \bm{P} \mapsto \frac{\bm{P}}{A} \,,
    \quad
    \bm{J} \mapsto \frac{L \bm{J}}{A c} \,,
    \quad
    Q \mapsto \frac{Q}{A^2}
    \quad \text{and} \quad
    \sigma \mapsto \frac{L \sigma}{A^2 c} \,.
\end{equation}
Finally, the parameters are re-scaled as
\begin{equation}
    \omega_p^2 \mapsto \frac{L^2 \omega_p^2}{c^2} \,,
    \quad 
    \omega_0^2 \mapsto \frac{L^2 \omega_0^2}{c^2} \,,
    \quad
    \omega_v^2 \mapsto \frac{L^2 \omega_v^2}{c^2} \,,
    \quad \text{and} \quad
    a \mapsto A^2 a \,.
\end{equation}
Note, $\epsilon_\infty$ and $\theta$ were already dimensionless and remain the same. Under these assumptions, the dimensionless system is written
\begin{equation} \label{eq:nondim_nonlinear_maxwell}
\begin{aligned}
    \partial_t \bm{B} &= - \nabla \times \bm{E}, \quad 
    \partial_t \bm{D} = \nabla \times \bm{B} \\
    \bm{D} &= \epsilon_\infty \bm{E} + \bm{P} + a(1 - \theta) | \bm{E} |^2 \bm{E} + a \theta Q \bm{E} \\
    \partial_t \bm{P} &= \bm{J}, \quad 
    \partial_t \bm{J} = \omega_p^2 \bm{E} - \omega_0^2 \bm{P} \\
    \partial_t Q &= \sigma, \quad 
    \partial_t \sigma = \omega_v^2 | \bm{E} |^2 - \omega_v^2 Q \,.
\end{aligned}
\end{equation}
Hence, there are six independent parameters in this model: $(\epsilon_\infty, a, \theta, \omega_p, \omega_0, \omega_v)$. These dimensionless units will be used throughout the remainder of this paper. Notationally, this is accomplished by simply letting $\epsilon_0 = \mu_0 = 1$ in the equations. 

%% file: feec_overview.tex
\section{A brief overview of FEEC methods and the discretization of functional derivatives} \label{sec:general_feec_overview}

Let $\Omega \subset \mathbb{R}^3$ be the physical domain. The central idea of a FEEC method is that the follow diagram of vector spaces (linked together by the standard vector calculus differential operators) should commute:
\begin{equation}
	\begin{tikzcd}
		V^0 = H^1_0(\Omega) \arrow{r}{\nabla} \arrow{d}{\Pi^0} 
			& V^1 = H_0(\text{curl}, \Omega)  \arrow{r}{\nabla \times} \arrow{d}{\Pi^1} 
				& V^2 = H_0(\text{div}, \Omega) \arrow{r}{\nabla \cdot} \arrow{d}{\Pi^2} 
					& V^3 = L^2(\Omega) \arrow{d}{\Pi_3} \\
		V^0_h \arrow{r}{\nabla_h} 
			& V^1_h \arrow{r}{\nabla \times_h} 
				& V^2_h \arrow{r}{\nabla \cdot_h} 
					& V^3_h
	\end{tikzcd}
\end{equation}
where $V^\ell_h \subset V^\ell$ are the finite element subspaces and $\Pi^\ell: V^\ell \to V^\ell_h$ are projection operators. The commutativity of this diagram ensures that the vector calculus differential operators in the finite element subspace retain their nilpotency conditions as in the continuous setting: i.e. $\text{ran}(\nabla_h) \subset \text{ker}(\nabla \times_h)$ and $\text{ran}(\nabla \times_h) \subset \text{ker}(\nabla \cdot_h)$. 

\subsection{FEEC method: A notational overview}
It would be cumbersome and distract from the purpose of this article to discuss the fine details underlying the FEEC method and its construction. Instead, the interested reader is directed to \cite{ArnoldDouglasN2010Feec} for a discussion of the theory of FEEC methods, and to \cite{campos-pinto_and_guclu_2021, güçlü2022broken} for a discussion of extensions of the method to admit broken basis functions. In what follows, only as much detail is included as is needed to establish notation. The general discretization approach described in this section should be compatible with any FEEC or broken-FEEC method. This work seeks to follow the notational conventions used in \cite{campos-pinto_and_guclu_2021, güçlü2022broken}. 

\begin{definition}
Let the \textbf{continuous spaces of $\ell$-forms} be $V^\ell \subset L^2 \Lambda^\ell(\Omega)$ (that is $V^\ell$ is a (possibly more regular) subspace of the square-integrable $\ell$-forms), let $V^\ell_h$ be the \textbf{finite element space} approximating $V^\ell$ ($V^\ell_h \subset V^\ell$ if a conforming FEEC method is being used), and let $\mathcal{C}_\ell \subset \mathbb{R}^{N_\ell}$ be the space of \textbf{Galerkin coefficients} where $N^\ell \in \mathbb{N}^+$ is the number of degrees of freedom associated with $V^\ell_h$. 

The \textbf{degrees of freedom operators} are linear operators such that $\bm{\sigma}^\ell = (\sigma^\ell_i)_{1 \leq i \leq N_\ell}: V^\ell \to \mathcal{C}_\ell$. These are generally associated with an integrals over given geometric entities of the domain: nodes, edges, faces, and cells. The \textbf{Interpolation operators} are defined such that ${\mathcal{I}}^\ell = \left( \left. {\bm{\sigma}}^\ell \right|_{V_h^\ell} \right)^{-1} : \mathcal{C}_\ell \to V^\ell_h$. The \textbf{basis functions} $\{ {\Lambda}_{\ell,i} \}_{i=1}^{N_\ell}$ are such that $\forall \bmsf{c} = (\mathsf{c}_i)_{1 \leq i \leq N_\ell} \in \mathcal{C}_\ell$, 
\begin{equation}
    \mathcal{I}^\ell \bmsf{c} = \sum_{i = 1}^{N_\ell} \mathsf{c}_i \Lambda_{\ell,i} \,.
\end{equation}
One defines the \textbf{projection operators} $\Pi^{\ell} = \mathcal{I}^{\ell} \circ \bm{\sigma}^{\ell}: V^{\ell} \to V^{\ell}_h$. These are sometimes referred to as the geometric or conforming projection operators. Hence,
\begin{equation}
    \Pi^{\ell} \phi = \sum_{i=1}^{N_{\ell}} \sigma^{\ell}_i(\phi) \Lambda_{\ell,i} \,.
\end{equation}
One requires that $\sigma^{\ell}_i ( \Pi^{\ell} \phi) = \sigma^{\ell}_i (\phi)$ so that this operator is indeed a projection. The \textbf{discrete exterior derivatives} are matrices $\mathbbm{d}_\ell: \mathcal{C}_\ell \to \mathcal{C}_{\ell + 1}$ defined by
\begin{equation}
	(\mathbbm{d}_\ell)_{ij} = \sigma^{\ell+1}_i ( \mathsf{d}_\ell \Lambda_{\ell,j} ) \,.
\end{equation}

The essential features of the \textbf{finite element de Rham complex} may be succinctly summarized in the following commuting diagram:
\begin{equation}
	\begin{tikzcd}
	\cdots \arrow{r} & 
		V^{\ell} \arrow{r}{\mathsf{d}_{\ell}} \arrow{d}{\bm{\sigma}^{\ell}} \arrow[bend right=35,swap]{dd}{\Pi^{\ell}} & 
			V^{{\ell}+1} \arrow[swap]{d}{\bm{\sigma}^{{\ell}+1}} \arrow[bend left=35]{dd}{\Pi^{{\ell}+1}} \arrow{r} & 
				\cdots \\
	& \mathcal{C}_{\ell} \arrow{r}{\mathbbm{d}_{\ell}} \arrow{d}{\mathcal{I}^{\ell}} & 
		\mathcal{C}_{{\ell}+1} \arrow[swap]{d}{\mathcal{I}^{{\ell}+1}} & \\
	\cdots \arrow{r} & 
		V^{\ell}_h \arrow{r}{\mathsf{d}_{\ell}} & 
			V^{{\ell}+1}_h \arrow{r} & 
				\cdots
	\end{tikzcd}
\end{equation}
For the diagram of finite element spaces to be commutative, one requires that
\begin{equation}
	\mathbbm{d}_{\ell} \bm{\sigma}^{\ell} = \bm{\sigma}^{{\ell}+1} \mathsf{d}_{\ell}, 
	\quad 
	\mathsf{d}_{\ell} \mathcal{I}^{\ell} = \mathcal{I}^{{\ell}+1} \mathbbm{d}_{\ell}, 
	\quad \text{and} \quad
	\mathsf{d}_{\ell} \Pi^{\ell} = \Pi^{{\ell}+1} \mathsf{d}_{\ell} \,.
\end{equation}
\end{definition}
Commutativity of the operators ensures the preservation of vector calculus identities at the discrete level: i.e. commutativity ensures
\begin{equation}
	\text{ran}(\mathsf{d}_\ell) \subset \text{ker}(\mathsf{d}_{\ell + 1}) \implies \text{ran}(\mathbbm{d}_\ell) \subset \text{ker}(\mathbbm{d}_{\ell + 1}).
\end{equation}
One further assumes that the projection operator is a bounded approximation of the identity in $V^\ell$:
\begin{equation}
    \| I - \Pi^\ell \| := \sup_{u \in V^\ell} \frac{\| u - \Pi^\ell u \|_{V^\ell}}{\| u \|_{V^\ell}} = O(h^{p+1}) \,,
\end{equation}
where (roughly speaking) $h$ is interpreted as the grid size and $p$ as the order of polynomial used for interpolation. Whether this is satisfied is of course dependent on the choice of $V^\ell$, but this work will not be concerned with these details. It is sufficient to note that ensuring the boundedness of the conforming projection operators requires that one restrict to sufficiently regular domain spaces on which the conforming degrees of freedom operators are bounded \cite{campos-pinto_and_guclu_2021}. 

\subsubsection{The dual de Rham complex} \label{section:dual_de-rham}

The abstract formulation of the Hamiltonian structure to be discussed subsequently in section \ref{subsection:general_nonlinear_maxwell} requires that one consider a dual de Rham complex (the complex for the adjoint differential operators). Once the primal complex has been defined, one defines the dual complex in a subsidiary manner entirely in terms of quantities previously defined. 

\begin{definition}
The \textbf{dual degrees of freedom} $\bm{\sigma}_{\ell}^*: (V^{\ell})^* \to \mathcal{C}_{\ell}^*$ are defined such that 
\begin{equation}
	\sigma_{{\ell},\bm{I}}^*(u) = \left\langle u, \Lambda_{{\ell},\bm{I}} \right\rangle
	\quad \text{for} \quad
	u \in (V^\ell)^*
\end{equation} 
where $\langle \cdot, \cdot \rangle: (V^\ell)^* \times V^\ell \to \mathbb{R}$ denotes the evaluation pairing of a vector space with its dual. One defines the \textbf{dual basis}, $\{ \Lambda_{{\ell},\bm{I}}^*\}_{\bm{I} \in \mathcal{M}_\ell}$, such that $\left( \Lambda_{{\ell},\bm{I}}, \Lambda_{{\ell},\bm{J}}^* \right) = \delta_{\bm{I}, \bm{J}}$. The \textbf{dual interpolation operator} $\mathcal{I}_{\ell}^*: \mathcal{C}_{\ell}^* \to (V_h^{\ell})^*$ is defined such that
\begin{equation}
	\mathcal{I}_{\ell}^* \bmsf{c}_* = \sum_{\bm{I} \in \mathcal{M}_\ell} (\msf{c}_*)_i \left( \Lambda_{{\ell},i}^*, \cdot \right).
\end{equation}
The \textbf{mass matrix} is defined $(\mathbb{M}_{\ell})_{\bm{I}, \bm{J}} = ( \Lambda_{{\ell},\bm{I}}, \Lambda_{{\ell},\bm{J}} ): \mathcal{C}_\ell \to \mathcal{C}_\ell^*$. The \textbf{dual projection operator} is defined such that 
\begin{equation}
	\Pi_{\ell}^* = \mathcal{I}_{\ell}^* \circ \bm{\sigma}_{\ell}^*: (V^{\ell})^* \to (V^{\ell}_h)^*.
\end{equation}
One can see that this is simply the $L^2$ projection. 
\end{definition}

\begin{prop} \label{prop:primal_dual_relationship}
The mass matrix provides the change of basis from $\mathcal{C}_{\ell}$ to $\mathcal{C}_{\ell}^*$. That is, if $u \in V^\ell_h$ and one denotes its corresponding element in the dual space $u^*_h = (u_h, \cdot) \in (V^\ell_h)^*$, then
\begin{equation}
	\mathbb{M}_\ell \bm{\sigma}_\ell(u_h) =  \bm{\sigma}_\ell^*(u^*_h).
\end{equation}
\end{prop}
\noindent \textit{Proof:} 
\begin{equation}
	\begin{aligned}
		\bm{\sigma}_\ell^*(u^*_h) 
			&= \left\langle u^*_h, \Lambda_{{\ell},\bm{I}} \right\rangle
				= \left( u_h,  \Lambda_{{\ell},\bm{I}} \right)
				= \sum_{\bm{J}} \left( \Lambda_{\ell, \bm{I}}, \Lambda_{\ell, \bm{J}} \right) \msf{u}_{\bm{J}}
				= \mathbb{M}_\ell \bm{\sigma}_\ell(u_h).
	\end{aligned}
\end{equation}
\qed

The primal and dual de Rham complexes together are frequently called the double de Rham complex. The discrete double de Rham complex may be schematically represented by
\begin{equation}
	\begin{tikzcd}[column sep=4em, row sep=2em]
		V^{\ell} \arrow{d}{\mathsf{d}_{\ell}} \arrow{r}{\Pi^\ell}
			& V^\ell_h \arrow{d}{ \mathsf{d}_{\ell,h}} \arrow{r}{ \mathcal{R}_{\ell} }
				& (V^\ell_h)^* \arrow[shift left=1.5]{l}{ \mathcal{R}_{\ell}^{-1} } 
					& (V^{\ell})^* \arrow{l}{ \Pi_\ell^* }
		\\
		V^{{\ell}+1} \arrow{r}{\Pi^{{\ell}+1}} 
			& V^{{\ell}+1}_h \arrow{r}{ \mathcal{R}_{{\ell}+1} } 
				& (V^{{\ell}+1}_h)^* \arrow[swap]{u}{ \mathsf{d}_{\ell+1,h}^* }  \arrow[shift left=1.5]{l}{ \mathcal{R}_{{\ell}+1}^{-1}} 
					& (V^{{\ell}+1})^* \arrow[swap]{u}{ \mathsf{d}^*_{{\ell}+1} } \arrow{l}{ \Pi_{\ell+1}^*} 
	\end{tikzcd} \\
\end{equation}
where $\mathcal{R}_\ell: V^\ell_h \to (V^\ell_h)^*$ is the isomorphism guaranteed by the Reisz representation theorem, or, at the coefficient level,
\begin{equation}
	\begin{tikzcd}[column sep=4em, row sep=2em]
		V^{\ell} \arrow{d}{\mathsf{d}_{\ell}} \arrow{r}{\bm{\sigma}^{\ell}}
			& \mathcal{C}_{\ell} \arrow{d}{ \mathbbm{d}_{{\ell}}} \arrow{r}{ \mathbb{M}_{\ell} }
				& \mathcal{C}_{\ell}^* \arrow[shift left=1.5]{l}{ \mathbb{M}_{\ell}^{-1} } 
					& (V^{\ell})^* \arrow{l}{ \bm{\sigma}_{{\ell}}^* }
		\\
		V^{{\ell}+1} \arrow{r}{\bm{\sigma}^{{\ell}+1}} 
			& \mathcal{C}_{{\ell}+1}  \arrow{r}{ \mathbb{M}_{{\ell}+1} } 
				& \mathcal{C}_{{\ell}+1}^* \arrow[swap]{u}{ \mathbbm{d}_{{\ell}}^* }  \arrow[shift left=1.5]{l}{ \mathbb{M}_{{\ell}+1}^{-1}} 
					& (V^{{\ell}+1})^* \arrow[swap]{u}{ \mathsf{d}^*_{{\ell}+1} } \arrow{l}{ \bm{\sigma}_{{\ell}+1}^*} 
	\end{tikzcd}
\end{equation}
where each diagram commutes. One can see that the discrete integration by parts formula is given by $(\mathbbm{d}^*_{{\ell}})^T \mathbb{M}_{{\ell}} = \mathbb{M}_{{\ell}+1} \mathbbm{d}_{\ell}$. 

\subsection{Discrete functional derivatives}
As Hamiltonian field theories are stated in terms of functional derivatives, in order to discretize the Hamiltonian structure directly, it is necessary to consider what happens when a functional is restricted to a finite dimensional subspace (e.g. a finite element subspace). Let $(U, \| \cdot \|_U)$ be a normed vector space. For any functional $K: U \to \mathbb{R}$, one defines its Fr{\'e}chet derivative at $u \in U$ (if it exists) to be the linear functional $DK[u]$ such that
\begin{equation}
	\frac{ \| K[u + \delta u] - K[u] - DK[u] \delta u \|_U }{ \| \delta u \|_U } = O( \| \delta u \|_U ).
\end{equation}
This paper will only consider Fr{\'e}chet differentiable functionals. When $U = \mathbb{R}^n$, this reduces to the usual definition of the Jacobian matrix. Hence, this definition generalizes the notion of derivatives to arbitrary normed vector spaces (e.g. function spaces). Two common notations for the functional derivative $DK[u]$ and $\delta K/\delta u$ will be used interchangeably. 

Derivatives with respect to functions which are Galerkin expansions in the finite element basis reduce to finite dimensional derivatives with respect to the Galerkin coefficients. 
\begin{theorem} \label{thm:disc_func_deriv}
Let $K: V^{\ell} \to \mathbb{R}$ be an arbitrary functional on the continuous ${\ell}$-forms and let $\msf{K} := K \circ \mathcal{I}^{\ell}: \mathcal{C}_{\ell} \to \mathbb{R}$ represent the discrete analog of the functional $K$. Moreover, define $\bmsf{u} = \bm{\sigma}^{\ell}(u)$ and $\bmsf{v} = \bm{\sigma}^{\ell}(v)$. Then 
\begin{equation}
	\begin{aligned}
		\left\langle D(K \circ \Pi^{\ell})[u], v \right\rangle_{(V^{\ell})^*, V^{\ell}} 
			&= \left\langle D(K \circ \mathcal{I}^{\ell} )[ \bm{\sigma}^{\ell}(u) ], \bm{\sigma}^{\ell}(v) \right\rangle_{\mathcal{C}_{\ell}^*, \mathcal{C}_{\ell}} \\
			&= \left\langle D\msf{K}[ \bmsf{u} ], \bmsf{v} \right\rangle_{\mathcal{C}_{\ell}^*, \mathcal{C}_{\ell}} 
				=\left( \frac{\partial \msf{K}}{\partial \bmsf{u}} \right)^T \bmsf{v}.
	\end{aligned}
\end{equation}
Similarly, if one has a functional of the dual de Rham sequence, $K_*: (V^{\ell})^* \to \mathbb{R}$, defining $\msf{K}_* = K \circ \mathcal{I}_{\ell}^*$, $\bmsf{u}_* = \bm{\sigma}_{\ell}^*(u)$, and $\bmsf{v}_* = \bm{\sigma}_{\ell}^*(v)$, then
\begin{equation}
	\begin{aligned}
		\left\langle D(K_* \circ \Pi_{\ell}^*)[u], v \right\rangle_{V^{\ell}, (V^{\ell})^*} 
			&= \left\langle D(K_* \circ \mathcal{I}_{\ell}^* )[ \bm{\sigma}_{\ell}^*(u) ], 
				\bm{\sigma}_{\ell}^*(v) \right\rangle_{\mathcal{C}_{\ell}, \mathcal{C}_{\ell}^*} \\
			&= \left\langle D\msf{K}_*[ \bmsf{u}_* ], \bmsf{v}_* \right\rangle_{\mathcal{C}_{\ell}, \mathcal{C}_{\ell}^*} 
				= \left( \frac{\partial \msf{K}_*}{\partial \bmsf{u}_*} \right)^T \bmsf{v}_*.
	\end{aligned}
\end{equation}
\end{theorem}
\noindent \textit{Proof:} The result follows from the chain rule and the linearity of the degrees of freedom operator. By the chain rule, one has
\begin{equation}
	\left\langle D(K \circ \Pi^{\ell})[u], v \right\rangle_{(V^{\ell})^*, V^{\ell}} 
			= \left\langle D(K \circ \mathcal{I}^{\ell} )[ \bm{\sigma}^{\ell}(u) ] \circ D \bm{\sigma}^\ell[u], v \right\rangle_{(V^\ell)^*, V^{\ell}}.
\end{equation}
But by the linearity of $\bm{\sigma}^\ell$, $D \bm{\sigma}^\ell[u] = \bm{\sigma}^\ell$. Hence, one finds
\begin{equation}
    \left\langle D(K \circ \Pi^{\ell})[u], v \right\rangle_{(V^{\ell})^*, V^{\ell}} 
		=  \left\langle D(K \circ \mathcal{I}^{\ell} )[ \bm{\sigma}^{\ell}(u) ], \bm{\sigma}^{\ell}(v) \right\rangle_{\mathcal{C}_{\ell}^*, \mathcal{C}_{\ell}}.
\end{equation}
The remainder of the result follows from simple notational manipulation. The result for $K_*$ follows from an entirely analogous argument.
\qed

Hence, one may directly translate continuous functional derivatives into discrete functional derivatives. Moreover, by appropriate interpretation of the discrete representation of the variables on the dual sequence, one may also prescribe discrete versions of functional derivatives with respect to variables in the dual space. 

\subsubsection{Adjoint of degrees of freedom operator}
This section proves a technical result which is needed to discretize the Poisson bracket. The ``$\dagger$" is used to denote operator adjoints because the notation ``$*$" is already employed. 

\begin{lemma} \label{lemma:disc_to_cont_duality}
Suppose $\bmsf{u}_* \in \mathcal{C}^*_\ell$ and $\bmsf{v} \in \mathcal{C}^\ell$. Then
\begin{equation}
	\left\langle \bmsf{u}_*, \bmsf{v} \right\rangle_{C^*_\ell, C^\ell} 
		= \left\langle \mathcal{I}_\ell^* \bmsf{u}_*, \mathcal{I}^\ell \bmsf{v} \right\rangle_{(V^\ell)^*, V^\ell}.
\end{equation}
\end{lemma}
\noindent \textit{Proof:} The result follows by simple application of the definitions:
\begin{equation}
	\begin{aligned}
	       \left\langle \mathcal{I}_\ell^* \bmsf{u}_*, \mathcal{I}^\ell \bmsf{v} \right\rangle_{(V^\ell)^*, V^\ell} 
			&= \left\langle \sum_{i=1}^{N_{\ell}} (\msf{u}_*)_i \left( \Lambda_{{\ell},i}^*, \cdot \right), 
				\sum_{j=1}^{N_{\ell}} \msf{v}_j \Lambda_{{\ell},j} \right\rangle_{(V^\ell)^*, V^\ell} \\
			&= \sum_{i,j=1}^{N_{\ell}} (\msf{u}_*)_i \msf{v}_j \left( \Lambda_{{\ell},i}^*, \Lambda_{{\ell},j} \right) 
				= \bmsf{u}_*^T \bmsf{v} := \left\langle \bmsf{u}_*, \bmsf{v} \right\rangle_{C^*_\ell, C^\ell}.
	\end{aligned}
\end{equation}
\qed

\begin{theorem}
With respect to the evaluation pairing, $\left( \left. \bm{\sigma}^\ell \right|_{V^\ell} \right)^\dagger = \mathcal{I}_\ell^*$. Moreover, 
\begin{equation}
    \frac{ \left| \left\langle [ (\bm{\sigma}^\ell)^\dagger - \mathcal{I}_\ell^* ] \bmsf{u}_*, v \right\rangle \right|}{ \| \bmsf{u}_* \|_{C^*_\ell} \| v \|_{V^\ell}} 
    \leq \| I - \Pi_\ell \| \| \mathcal{I}_\ell^* \| 
    \quad \forall \bmsf{u} \in \mathcal{C}^*_\ell 
    \text{ and } \forall v \in V^\ell \,.
\end{equation}
Hence, $\mathcal{I}_\ell^*$ approximates $\bm{\sigma}_\ell^\dagger$ in the following sense:
\begin{equation}
	\| (\bm{\sigma}^\ell)^\dagger - \mathcal{I}_\ell^* \| 
		:= \sup_{ \substack{\| \bmsf{u} \|_{C^*_\ell} \leq 1 \\ \bmsf{u}_* \in \mathcal{C}^*_\ell } } 
		    \sup_{ \substack{\| v \|_{V^\ell} \leq 1 \\ v \in V^\ell}} 
		    \frac{ \left| \left\langle ( \bm{\sigma}_\ell^\dagger - \mathcal{I}_\ell^*) \bmsf{u}_*, v \right\rangle \right|}{ \| \bmsf{u}_* \|_{C^*_\ell} \| v \|_{V^\ell}} 
    		= O(h^{p+1}).
\end{equation}
An analogous result holds for $\mathcal{I}^\ell$ and $\bm{\sigma}_\ell^*$.
\end{theorem}
\noindent \textit{Proof:} Let $v_h \in V^\ell_h$. Then by the previous lemma, 
\begin{equation}
    \left\langle \bmsf{u}_*, \bm{\sigma}^\ell(v_h) \right\rangle_{C^*_\ell, C^\ell} 
    	= \left\langle \mathcal{I}_\ell^* \bmsf{u}_*, \mathcal{I}^\ell \circ \bm{\sigma}^\ell(v_h) \right\rangle_{(V^\ell)^*, V^\ell} 
   	= \left\langle \mathcal{I}_\ell^* \bmsf{u}_*, v_h \right\rangle_{(V^\ell)^*, V^\ell}
\end{equation}
since $\mathcal{I}^\ell \circ \bm{\sigma}^\ell = \Pi^\ell = I$ on $V^\ell_h$. This proves the first claim. 

Now, let $v \in V^\ell$ be arbitrary. Then
\begin{equation}
	\left\langle \bmsf{u}_*, \bm{\sigma}^\ell(v) \right\rangle_{C^*_\ell, C^\ell} 
		= \left\langle \mathcal{I}_\ell^* \bmsf{u}_*, \Pi^\ell v \right\rangle_{(V^\ell)^*, V^\ell}.
\end{equation}
Hence,
\begin{equation}
    \left\langle \bmsf{u}_*, \bm{\sigma}^\ell(v) \right\rangle_{C^*_\ell, C^\ell} - \left\langle \mathcal{I}_\ell^* \bmsf{u}_*, v \right\rangle_{(V^\ell)^*, V^\ell} 
    	= - \left\langle \mathcal{I}_\ell^* \bmsf{u}_*, (I - \Pi^\ell) v \right\rangle_{(V^\ell)^*, V^\ell}
\end{equation}
which implies
\begin{equation}
	\begin{aligned}
		\left| \left\langle \bmsf{u}_*, \bm{\sigma}^\ell(v) \right\rangle_{C^*_\ell, C^\ell} 
			- \left\langle \mathcal{I}_\ell^* \bmsf{u}_*, v \right\rangle_{(V^\ell)^*, V^\ell} \right| 
			&= \left| \left\langle \mathcal{I}_\ell^* \bmsf{u}_*, (I - \Pi^\ell) v \right\rangle_{(V^\ell)^*, V^\ell} \right| \\
			&\leq \| I - \Pi^\ell \| \| \mathcal{I}_\ell^* \bmsf{u}_* \|_{(V^\ell)*} \| v \|_{V^\ell} \\
			&\leq \| I - \Pi^\ell \| \| \mathcal{I}_\ell^* \| \| \bmsf{u}_* \|_{(V^\ell)*} \| v \|_{V^\ell} 
	\end{aligned}
\end{equation}
by repeated application of the triangle inequality. Since $\mathcal{I}_\ell^*: C_\ell^* \to (V^\ell_h)^*$ is a map between finite dimensional spaces, it is bounded. Moreover, by assumption, $\| I - \Pi^\ell \| = O(h^{p+1})$. The result follows. \qed

This combined with the notion of discrete functional derivatives in the previous section yields the following result.
\begin{theorem} \label{thm:disc_func_deriv_2}
Let $K: V^\ell \to \mathbb{R}$. Define $\mathsf{K} := K \circ \mathcal{I}^\ell : \mathcal{C}_\ell \to \mathbb{R}$. Then
\begin{equation}
	\frac{\delta (K \circ \Pi^\ell) }{\delta u} = (\bm{\sigma}^\ell)^\dagger \frac{\partial \mathsf{K}}{\partial \bmsf{u} } = \mathcal{I}^*_\ell \frac{\partial \mathsf{K}}{\partial \bmsf{u}} + O(h^{p+1})
\end{equation}
where $\bmsf{u} = \bm{\sigma}^\ell(u)$. If a functional depends only on the finite dimensional space $V^\ell_h$, i.e. $K: V^\ell_h \to \mathbb{R}$, then
\begin{equation}
	\frac{\delta K}{\delta u_h} = \mathcal{I}^*_\ell \frac{\partial \mathsf{K} }{\partial \bmsf{u} }.
\end{equation}
A similar result holds for $K_*: (V^\ell)^* \to \mathbb{R}$. 
\end{theorem}
\noindent \textit{Proof:} As before, from the chain rule, one has
\begin{equation}
	\begin{aligned}
		\left\langle D(K \circ \Pi^{\ell})[u], v \right\rangle_{(V^{\ell})^*, V^{\ell}} 
			&= \left\langle D(K \circ \mathcal{I}^{\ell} )[ \bm{\sigma}^{\ell}(u) ], \bm{\sigma}^{\ell}(v) \right\rangle_{\mathcal{C}_{\ell}^*, \mathcal{C}_{\ell}} \\
			&= \left\langle (\bm{\sigma}^\ell)^\dagger \left( D \msf{K}[ \bmsf{u} ] \right), v \right\rangle_{(V^\ell)^*, V^\ell} \,.
	\end{aligned}
\end{equation}
Hence, simply changing notation, one finds that
\begin{equation}
	\frac{\delta (K \circ \Pi^\ell) }{\delta u} = (\bm{\sigma}^\ell)^\dagger \frac{\partial \mathsf{K}}{\partial \bmsf{u} }.
\end{equation}
The result follows from the previous theorem. \qed

%% file: general_discrete_theory.tex
\section{A Hamiltonian structure-preserving discretization of Maxwell's equations in general media} \label{sec:general_disc_theory}

Before specializing the discussion to study structure-preserving discretizations of Maxwell's equations cubicly nonlinear media in the latter part of this paper, it is illustrative to first demonstrate the procedure in general nonlinear media. 

\subsection{An exterior calculus formulation of Maxwell's equations in nonlinear media} \label{subsection:general_nonlinear_maxwell}

Maxwell's equations may be succinctly written in the language of exterior calculus, a perspective which is likewise helpful in the design of structure-preserving discretizations. To motivate this paper's approach to discretizing Maxwell's equations, it is helpful to restate the Hamiltonian structure of the macroscopic Maxwell equations in the language of exterior calculus. Because it would distract from the purpose of this article to offer an exposition of exterior calculus and its application to physical modeling, the interested reader is directed to other sources \cite{flanders_1963, frankel_2011}. See \cite{BARHAM2023100016} for the full details regarding this manner of writing the Hamiltonian structure of Maxwell's equations. 

One begins by simply restating the macroscopic Maxwell equations in the language of exterior calculus. Letting $V^\ell$ be a vector space of differential $\ell$-forms, one models the various fields as coming from the following spaces:
\begin{equation}
	\rho_f^0 \in V^0 \,,
	\quad
	\bm{E}^1 \,, \bm{D}^1 \,, \bm{J}_f^1 \in V^1 \,,
	\quad \text{and} \quad
	\bm{B}^2 \,, \bm{H}^2 \in V^2 \,,
\end{equation}
where each field is given a superscript to indicate the degree of differential form. Maxwell's equations are then written as follows:
\begin{equation}
\begin{aligned}[c]
    \partial_t \bm{B}^2 &= - c \mathsf{d}_1 \bm{E}^1 \,, \\
    \partial_t \bm{D}^1 &= c \mathsf{d}^*_2 \bm{H}^2 - 4 \pi \bm{J}_f^1 \,,
\end{aligned}
\quad \text{and} \quad
\begin{aligned}[c]
    \mathsf{d}^*_1 \bm{D}^1 &= 4 \pi \rho_f^0 \,, \\
    \mathsf{d}_2 \bm{B}^2 &= 0 \,.
\end{aligned}
\end{equation}
The operator $\mathsf{d}_\ell: V^\ell \to V^{\ell+1}$ is called the exterior derivative and its adjoint $\mathsf{d}_\ell^*: V^\ell \to V^{\ell-1}$ is frequently called the codifferential operator. One might prefer to make $\rho_f^0$ be a $3$-form and $\bm{J}^1_f$ be a $2$-form (so that these may be interpreted as scalar and vector densities respectively). However, if one wishes to state Faraday's law in strong form and Amp{\`e}re's law in weak form, it is necessary to write the equations as currently stated. It is, of course, possible to state Faraday's law weakly and Amp{\`e}re's law strongly, but this paper has arbitrarily taken the prior convention. See \cite{BARHAM2023100016} for further discussion of various ways of stating Maxwell's equations in the language of exterior calculus. 

The manner of writing the Hamiltonian structure for Maxwell's equations which most transparently suggests the approach to Hamiltonian structure-preserving discretization employed in this paper requires that one interpret the macroscopic fields, $(\bm{D}, \bm{H})$, as being elements of the dual space. The dual space of differential $k$-forms is sometimes referred to as the space of $k$-currents in differential geometry \cite{deRham1984}. Since functional derivatives naturally live in the dual vector space, and the polarization and magnetization vectors are defined in terms of functional derivatives, it is natural to let $\bm{D}_*^1 \in (V^1)^*$ and $\bm{H}_*^2 \in (V^2)^*$. 

Let $\mathcal{R}_\ell: V^\ell \to (V^\ell)^*$ denote the isomorphism between a vector space and its dual guaranteed by the Riesz representation theorem (note that $V^\ell$ is Hilbert). Define
\begin{equation} \label{eq:continuous_constitutive_relation}
    \bm{D}^1_* = \mathcal{R}_1 \bm{E}^1 - 4 \pi \frac{\delta K}{\delta \bm{E}^1} \in (V^1)^* 
    \quad \text{and} \quad 
    \bm{H}^2_* = \mathcal{R}_2 \bm{B}^2 - 4 \pi \frac{\delta K}{\delta \bm{B}^2} \in (V^2)^*
\end{equation}
where the functional derivatives are in $(V^\ell)^*$. Then one may write the Hamiltonian as
\begin{equation}
	H[\bm{E}^1, \bm{B}^2] = K - \left\langle \frac{\delta K}{\delta \bm{E}^1}, \bm{E}^1 \right\rangle_{(V^1)^*, V^1} 
    		+ \frac{1}{8 \pi} \left[ \left( \bm{E}^1, \bm{E}^1 \right)_{L^2} + \left( \bm{B}^2, \bm{B}^2 \right)_{L^2} \right] \,,
\end{equation}
and the Poisson bracket may be written
\begin{equation}
	\{ F, G \} = 4 \pi c \bigg[ \left\langle \frac{\delta G}{\delta \bm{B}^2}, \mathsf{d}_1 \frac{\delta F}{\delta \bm{D}^1_*} \right\rangle_{(V^2)^*, V^2}  
     		- \left\langle \frac{\delta F}{ \delta \bm{B}^2 }, \mathsf{d}_1 \frac{\delta G}{\delta \bm{D}^1_*} \right\rangle_{(V^2)^*, V^2}  \bigg]
		+ \{F, G\}_{\chi} \,,
\end{equation}
where $\delta F/\delta \bm{D}^1_* \in (V^1)^{**} \sim V^1$ (because Hilbert spaces are reflexive). This admittedly abstract manner of writing the Hamiltonian and Poisson bracket is a convenient starting place to derive the spatially discrete Hamiltonian structure. 

\subsection{A FEEC discretization of Maxwell's equations in nonlinear media}

It is appropriate to spatially discretize Maxwell's equations using a finite element exterior calculus (FEEC) method \cite{ArnoldDouglasN2010Feec}. In the following, the model for the additional degrees of freedom, denoted $\{\chi_k\}$, is neglected and only the electromagnetic portion of the model is considered. The manner by which the additional degrees of freedom must be discretized is problem dependent, and must be handled on a case-by-case basis. The discretization of the special case of cubicly nonlinear media will demonstrate this. 

\subsubsection{Discretization via directly applying degrees of freedom operators}
Before discussing a Hamiltonian approach to discretizing the macroscopic Maxwell equations, it is helpful to first consider a direct approach in which one applies the degrees of freedom (and dual degrees of freedom) operators to the dynamical equations in order to obtain a spatially semi-discretized system. 

\begin{prop}
Denote the degrees of freedom associated with a given field by $\bm{\sigma}^\ell (\bm{A}^\ell ) = \bmsf{a}^\ell$, and similarly $\bm{\sigma}^\ell_* (\bm{A}^\ell ) = \bmsf{a}^\ell_*$. Upon applying $\bm{\sigma}_2$ to Faraday's law and $\bm{\sigma}_1^*$ to Amp{\`e}re's law, one finds
\begin{equation}
	\begin{aligned}
		\partial_t \bmsf{b}^2 &= - c \mathbbm{d}_1 \bmsf{e}^1 , \\
		\partial_t \bmsf{d}^1_* &= c  \mathbbm{d}_1^* \bmsf{h}^2_*  - 4 \pi \bm{\sigma}^1_* ( \bm{J}_{f,*}^1 ) \,,
	\end{aligned}
\end{equation}
where $\bm{J}_{f,*}^1 \in (V^1)^*$ is a general free current. 
\end{prop}
\noindent \textit{Proof:} This follows from a straightforward application of the commutativity of the exterior derivative with the degrees of freedom operators:
\begin{equation}
	\begin{aligned}
		\partial_t \bm{\sigma}^2 ( \bm{B}^2 ) 
			&= - c \bm{\sigma}^2 ( \mathsf{d}_1 \bm{E}^1 ) 
			= - c \mathbbm{d}_1 \bm{\sigma}^1 ( \bm{E}^1 ), \\
		\partial_t \bm{\sigma}^1_* ( \bm{D}^1_* ) 
			&= c \bm{\sigma}^1_* ( \mathsf{d}^*_1 \bm{H}^2_* ) - 4 \pi \bm{\sigma}^1_* ( \bm{J}_{f,*}^1 ) 
			\\
                &= c  \mathbbm{d}_1^* \bm{\sigma}^2_* ( \bm{H}^2_* ) - 4 \pi \bm{\sigma}^1_* ( \bm{J}_{f,*}^1) \,,
	\end{aligned}
\end{equation}
where, as a matter of notation, the star subscripts mean $\bm{D}_*^1 \in (V^1)^*$ and $\bm{H}^2_* \in (V^2)^*$. 
\qed

\vspace{0.5em}

Recalling that $\bmsf{a}^\ell_* = \mathbb{M}_\ell \bmsf{a}^\ell$, which is the discrete analog of the isomorphism between $V^\ell$ and $(V^\ell)^*$, one finds
\begin{equation}
	\begin{aligned}
		\partial_t \bmsf{b}^2 &= - c \mathbbm{d}_1 \bmsf{e}^1 , \\
		\partial_t \bmsf{d}^1 &= c \mathbb{M}_1^{-1} \mathbbm{d}_1^T \mathbb{M}_2 \bmsf{h}^2  - 4 \pi \mathbb{M}_1^{-1} \bm{\sigma}^1_* ( \bm{J}_{f,*}^1 ).
	\end{aligned}
\end{equation}
This is nearly identical to what is usually found in FEEC discretizations of Maxwell's equations \cite{güçlü2022broken}. The free current is left in terms of the dual degrees of freedom to emphasize that this external model must be projected into the finite element space (if it is not already an element of the appropriate finite element space). What remains uncertain from this discretization procedure is the appropriate discrete constitutive model relating $(\bmsf{d}^1, \bmsf{b}^2)$ to $(\bmsf{e}^1, \bmsf{h}^2)$ such that this semi-discrete system is Hamiltonian. It is therefore beneficial to construct a discretization procedure based on the continuous Hamiltonian structure. 

\subsubsection{Discretization of the Hamiltonian structure}
The Hamiltonian structure is discretized by restricting all the functionals to act on appropriate finite element spaces. 
\begin{theorem}
Suppose one lets $\msf{H}[ \bmsf{e}^1, \bmsf{b}^2] = H[\bm{E}^1_h, \bm{B}^2_h]$ where $\bm{E}^1_h \in V^1_h$ and $\bm{B}^2_h \in V^2_h$ and
\begin{equation}
	\bm{e}_h^1 = \sum_i \msf{e}_i \Lambda_{1,i}
	\quad \text{and} \quad
	\bm{b}_h^2 = \sum_i \msf{b}_i \Lambda_{2,i}.
\end{equation}
Define a discrete analog of $K$ similarly. It follows that the discrete Hamiltonian is given by
\begin{equation}
    \msf{H}[\bmsf{e}^1, \bmsf{b}^2] 
	= \msf{K} - \left( \frac{\partial \msf{K}}{\partial \bmsf{e}^1} \right)^T \bmsf{e}^1 
    	+ \frac{1}{8 \pi} \left[ (\bmsf{e}^1)^T \mathbb{M}_1 \bmsf{e}^1 + (\bmsf{b}^2 )^T \mathbb{M}_2 \bmsf{b}^2 \right].
\end{equation}
\end{theorem}
\noindent \textit{Proof:} One immediately has that
\begin{equation}
	\frac{1}{8 \pi} \left[ \left( \bm{E}^1_h, \bm{E}^1_h \right)_{L^2} + \left( \bm{B}^2_h, \bm{B}^2_h \right)_{L^2} \right] 
		= \frac{1}{8 \pi} \left[ (\bmsf{e}^1)^T \mathbb{M}_1 \bmsf{e}^1 + (\bmsf{b}^2 )^T \mathbb{M}_2 \bmsf{b}^2 \right].
\end{equation}
Moreover, by theorem \ref{thm:disc_func_deriv}, one finds that
\begin{equation}
	\left\langle \frac{\delta K}{\delta \bm{E}^1_h}, \bm{E}^1_h \right\rangle_{(V^1)^*, V^1} = \left( \frac{\partial \msf{K}}{\partial \bmsf{e}^1} \right)^T \bmsf{e}^1.
\end{equation}
The result follows. \qed

\begin{theorem} \label{thm:discrete_constitutive_relations}
Let $(\bm{D}^1_{h,*}, \bm{H}^2_{h,*})$ be defined as in equation \ref{eq:continuous_constitutive_relation} except with the continuous electromagnetic fields replaced with their finite element counterparts, $(\bm{E}^1_h, \bm{B}^2_h)$. Define $\bmsf{d}^1_* = \bm{\sigma}_1^*(\bm{D}^1_{h,*})$ and $\bmsf{h}^2_* = \bm{\sigma}_2^*(\bm{H}^2_{h,*})$. Then the discrete constitutive relation is given by
\begin{equation} \label{eq:disc_const_laws}
	\bmsf{d}^1_* = \mathbb{M}_1 \bmsf{e}^1 - 4 \pi \frac{\partial \msf{K}}{\partial \bmsf{e}^1} 
	\quad \text{and} \quad 
	\bmsf{h}^2_* = \mathbb{M}_2 \bmsf{b}^2 + 4 \pi \frac{\partial \msf{K}}{\partial \bmsf{b}^2}.
\end{equation}
\end{theorem}
\noindent \textit{Proof:} One has that
\begin{equation}
	\begin{aligned}
		(\bmsf{d}^1_*)_i &= \bm{\sigma}_{1,i}^*(\bm{D}^1_{h,*}) 
			= \left\langle \mathcal{R}_1 \bm{E}^1_h - 4 \pi \frac{\delta K}{\delta \bm{E}^1_h}, \Lambda_{1, i} \right\rangle_{(V^1)^*, V^1} \\
			&= \left( \bm{E}^1_h, \Lambda_{1, i} \right)_{L^2} 
				- 4 \pi \left\langle \mathcal{I}^*_1 \frac{\partial \msf{K}}{\partial \bmsf{e}^1}, \Lambda_{1,i} \right\rangle_{(V^1)^*, V^1} 
			= \left( \mathbb{M}_1 \bmsf{e}^1 - 4 \pi \frac{\partial \msf{K}}{\partial \bmsf{e}^1} \right)_i.
	\end{aligned}
\end{equation}
One proceeds similarly with $\bmsf{h}^2_*$. \qed

\begin{theorem}
Let $(\bm{D}^1_{h,*}, \bm{H}^2_{h,*})$ and $(\bmsf{d}^1_*, \bmsf{h}^2_*)$ be defined as in theorem \ref{thm:discrete_constitutive_relations}. Let $\msf{F}[ \bmsf{d}^1_*, \bmsf{b}^2] = F[\bm{D}^1_{h,*}, \bm{B}^2_h]$ and similarly for $G$. One finds that the discrete Poisson bracket is given by
\begin{equation}
	[ \msf{F}, \msf{G} ] = 4 \pi c \left[ \left( \frac{\partial \msf{G}}{\partial \bmsf{b}^2} \right)^T \mathbbm{d}_1 \frac{\partial \msf{F}}{\partial \bmsf{d}^1_*} 
					- \left( \frac{\partial \msf{F}}{\partial \bmsf{b}^2} \right)^T \mathbbm{d}_1 \frac{\partial \msf{G}}{\partial \bmsf{d}^1_*} \right].
\end{equation}
\end{theorem}
\noindent \textit{Proof:} Using theorem \ref{thm:disc_func_deriv_2} and lemma \ref{lemma:disc_to_cont_duality}, one has that
\begin{equation}
	\begin{aligned}
		\left\langle \frac{\delta G}{\delta \bm{B}^2_h}, \mathsf{d}_1 \frac{\delta F}{\delta \bm{D}^1_{h,*}} \right\rangle_{(V^2)^*, V^2} 
		&= \left\langle \mathcal{I}_2^* \frac{\partial \msf{G}}{\partial \bmsf{b}^2}, 
			\mathsf{d}_1 \mathcal{I}^1 \frac{\partial \msf{F}}{\partial \bmsf{d}^1_*} \right\rangle_{\mathcal{C}^*_2, \mathcal{C}^2} \\
		&= \left\langle \mathcal{I}_2^* \frac{\partial \msf{G}}{\partial \bmsf{b}^2}, 
			\mathcal{I}^2 \mathbbm{d}_1 \frac{\partial \msf{F}}{\partial \bmsf{d}^1_*} \right\rangle_{\mathcal{C}^*_2, \mathcal{C}^2} 
		= \left( \frac{\partial \msf{G}}{\partial \bmsf{b}^2} \right)^T \mathbbm{d}_1 \frac{\partial \msf{F}}{\partial \bmsf{d}^1_*}. 
	\end{aligned}
\end{equation}
Analogous manipulations complete the proof. \qed

As in the continuous setting, the Hamiltonian and Poisson bracket are defined in terms of different variables. Hence, a change of variables is needed to obtain the equations of motion. 
\begin{lemma}
Let $\bmsf{e}^1 = \bmsf{e}^1(\bmsf{d}^1_*, \bmsf{b}^2)$ defined implicitly by the constitutive relations in equation \eqref{eq:disc_const_laws}. Then
\begin{equation}
	\frac{\partial \bmsf{e}^1}{\partial \bmsf{d}^1_*} = \left( \mathbb{M}_1 - 4 \pi \frac{\partial^2 \msf{K}}{\partial \bmsf{e}^1 \partial \bmsf{e}^1} \right)^{-1}
        \quad \text{and} \quad
	\frac{\partial \bmsf{e}^1}{\partial \bmsf{b}^2} 
	= 4 \pi \left( \mathbb{M}_1 - 4 \pi \frac{\partial^2 \msf{K}}{\delta \bmsf{e}^1 \partial \bmsf{e}^1}  \right)^{-1} \frac{\partial^2 \msf{K}}{\partial \bmsf{b}^2 \partial \bmsf{e}^1} \,.
\end{equation}
\end{lemma}
\noindent \textit{Proof:} Let $\Phi[ \bmsf{e}^1, \bmsf{b}^2] = (\bmsf{d}^1_*, \bmsf{b}^2)$. That is,
\begin{equation*}
	\Phi[ \bmsf{e}^1, \bmsf{b}^2] = \left( \mathbb{M}_1 \bmsf{e}^1 - 4 \pi \frac{\partial \msf{K}}{\partial \bmsf{e}^1}, \bmsf{b}^2 \right) \,.
\end{equation*}
Assume that $\msf{K}$ is such that $\Phi$ is a diffeomorphism. Hence, one also has that $\Phi^{-1}[ \bmsf{d}^1_*, \bmsf{b}^2] = (\bmsf{e}^1, \bmsf{b}^2)$. Differentiating, one finds that
\begin{equation*}
\begin{aligned}
    D \Phi [\bmsf{e}^1, \bmsf{b}^2](\delta \bmsf{e}^1, \delta \bmsf{b}^2) 
            &= 
    \begin{pmatrix}
        D_1 \Phi_1[\bmsf{e}^1,\bmsf{b}^2] & D_2 \Phi_1[\bmsf{e}^1, \bmsf{b}^2] \\
        0 & \mathbb{I}
    \end{pmatrix}
    \begin{pmatrix}
        \delta \bmsf{e}^1 \\
        \delta \bmsf{b}^2
    \end{pmatrix} \\
    &=
    \begin{pmatrix}
        \mathbb{M}_1 - 4 \pi \frac{\partial^2 \msf{K}}{\partial \bmsf{e}^1 \partial \bmsf{e}^1} & - 4 \pi \frac{\partial^2 \msf{K}}{\partial \bmsf{b}^2 \partial \bmsf{e}^1} \\
        0 & \mathbb{I}
    \end{pmatrix}
    \begin{pmatrix}
        \delta \bmsf{e}^1 \\
        \delta \bmsf{b}^2
    \end{pmatrix} \,.
\end{aligned}
\end{equation*}
Hence, it follows that
\begin{equation*}
    D \Phi^{-1}[\bmsf{d}^1_*,\bmsf{b}^2](\delta \bmsf{d}^1_*, \delta \bmsf{b}^2) 
    = 
    \begin{pmatrix}
	D_1 \Phi_1^{-1} & - D_1 \Phi_1^{-1} D_2 \Phi_1 \\
	0 & \mathbb{I}
    \end{pmatrix}
    \begin{pmatrix}
	\delta \bmsf{d}^1_* \\
	\delta \bmsf{b}^2
    \end{pmatrix}.
\end{equation*}
Computing the entries of this matrix, one finds
\begin{equation*}
    \frac{\partial \bmsf{e}^1}{\partial \bmsf{d}^1_*} = \left( \mathbb{M}_1 - 4 \pi \frac{\partial^2 \msf{K}}{\partial \bmsf{e}^1 \partial \bmsf{e}^1}  \right)^{-1}
    \text{and} \quad
    \frac{\partial \bmsf{e}^1}{\partial \bmsf{b}^2} = 4 \pi \left( \mathbb{M}_1 - 4 \pi \frac{\partial^2 \msf{K}}{\partial \bmsf{e}^1 \partial \bmsf{e}^1}  \right)^{-1} \frac{\partial^2 \msf{K}}{\partial \bmsf{b}^2 \partial \bmsf{e}^1} \,.
\end{equation*}
\qed

\begin{theorem}
Let $\overline{\msf{H}}[\bmsf{d}_*^1, \bmsf{b}^2] = H[\bmsf{e}^1, \bmsf{b}^2]$. Then
\begin{equation} \label{vec_ham_derivs}
    \frac{\partial \overline{\msf{H}}}{\partial \bmsf{d}^1_*} = \frac{\bmsf{e}^1}{4 \pi} 
    \quad \text{and} \quad 
    \frac{\partial \overline{\msf{H}}}{\partial \bmsf{b}^2} = \frac{\bmsf{h}^2_*}{4 \pi} \,.
\end{equation}
\end{theorem}
\noindent \textit{Proof:} Taking derivatives of $\msf{H}$ with respect to $(\bmsf{e}^1, \bmsf{b}^2)$, one finds
\begin{equation*}
    \frac{\partial \msf{H}}{\partial \bmsf{e}^1 } = \left( \mathbb{M}_1 - 4 \pi \frac{\partial^2 \msf{K}}{\partial \bmsf{e}^1 \partial \bmsf{e}^1} \right) \frac{\bmsf{e}^1}{4 \pi} 
    \quad\text{and} \quad
    \frac{\partial \msf{H}}{\partial \bmsf{b}^2 } = \frac{\partial \msf{K}}{\partial \bmsf{b}^2} - \left( \frac{\partial \msf{K}}{\partial \bmsf{b}^2 \partial \bmsf{e}^1} \right)^* \bmsf{e}^1 + \frac{\mathbb{M}_2 \bmsf{b}^2}{4 \pi} \,.
\end{equation*}
The chain rule implies
\begin{equation*}
    \frac{\partial \overline{\msf{H}}}{\partial \bmsf{d}_*^1} = \left( \frac{\partial \bmsf{e}^1 }{\partial \bmsf{d}^1_* } \right)^* \frac{\partial \msf{H}}{\partial \bmsf{e}^1} 
    = \left( \frac{\partial \bmsf{e}^1}{\partial \bmsf{d}^1_*} \right)^* \left( \mathbb{M}_1 - 4 \pi \frac{\partial \msf{K} }{\partial \bmsf{e}^1 \partial \bmsf{e}^1 } \right) \frac{\bmsf{e}^1}{4 \pi} 
    = \frac{\bmsf{e}^1}{4 \pi}.
\end{equation*}
Likewise, 
\begin{align*}
	\frac{\partial \overline{\msf{H}}}{\partial \bmsf{b}^2} 
	&= \frac{\partial \msf{H}}{\partial \bmsf{b}^2} + \left( \frac{\partial \bmsf{e}^1}{\partial \bmsf{b}^2} \right)^* \frac{\partial \msf{H}}{\partial \bmsf{e}^1} \\
	&= \frac{\partial \msf{K}}{\partial \bmsf{b}^2} - \left( \frac{\partial^2 \msf{K}}{\partial \bmsf{b}^2 \partial \bmsf{e}^1} \right)^* \bmsf{e}^1 + \frac{\mathbb{M}_2 \bmsf{b}^2}{4 \pi} 
        + \left( \frac{\partial \bmsf{e}^1}{\partial \bmsf{b}^2} \right)^* \left( \mathbb{M}_1 - 4 \pi \frac{\partial \msf{K}}{\partial \bmsf{e}^1 \partial \bmsf{e}^1} \right) \frac{\bmsf{e}^1}{4 \pi} \\
	&= \frac{ \mathbb{M}_2 \bmsf{b}^2}{4 \pi} + \frac{\partial \msf{K}}{\partial \bmsf{b}^2} = \frac{\bmsf{h}^2_*}{4 \pi}.
\end{align*}
\qed

This allows one to finally obtain equations of motion from the spatially semi-discretized Hamiltonian structure. 
\begin{theorem}
The spatially discrete macroscopic Maxwell equations are given by
\begin{equation}
	\begin{split}
		\partial_t \bmsf{d}^1_* &= c \mathbbm{d}^T_1 \bmsf{h}^2_*, \\
		\partial_t \bmsf{b}^2 &= -c \mathbbm{d}_1 \bmsf{e}^1,
	\end{split}
	\quad \text{where} \quad
        \begin{split}
            \bmsf{d}^1_* = \mathbb{M}_1 \bmsf{e}^1 - 4 \pi \frac{\partial \msf{K}}{\partial \bmsf{e}^1} \\
            \bmsf{h}^2_* = \mathbb{M}_2 \bmsf{b}^2 + 4 \pi \frac{\partial \msf{K}}{\partial \bmsf{b}^2}.   
        \end{split}
\end{equation}
Moreover, one has that
\begin{equation}
    \partial_t ( \mathbbm{d}_0^T \bmsf{d}^1_*) = 0 \quad \text{and} \quad \partial_t (\mathbbm{d}_2 \bmsf{b}^2) = 0 \,,
\end{equation}
which are the discrete analogs of Gauss's laws. 
\end{theorem}
\noindent \textit{Proof:} The equations of motion are obtained as follows:
\begin{equation}
	\begin{split}
		\dot{F} = [ \msf{F}, \overline{\msf{H}} ] = 4 \pi c \left[ \left( \bmsf{h}^2_* \right)^T \mathbbm{d}_1 \frac{\partial \msf{F}}{\partial \bmsf{d}^1_*} 
						- \left( \frac{\partial \msf{F}}{\partial \bmsf{b}^2} \right)^T \mathbbm{d}_1 \bmsf{e}^1 \right]
	\end{split}
	\implies
	\begin{split}
		\partial_t \bmsf{d}^1_* &= c \mathbbm{d}^T_1 \bmsf{h}^2_* \\
		\partial_t \bmsf{b}^2 &= -c \mathbbm{d}_1 \bmsf{e}^1.
	\end{split}
\end{equation}
The discrete Gauss's laws are implied as degeneracies of the Poisson bracket, Casimir invariants, since $\mathbbm{d}_0^T \mathbbm{d}_1^T = 0$ and $\mathbbm{d}_2 \mathbbm{d}_1 = 0$ (because $\text{ran}(\mathbbm{d}_\ell) \subset \text{ker}(\mathbbm{d}_{\ell+1})$ in a FEEC discretization). \qed 

\vspace{0.5em}

When $\msf{K} = 0$, the resulting semi-discrete equations are exactly what one obtains from traditional FEEC discretization of Maxwell's equations. What is gained from this approach is a prescription for the proper discretization of the constitutive laws. That is, given $K = K[\bm{E}^1, \bm{B}^2]$, one directly plugs in finite element expansions of the fields to obtain the discrete functional. Then, one takes derivatives with respect to the coefficients of the discretized functional to obtain the polarization and magnetization. 

%% file: discrete_model.tex
\section{A Hamiltonian structure-preserving discretization of Maxwell's equations in cubicly nonlinear media} \label{sec:disc_cubic_maxwell}

The spatial and temporal discretization of Maxwell's equations in cubicly nonlinear media are now investigated in detail. First, the spatial discretization is derived in the full three-dimensional case, before briefly sketching the details in two- and one-dimensional versions of the model. This work only consider simulations in the two- and one-dimensional settings. Temporal discretization is accomplished via Hamiltonian splitting to ensure that the time-stepper is symplectic, conserves Gauss's laws, and (being symplectic) conserves a discrete energy on average \cite{geometric_numerical_integration}. Because the time-stepper is explicit, a CFL condition is derived to ensure stability of the time-step. Finally, it is shown that the dissipative dynamics may be accommodated by a straightforward modification of the time-step.

\subsection{Discretization of full three-dimensional model}

Following the approach outlined in the previous section, the only choice that must be made in prescribing the spatial discretization for Maxwell's equations in nonlinear media is which finite element space in the finite element de Rham complex each field is to be approximated in. Vector fields are taken from the vector FEM spaces, $V_h^1$ or $V_h^2$, and scalar fields from the scalar FEM spaces, $V_h^0$ or $V_h^3$. It is also necessary to determine which fields are to be modeled on the primal de Rham complex, and which are to be modeled on the dual complex. This helps to determine which differential operators are discretized strongly and which are discretized weakly. One can see from the Poisson bracket that each field is paired with a corresponding dual field: $(\bm{D},\bm{B})$, $(\bm{P},\bm{J})$, and $(Q, \sigma)$. By examining the equations of motion, the Poisson bracket, and the Hamiltonian, one may infer the following:
\begin{multline} \label{eq:disc_3d_fields}
	\bm{D}_h = \sum_{i} \tilde{\msf{D}}_i \tilde{\bm{\Lambda}}_{1,i}(\bm{x}),
	\quad
	\bm{E}_h = \sum_i \msf{E}_i \bm{\Lambda}_{1,i}(\bm{x}),
  	\quad
	\bm{B}_h = \sum_i \msf{B}_i \Lambda_{2,i}(\bm{x}),
	\quad
	\bm{P}_h = \sum_{i} \msf{P}_i \bm{\Lambda}_{1,i}(\bm{x}), \\
	\bm{J}_h = \sum_{i} \tilde{\msf{J}}_i \tilde{\bm{\Lambda}}_{1,i}(\bm{x}), 
	\quad
	Q_h = \sum_i \msf{Q}_i \Lambda_{3,i}(\bm{x}),
	\quad \text{and} \quad
	\sigma_h = \sum_i \tilde{\msf{\upsigma}}_i \tilde{\Lambda}_{3,i}(\bm{x}).
\end{multline}
That is, $\bm{E}_h, \bm{P}_h \in V_h^1$, $\bm{D}_h, \bm{J}_h \in (V_h^1)^*$, $\bm{B}_h \in V_h^2$, $Q_h \in V_h^3$, and $\sigma_h \in (V_h^3)^*$. There is some flexibility in which finite element spaces one chooses each field to be modeled in, e.g. whether Faraday's law is weak and Amp{\`e}re's law is strong or vice versa, but the Poisson bracket implies that certain variables need to be dual to each other for consistency. However, once the above choice of finite element spaces has been made, the spatial discretization is fully determined: the Hamiltonian structure is discretized by restricting all functionals to act on fields in the appropriate finite element subspaces:
\begin{equation} \label{eq:disc_3d_functionals}
	F[\bm{D}_h, \bm{B}_h, \bm{P}_h, \bm{J}_h, Q_h, \sigma_h] = \msf{F}[ \tilde{\bmsf{D}}, \bmsf{B}, \bmsf{P}, \tilde{\bmsf{J}}, \bmsf{Q}, \tilde{\bmsf{\upsigma}}] \,,
\end{equation}
where the sans-serif notation on the righthand side is used consistently throughout this paper to denote the coefficient vectors and functions of these coefficient vectors. 

\begin{theorem} \label{eq:cubic_3d_disc_ham_system}
Modeling the fields as in equation \eqref{eq:disc_3d_fields} and reducing functionals of the fields to functions of the coefficient vectors as in equation \eqref{eq:disc_3d_functionals}, one obtains a finite dimensional Hamiltonian system. The discrete Poisson bracket is
\begin{multline} \label{eq:cubic_3d_2d_pb}
    [ \msf{F}, \msf{G} ] = 
    - \left[ \left( \frac{\partial \msf{F} }{\partial \bmsf{B} } \right)^T \mathbbm{d}_1 \frac{\partial \msf{G}}{\partial \tilde{\bmsf{D}} }
    - \left( \frac{\partial \msf{G} }{\partial \bmsf{B} } \right)^T \mathbbm{d}_1 \frac{\partial \msf{F}}{\partial \tilde{\bmsf{D}} } \right]
    + \omega_p^2 \left( 
    \left( \frac{\partial \msf{F} }{\partial \bmsf{P} } \right)^T \frac{\partial \msf{G} }{\partial \tilde{\bmsf{J}} }
    - \left( \frac{\partial \msf{G} }{\partial \bmsf{P} } \right)^T \frac{\partial \msf{F} }{\partial \tilde{\bmsf{J}} } 
    \right) \\
    +  \frac{2 \omega_v^2}{a \theta} \left( 
    \left( \frac{\partial \msf{F}}{\delta \bmsf{Q} } \right)^T \frac{\partial \msf{G} }{\partial \tilde{\bmsf{\upsigma}} } 
    - \left( \frac{\partial \msf{G}}{\delta \bmsf{Q} } \right)^T \frac{\partial \msf{F} }{\partial \tilde{\bmsf{\upsigma}} } 
    \right) \,,
\end{multline}
and the discrete Hamiltonian is
\begin{multline}
    \msf{H} [\bmsf{E}, \bmsf{B}, \bmsf{P}, \tilde{\bmsf{J}}, \bmsf{Q}, \tilde{\bmsf{\upsigma}}] 
    = 
    \frac{1}{2} 
    \bigg[ 
    \tilde{\bmsf{E}}_*^T \bmsf{E} 
    + \frac{\omega_0^2}{\omega_p^2} \bmsf{P}^T \mathbb{M}_1 \bmsf{P} 
    + \frac{1}{\omega_p^2}  \tilde{\bmsf{J}}^T \mathbb{M}_1^{-1} \tilde{\bmsf{J}} \\
    + \frac{a \theta}{2} \bmsf{Q}^T \mathbb{M}_3 \bmsf{Q} 
    + \frac{a \theta}{2 \omega_v^2} \tilde{\bmsf{\upsigma}}^T \mathbb{M}_3^{-1} \tilde{\bmsf{\upsigma}}
    + \bmsf{B}^T \mathbb{M}_2 \bmsf{B} 
    \bigg] \,,
\end{multline}
where
\begin{equation}
    (\tilde{\msf{E}}_*)_i = \left(\Lambda_{1,i}, \left( \epsilon_\infty 
    + \frac{3 a ( 1 - \theta) | \bm{E}_h |^2}{2} + a \theta Q_h \right) \bm{E}_h \right)_{L^2(\Omega)} \,.
\end{equation}
The discrete constitutive relation is given by
\begin{equation}
    \tilde{\msf{D}}_i = \left( \Lambda_{1,i}, \left( \epsilon_\infty \bm{E}_h + \bm{P}_h + a(1 - \theta) | \bm{E}_h |^2 \bm{E}_h + a \theta Q_h \bm{E}_h \right) \right)_{L^2(\Omega)} \,.
\end{equation}
\end{theorem}

For notational convenience, let $F_h = \left. F \right|_{V_h}$, i.e.\ $F_h$ is a functional which has been restricted to act on the finite element subspaces. As in the continuous setting, the Hamiltonian and Poisson bracket are most conveniently expressed in terms of different coordinates. Hence, one must perform a change of coordinates:
\begin{multline}
    \overline{H}_h[\bm{D}_h, B_h, \bm{P}_h, \bm{J}_h, Q_h, \sigma_h] 
    = H_h[\bm{E}_h, B_h, \bm{P}_h, \bm{J}_h, Q_h, \sigma_h] \\
    \quad \text{and} \quad
    \overline{\msf{H}} [\tilde{\bmsf{D}}, \bmsf{B}, \bmsf{P}, \tilde{\bmsf{J}}, \bmsf{Q}, \tilde{\bmsf{\upsigma}}]
    = \msf{H} [\bmsf{E}, \bmsf{B}, \bmsf{P}, \tilde{\bmsf{J}}, \bmsf{Q}, \tilde{\bmsf{\upsigma}}] \,.
\end{multline}
In order to recover the equations of motion, one must take derivatives of the Hamiltonian.

\begin{theorem}
The derivatives of the discrete Hamiltonian are given by
\begin{multline}
    \frac{\partial \overline{\msf{H}}}{\partial \tilde{\bmsf{D}} } = \bmsf{E},
    \quad
    \frac{\partial \overline{\msf{H}}}{\partial \bmsf{B}} = \mathbb{M}_2 \bmsf{B},
    \quad
    \frac{\partial \overline{\msf{H}}}{\partial \bmsf{P}} 
    = \mathbb{M}_1 \left( \frac{\omega_0^2}{\omega_P^2} \bmsf{P} - \bmsf{E} \right), 
    \quad 
    \frac{\partial \overline{\msf{H}}}{\partial \tilde{\bmsf{J}}} = \omega_p^{-2} \mathbb{M}_1^{-1} \tilde{\bmsf{J}}, \\
    \quad
    \frac{\partial \overline{\msf{H}}}{\partial \bmsf{Q}} 
    = \frac{a \theta}{2} \left( \mathbb{M}_3 \bmsf{Q} - \left( \bm{\Lambda}_{3}, | \bm{E}_h |^2 \right)_{L^2(\Omega)} \right),
    \quad \text{and} \quad
    \frac{\partial \overline{\msf{H}}}{\partial \tilde{\bmsf{\upsigma}}}
    = \frac{a \theta}{2 \omega_v^2} \mathbb{M}_3^{-1} \tilde{\bmsf{\upsigma}} \,.
\end{multline}
\end{theorem}
\noindent \textit{Proof:} One may replicate the procedure for taking derivatives of the Hamiltonian in the continuous case in section \ref{sec:cts_cubic_maxwell}, to find that
\begin{multline}
	\frac{\delta \overline{H}_h}{\delta \bm{D}_h} = \bm{E}_h \,,
	\quad
	\frac{\delta \overline{H}_h}{\delta B_h} = B_h \,,
	\quad
	\frac{\delta \overline{H}_h}{\delta \bm{P}_h} = \left( \frac{\omega_0^2}{\omega_P^2} \bm{P}_h - \bm{E}_h \right) \,, \\
	\quad 
	\frac{\delta \overline{H}_h}{\delta \bm{J}_h} = \omega_p^{-2} \bm{J}_h,
	\quad
	\frac{\delta \overline{H}_h}{\delta Q_h} = \frac{a \theta}{2} \left( Q_h - | \bm{E}_h |^2 \right) \,,
	\quad \text{and} \quad
	\frac{\delta \overline{H}_h}{\delta \sigma_h} = \frac{a \theta}{2 \omega_v^2} \sigma_h \,.
\end{multline}
The derivatives with respect to the coefficients are obtained as follows:
\begin{equation}
	\left( \frac{\delta \overline{H}_h}{\delta \bm{D}_h}, \delta \bm{D}_h \right)_{L^2(\Omega)} 
		= \left( \sum_i \msf{E}_i \bm{\Lambda}_{1,i}(\bm{x}), \sum_{j} \delta \tilde{\msf{D}}_j \tilde{\bm{\Lambda}}_{1,j}(\bm{x}) \right)_{L^2(\Omega)}
		= \bmsf{E}^T \delta \tilde{\bmsf{D}}
		= \left( \frac{\partial \overline{\msf{H}}}{\partial \tilde{\bmsf{D}} } \right)^T \delta \tilde{\bmsf{D}} \,,
\end{equation}
since $( \bm{\Lambda}_{1,i}, \tilde{\bm{\Lambda}}_{1,j} )_{L^2(\Omega)} = \delta_{ij}$. Similarly,
\begin{multline}
	\left( \frac{\delta \overline{H}_h}{\delta Q_h}, \delta Q_h \right)_{L^2(\Omega)} 
		= \left( \frac{a \theta}{2} \left( Q_h - | \bm{E}_h |^2 \right), 
			\sum_{i} \delta \msf{Q}_i \bm{\Lambda}_{3,i}(\bm{x}) \right)_{L^2(\Omega)} \\
		= \sum_i \frac{a \theta}{2} \left( (\mathbb{M}_3 \bmsf{Q})_i 
			- \left( \Lambda_{3,i}, | \bm{E}_h |^2 \right)_{L^2(\Omega)} \right) \delta \msf{Q}_i
		= \left( \frac{\partial \overline{\msf{H}}}{\partial \bmsf{Q} } \right)^T \delta \bmsf{Q}.
\end{multline}
The rest are computed in a like manner. \qed

Notice, a derivative with respect to a variable on the primal complex yields a variable on the dual complex and vice versa. This is because the derivative of a functional on a vector space yields an element of the dual space: i.e. if $V$ is a vector space, then $f: V \to \mathbb{R} \implies Df(v) \in V^*$ for $v \in V$. At the coefficient level, $\mathcal{C}_\ell$ and $\mathcal{C}_{\ell}^*$ are in duality with the duality pairing specified by the mass matrix weighted inner product. 

It is now straightforward to obtain the equations of motion from the Poisson bracket and Hamiltonian. 
\begin{theorem}
With the discrete Poisson bracket given in equation \ref{eq:cubic_2d_pb} and Hamiltonian given in equation \ref{eq:cubic_2d_ham}, one obtains the following equations of motion:
\begin{equation}
    \begin{split}
    \partial_t \tilde{\bmsf{D}} &= \{ \tilde{\bmsf{D}}, \overline{\msf{H}} \} = \mathbbm{d}_1^T \mathbb{M}_2 \bmsf{B} \\
    \partial_t \bmsf{B} &= \{ \bmsf{B}, \overline{\msf{H}} \} = - \mathbbm{d}_1 \bmsf{E} \\
    \partial_t \bmsf{P} &= \{ \bmsf{P}, \overline{\msf{H}} \} = \mathbb{M}_1^{-1} \tilde{\bmsf{J}}
    \end{split}
    \quad \text{and} \quad
    \begin{split}
    \partial_t \tilde{\bmsf{J}} &= \{ \tilde{\bmsf{J}}, \overline{\msf{H}} \} = \mathbb{M}_1 ( \omega_p^2 \bmsf{E} - \omega_0^2 \bmsf{P} ) \\
    \partial_t \bmsf{Q} &= \{ \bmsf{Q}, \overline{\msf{H}} \} = \mathbb{M}_3^{-1} \tilde{\bmsf{\upsigma}} \\
    \partial_t \tilde{\msf{\upsigma}}_i &= \{ \tilde{\bmsf{\upsigma}}, \overline{\msf{H}} \} = \omega_v^2 \left[ \left( \Lambda_{3,i}, | \bm{E}_h |^2 \right)_{L^2(\Omega)}  - (\mathbb{M}_3)_{ij} \msf{Q}_j \right] \,.
    \end{split}
\end{equation}
\end{theorem}
\noindent \textit{Proof:} For any observable $F$, its evolution is given by $\dot{F} = \{F, \overline{H} \}$. \qed \\[0.5em]
\noindent The discrete equations of motion may be slightly refactored to be in a more convenient form for computation:
\begin{equation}
    \left\{
    \begin{aligned}
        \partial_t \bmsf{D} &= \mathbb{M}_1^{-1} \mathbbm{d}_1^T \mathbb{M}_2 \bmsf{B} \\
	\partial_t \bmsf{B} &= - \mathbbm{d}_1 \bmsf{E} \\
	\partial_t \bmsf{P} &= \bmsf{J}
    \end{aligned}
    \right. 
    \quad \text{and} \quad
    \left\{
    \begin{aligned}
	\partial_t \bmsf{J} &= \omega_p^2 \bmsf{E} - \omega_0^2 \bmsf{P} \\
	\partial_t \bmsf{Q} &= \bmsf{\upsigma} \\
	\partial_t \msf{\upsigma}_i &= \omega_v^2 \left[ (\mathbb{M}_2^{-1})_{ij} \left( \Lambda_{2,j}, | \bm{E}_h |^2 \right)_{L^2(\Omega)}  - \msf{Q}_i \right].
    \end{aligned}
    \right.
\end{equation}
One has to perform two matrix inversions each time-step. This is the most convenient way to formulate this system because it ensures one may easily assemble and evaluate the fields. The fields need to be evaluated at each time-step in order to compute $\bmsf{E}$ by solving the nonlinear equation:
\begin{equation}
    \tilde{\msf{D}}_i = \left( \Lambda_{1,i}, \bm{P}_h + \left[ \epsilon_\infty + a \left( (1 - \theta) | \bm{E}_h |^2 + \theta Q_h \right) \right] \bm{E}_h \right)_{L^2(\Omega)} \,.
\end{equation}
This is accomplished with Picard iteration. 

The spatially discrete equations of motion possess the same conservation laws as the continuous system. 
\begin{prop}
The following conservation laws hold:
\begin{equation}
	\partial_t \mathbbm{d}_0^T \tilde{\bmsf{D}} = 0
	\quad \text{and} \quad
	\partial_t \mathbbm{d}_2 \bmsf{B} = 0 \,.
\end{equation}
These are discrete analogs of Gauss's laws. 
\end{prop}
\noindent \textit{Proof:} These are a consequence of the fact that for a FEEC method, $\mathbbm{d}_1 \mathbbm{d}_0 = 0$ and $\mathbbm{d}_2 \mathbbm{d}_1 = 0$. \qed

\subsection{Discrete one- and two-dimensional models}
The discretization procedure for the one- and two-dimensional models is now briefly sketched. The results are almost identical to the three-dimensional case. 

\subsubsection{Discrete one-dimensional model}
In the one-dimensional case, one defines the following Galerkin expansions for each field:
\begin{multline} \label{eq:1d_galerkin_representation}
    D_h = \sum_{i} \tilde{\msf{D}}_i \tilde{\Lambda}_{0,i}(z) \,,
    \quad
    B_h = \sum_i \msf{B}_i \Lambda_{1,i}(z) \,,
    \quad
    P_h = \sum_{i} \msf{P}_i \Lambda_{0,i}(z) \,,
    \quad
    E_h = \sum_i \msf{E}_i \Lambda_{0,i}(z) \,, \\
    J_h = \sum_{i} \tilde{\msf{J}}_i \tilde{\Lambda}_{0,i}(z) \,,
    \quad
    Q_h = \sum_i \msf{Q}_i \Lambda_{1,i}(z) \,,
    \quad \text{and} \quad
    \sigma_h = \sum_i \tilde{\msf{\upsigma}}_i \tilde{\Lambda}_{1,i}(z) \,.
\end{multline}
That is, $E_h, P_h \in V_h^0$, $D_h, J_h \in (V_h^0)^*$, $B_h, Q_h \in V_h^1$, and $\sigma_h \in (V_h^1)^*$. 

\begin{theorem} \label{eq:cubic_1d_disc_ham_system}
If one discretizes the functionals in the Hamiltonian formulation of the one-dimensional nonlinear Maxwell system using the Galerkin representations given in equation \eqref{eq:1d_galerkin_representation}, the discrete Poisson bracket is
\begin{multline} \label{eq:cubic_1d_pb}
    [ \msf{F}, \msf{G} ] = 
    - \left[ \left( \frac{\partial \msf{F} }{\partial \bmsf{B} } \right)^T \mathbbm{d}_0 \frac{\partial \msf{G}}{\partial \tilde{\bmsf{D}} }
    - \left( \frac{\partial \msf{G} }{\partial \bmsf{B} } \right)^T \mathbbm{d}_0 \frac{\partial \msf{F}}{\partial \tilde{\bmsf{D}} } \right]
    + \omega_p^2 \left( 
    \left( \frac{\partial \msf{F} }{\partial \bmsf{P} } \right)^T \frac{\partial \msf{G} }{\partial \tilde{\bmsf{J}} }
    - \left( \frac{\partial \msf{G} }{\partial \bmsf{P} } \right)^T \frac{\partial \msf{F} }{\partial \tilde{\bmsf{J}} } 
    \right) \\
    +  \frac{2 \omega_v^2}{a \theta} \left( 
    \left( \frac{\partial \msf{F}}{\delta \bmsf{Q} } \right)^T \frac{\partial \msf{G} }{\partial \tilde{\bmsf{\upsigma}} } 
    - \left( \frac{\partial \msf{G}}{\delta \bmsf{Q} } \right)^T \frac{\partial \msf{F} }{\partial \tilde{\bmsf{\upsigma}} } 
    \right) \,,
\end{multline}
and the discrete Hamiltonian is
\begin{multline} \label{eq:cubic_1d_ham}
    \msf{H} [\bmsf{E}, \bmsf{B}, \bmsf{P}, \tilde{\bmsf{J}}, \bmsf{Q}, \tilde{\bmsf{\upsigma}}] 
    = 
    \frac{1}{2} 
    \bigg[ 
    \tilde{\bmsf{E}}_*^T \bmsf{E} 
    + \frac{\omega_0^2}{\omega_p^2} \bmsf{P}^T \mathbb{M}_0 \bmsf{P} 
    + \frac{1}{\omega_p^2}  \tilde{\bmsf{J}}^T \mathbb{M}_0^{-1} \tilde{\bmsf{J}} \\
    + \frac{a \theta}{2} \bmsf{Q}^T \mathbb{M}_1 \bmsf{Q} 
    + \frac{a \theta}{2 \omega_v^2} \tilde{\bmsf{\upsigma}}^T \mathbb{M}_1^{-1} \tilde{\bmsf{\upsigma}}
    + \bmsf{B}^T \mathbb{M}_1 \bmsf{B} 
    \bigg] \,,
\end{multline}
where
\begin{equation}
    (\tilde{\msf{E}}_*)_i = \left(\Lambda_{0,i}, \left( \epsilon_\infty 
    + \frac{3 a ( 1 - \theta) | \bm{E}_h |^2}{2} + a \theta Q_h \right) \bm{E}_h \right)_{L^2(\Omega)} \,.
\end{equation}
Moreover, one finds that
\begin{equation}
    \tilde{\msf{D}}_i = \left( \Lambda_{0,i}, \bm{P}_h + \left[ \epsilon_\infty + a \left( (1 - \theta) | \bm{E}_h |^2 + \theta Q_h \right) \right] \bm{E}_h \right)_{L^2(\Omega)} \,.
\end{equation}
\end{theorem}

It is then possible to obtain evolution equations from this spatially discretized Hamiltonian system in the same manner as in the three-dimensional case. 
\begin{theorem}
The derivatives of the discrete Hamiltonian are given by
\begin{multline}
    \frac{\partial \overline{\msf{H}}}{\partial \tilde{\bmsf{D}} } = \bmsf{E},
    \quad
    \frac{\partial \overline{\msf{H}}}{\partial \bmsf{B}} = \mathbb{M}_1 \bmsf{B},
    \quad
    \frac{\partial \overline{\msf{H}}}{\partial \bmsf{P}} 
    = \mathbb{M}_0 \left( \frac{\omega_0^2}{\omega_P^2} \bmsf{P} - \bmsf{E} \right), 
    \quad 
    \frac{\partial \overline{\msf{H}}}{\partial \tilde{\bmsf{J}}} = \omega_p^{-2} \mathbb{M}_0^{-1} \tilde{\bmsf{J}}, \\
    \quad
    \frac{\partial \overline{\msf{H}}}{\partial \bmsf{Q}} 
    = \frac{a \theta}{2} \left( \mathbb{M}_1 \bmsf{Q} - \left( \bm{\Lambda}_{1}, | \bm{E}_h |^2 \right)_{L^2(\Omega)} \right),
    \quad \text{and} \quad
    \frac{\partial \overline{\msf{H}}}{\partial \tilde{\bmsf{\upsigma}}}
    = \frac{a \theta}{2 \omega_v^2} \mathbb{M}_1^{-1} \tilde{\bmsf{\upsigma}} \,.
\end{multline}
\end{theorem}
\begin{theorem}
The Hamiltonian system described in theorem \ref{eq:cubic_1d_disc_ham_system} gives rise to the following equations of motion:
\begin{equation}
    \begin{split}
    \partial_t \tilde{\bmsf{D}} &= \{ \tilde{\bmsf{D}}, \overline{\msf{H}} \} = \mathbbm{d}_1^T \mathbb{M}_1 \bmsf{B} \\
    \partial_t \bmsf{B} &= \{ \bmsf{B}, \overline{\msf{H}} \} = - \mathbbm{d}_0 \bmsf{E} \\
    \partial_t \bmsf{P} &= \{ \bmsf{P}, \overline{\msf{H}} \} = \mathbb{M}_0^{-1} \tilde{\bmsf{J}}
    \end{split}
    \quad \text{and} \quad
    \begin{split}
    \partial_t \tilde{\bmsf{J}} &= \{ \tilde{\bmsf{J}}, \overline{\msf{H}} \} = \mathbb{M}_0 ( \omega_p^2 \bmsf{E} - \omega_0^2 \bmsf{P} ) \\
    \partial_t \bmsf{Q} &= \{ \bmsf{Q}, \overline{\msf{H}} \} = \mathbb{M}_1^{-1} \tilde{\bmsf{\upsigma}} \\
    \partial_t \tilde{\msf{\upsigma}}_i &= \{ \tilde{\bmsf{\upsigma}}, \overline{\msf{H}} \} = \omega_v^2 \left[ \left( \Lambda_{1,i}, | \bm{E}_h |^2 \right)_{L^2(\Omega)}  - (\mathbb{M}_1)_{ij} \msf{Q}_j \right] \,.
    \end{split}
\end{equation}
\end{theorem}

The spatially discrete equations of motion possess the same conservation laws as the continuous system. From appendix \ref{appendix:integral_formulas}, one has that
\begin{equation} \label{eq:1d_disc_casimir}
    \int_\Omega D_h \mathsf{d} z = \mathbbm{1}^T \tilde{\bmsf{D}}
    \quad \text{and} \quad
    \int_\Omega B_h \mathsf{d} z = \mathbbm{1}^T \bmsf{B} \,,
\end{equation}
where $\mathbbm{1}_i = 1$ for all $i$. 
\begin{prop}
The following conservation laws hold:
\begin{equation}
    \partial_t \mathbbm{1}^T \tilde{\bmsf{D}} = 0
    \quad \text{and} \quad
    \partial_t \mathbbm{1}^T \bmsf{B} = 0 \,.
\end{equation}
\end{prop}
\noindent \textit{Proof:} This follows from the fact that $\mathbbm{1} \in \text{ker}(\mathbbm{d}_0)$ and $\mathbbm{1} \in \text{ker}(\mathbbm{d}_0^T)$. \qed

\subsubsection{Discrete two-dimensional model}

There are two two-dimensional FEEC sequences which are dual to each other: the $\bmsf{curl}$-$\msf{div}$ sequence and the $\bmsf{grad}$-$\msf{curl}$ sequence. Because the two-dimensional nonlinear Maxwell model is formulated with the scalar curl, one chooses the $\bmsf{grad}$-$\msf{curl}$ sequence to be the primal sequence. All vector quantities should be modeled as $1$-forms, and $(B, \sigma, Q)$ as $2$-forms. One further elects to let $(\bm{D}, \bm{J}, \sigma)$ to be taken from the dual de Rham sequence, and $(\bm{E}, B, \bm{P}, Q)$ to be taken from the primal sequence. This yields the following Galerkin expansions for each field:
\begin{multline} \label{eq:2d_galerkin_representation}
    \bm{D}_h = \sum_{i} \tilde{\msf{D}}_i \tilde{\bm{\Lambda}}_{1,i}(\bm{x}),
    \quad
    B_h = \sum_i \msf{B}_i \Lambda_{2,i}(\bm{x}),
    \quad
    \bm{P}_h = \sum_{i} \msf{P}_i \bm{\Lambda}_{1,i}(\bm{x}), 
    \quad
    \bm{E}_h = \sum_i \msf{E}_i \bm{\Lambda}_{1,i}(\bm{x}), \\
    \bm{J}_h = \sum_{i} \tilde{\msf{J}}_i \tilde{\bm{\Lambda}}_{1,i}(\bm{x}), 
    \quad
    Q_h = \sum_i \msf{Q}_i \Lambda_{2,i}(\bm{x}),
    \quad \text{and} \quad
    \sigma_h = \sum_i \tilde{\msf{\upsigma}}_i \tilde{\Lambda}_{2,i}(\bm{x}) \,.
\end{multline}
\begin{theorem}
Using the Galerkin expansions given in equation \eqref{eq:2d_galerkin_representation}, one finds that the discrete Poisson bracket is
\begin{multline} \label{eq:cubic_2d_pb}
    [ \msf{F}, \msf{G} ] = 
	- \left[ \left( \frac{\partial \msf{F} }{\partial \bmsf{B} } \right)^T \mathbbm{d}_1 \frac{\partial \msf{G}}{\partial \tilde{\bmsf{D}} }
	- \left( \frac{\partial \msf{G} }{\partial \bmsf{B} } \right)^T \mathbbm{d}_1 \frac{\partial \msf{F}}{\partial \tilde{\bmsf{D}} } \right]
        + \omega_p^2 \left( \left( \frac{\partial \msf{F} }{\partial \bmsf{P} } \right)^T \frac{\partial \msf{G} }{\partial \tilde{\bmsf{J}} }
	- \left( \frac{\partial \msf{G} }{\partial \bmsf{P} } \right)^T \frac{\partial \msf{F} }{\partial \tilde{\bmsf{J}} } \right) \\
	+ \frac{2 \omega_v^2}{a \theta} \left( 
	\left( \frac{\partial \msf{F}}{\delta \bmsf{Q} } \right)^T \frac{\partial \msf{G} }{\partial \tilde{\bmsf{\upsigma}} } 
	- \left( \frac{\partial \msf{G}}{\delta \bmsf{Q} } \right)^T \frac{\partial \msf{F} }{\partial \tilde{\bmsf{\upsigma}} } \right) \,,
\end{multline}
and the discrete Hamiltonian is
\begin{multline} \label{eq:cubic_2d_ham}
    \msf{H} [\bmsf{E}, \bmsf{B}, \bmsf{P}, \tilde{\bmsf{J}}, \bmsf{Q}, \tilde{\bmsf{\upsigma}}] 
    = \frac{1}{2} 
    \bigg[ \bigg( \tilde{\bmsf{E}}_*^T \bmsf{E} 
    + \frac{\omega_0^2}{\omega_p^2} \bmsf{P}^T \mathbb{M}_1 \bmsf{P} 
    + \frac{1}{\omega_p^2}  \tilde{\bmsf{J}}^T \mathbb{M}_1^{-1} \tilde{\bmsf{J}} \\
    + \frac{a \theta}{2} \bmsf{Q}^T \mathbb{M}_2 \bmsf{Q} 
    + \frac{a \theta}{2 \omega_v^2} \tilde{\bmsf{\upsigma}}^T \mathbb{M}_2^{-1} \tilde{\bmsf{\upsigma}} \bigg) 
    + \bmsf{B}^T \mathbb{M}_2 \bmsf{B} \bigg] \,,
\end{multline}
where
\begin{equation}
    (\tilde{\msf{E}}_*)_i = \left(\Lambda_{1,i}, \left( \epsilon_\infty 
    + \frac{3 a ( 1 - \theta) | \bm{E}_h |^2}{2} + a \theta Q_h \right) \bm{E}_h \right)_{L^2(\Omega)} \,.
\end{equation}
Moreover, one finds that
\begin{equation}
    \tilde{\msf{D}}_i = \left( \Lambda_{0,i}, \bm{P}_h + \left[ \epsilon_\infty + a \left( (1 - \theta) | \bm{E}_h |^2 + \theta Q_h \right) \right] \bm{E}_h \right)_{L^2(\Omega)} \,.
\end{equation}
\end{theorem}

It is then possible to obtain evolution equations from this spatially discretized Hamiltonian system in the same manner as in the three-dimensional case. 
\begin{theorem}
The derivatives of the discrete Hamiltonian are given by
\begin{multline}
    \frac{\partial \overline{\msf{H}}}{\partial \tilde{\bmsf{D}} } = \bmsf{E} \,,
    \quad
    \frac{\partial \overline{\msf{H}}}{\partial \bmsf{B}} = \mathbb{M}_2 \bmsf{B} \,,
    \quad
    \frac{\partial \overline{\msf{H}}}{\partial \bmsf{P}} 
    = \mathbb{M}_1 \left( \frac{\omega_0^2}{\omega_P^2} \bmsf{P} - \bmsf{E} \right) \,,
    \quad 
    \frac{\partial \overline{\msf{H}}}{\partial \tilde{\bmsf{J}}} = \omega_p^{-2} \mathbb{M}_1^{-1} \tilde{\bmsf{J}} \,, \\
    \quad
    \frac{\partial \overline{\msf{H}}}{\partial \bmsf{Q}} 
    = \frac{a \theta}{2} \left( \mathbb{M}_2 \bmsf{Q} - \left( \bm{\Lambda}_{2}, | \bm{E}_h |^2 \right)_{L^2(\Omega)} \right) \,,
    \quad \text{and} \quad
    \frac{\partial \overline{\msf{H}}}{\partial \tilde{\bmsf{\upsigma}}}
    = \frac{a \theta}{2 \omega_v^2} \mathbb{M}_2^{-1} \tilde{\bmsf{\upsigma}} \,.
\end{multline}
\end{theorem}
\noindent It is now straightforward to obtain the equations of motion from the Poisson bracket and Hamiltonian. 
\begin{theorem}
With the discrete Poisson bracket given in equation \ref{eq:cubic_2d_pb} and Hamiltonian given in equation \ref{eq:cubic_2d_ham}, one obtains the following equations of motion:
\begin{equation}
    \begin{split}
    \partial_t \tilde{\bmsf{D}} &= \{ \tilde{\bmsf{D}}, \overline{\msf{H}} \} = \mathbbm{d}_1^T \mathbb{M}_2 \bmsf{B} \\
    \partial_t \bmsf{B} &= \{ \bmsf{B}, \overline{\msf{H}} \} = - \mathbbm{d}_1 \bmsf{E} \\
    \partial_t \bmsf{P} &= \{ \bmsf{P}, \overline{\msf{H}} \} = \mathbb{M}_1^{-1} \tilde{\bmsf{J}}
    \end{split}
    \quad \text{and} \quad
    \begin{split}
    \partial_t \tilde{\bmsf{J}} &= \{ \tilde{\bmsf{J}}, \overline{\msf{H}} \} = \mathbb{M}_1 ( \omega_p^2 \bmsf{E} - \omega_0^2 \bmsf{P} ) \\
    \partial_t \bmsf{Q} &= \{ \bmsf{Q}, \overline{\msf{H}} \} = \mathbb{M}_2^{-1} \tilde{\bmsf{\upsigma}} \\
    \partial_t \tilde{\msf{\upsigma}}_i &= \{ \tilde{\bmsf{\upsigma}}, \overline{\msf{H}} \} = \omega_v^2 \left[ \left( \Lambda_{2,i}, | \bm{E}_h |^2 \right)_{L^2(\Omega)}  - (\mathbb{M}_2)_{ij} \msf{Q}_j \right] \,.
    \end{split}
\end{equation}
\end{theorem}

The spatially discrete equations of motion possess the same conservation laws as the continuous system. 
\begin{prop}
The following conservation laws hold:
\begin{equation}
	\partial_t \mathbbm{d}_0^T \tilde{\bmsf{D}} = 0
	\quad \text{and} \quad
	\partial_t \mathbbm{1}^T \bmsf{B} = 0.
\end{equation}
\end{prop}
\noindent \textit{Proof:} The first conservation law is a consequence of the fact that $\mathbbm{d}_1 \mathbbm{d}_0 = 0$. Note that
\begin{equation} \label{eq:disc_casimir_2d}
	\partial_t \bmsf{B} = - \mathbbm{d}_1 \bmsf{E}
	\iff
	\partial_t B_h = - \msf{curl}_h \bm{E}_h.
\end{equation}
Hence, by lemma \ref{eq:integral_formulas} in appendix \ref{appendix:integral_formulas} and letting $R$ be a matrix which applies a $90^\circ$ rotation, it follows that
\begin{equation}
    \partial_t \mathbbm{1}^T \bmsf{B} 
    = \partial_t \int_\Omega B_h \mathsf{d}^2 \bm{x} 
    = - \int_\Omega \msf{curl}_h \bm{E}_h \mathsf{d}^2 \bm{x} 
    = - \int_{\partial \Omega} R \bm{E}_h \cdot \mathsf{d} \bm{x} 
    = 0
\end{equation}
because of the assumption of homogeneous boundary conditions. \qed

\subsection{Temporal discretization} \label{section:temporal_disc}
The temporal discretization scheme is written out for the discretized three-dimensional model. As the one- and two-dimensional models are formally the same, an identical approach may be taken in those cases. Temporal discretization is accomplished via Hamiltonian splitting. It turns out that the Hamiltonian splits such that one can obtain two exactly integrable subsystems. Let
\begin{equation}
	 \msf{H} [\bmsf{E}, \bmsf{B}, \bmsf{P}, \tilde{\bmsf{J}}, \bmsf{Q}, \tilde{\bmsf{\upsigma}}] 
		= 
	 \msf{H}_{\bmsf{E}} [\bmsf{E}, \bmsf{P}, \bmsf{Q}] +  \msf{H}_{\bmsf{B}} [\bmsf{B}, \tilde{\bmsf{J}}, \tilde{\bmsf{\upsigma}}].
\end{equation}
where
\begin{equation}
	 \msf{H}_{\bmsf{E}} [\bmsf{E}, \bmsf{P}, \bmsf{Q}]
	 =
	 \frac{1}{2} 
			\bigg[ 
			 \bigg( \tilde{\bmsf{E}}_*^T \bmsf{E} 
				+ \frac{\omega_0^2}{\omega_p^2} \bmsf{P}^T \mathbb{M}_1 \bmsf{P} 
						+ \frac{a \theta}{2} \bmsf{Q}^T \mathbb{M}_3 \bmsf{Q} \bigg) 
			\bigg]
\end{equation}
and
\begin{equation}
	\msf{H}_{\bmsf{B}} [\bmsf{B}, \tilde{\bmsf{J}}, \tilde{\bmsf{\upsigma}}]
	=
	\frac{1}{2} \bigg[ \bigg( \frac{1}{\omega_p^2}  \tilde{\bmsf{J}}^T \mathbb{M}_1^{-1} \tilde{\bmsf{J}}
	+ \frac{a \theta}{2 \omega_v^2} \tilde{\bmsf{\upsigma}}^T \mathbb{M}_3^{-1} \tilde{\bmsf{\upsigma}} \bigg) 
	+ \bmsf{B}^T \mathbb{M}_2 \bmsf{B} \bigg].
\end{equation}
This yields two exactly integrable subsystems:
\begin{equation}
    \begin{cases}
        \partial_t \tilde{\bmsf{D}} &= \mathbbm{d}_1^T \mathbb{M}_2 \bmsf{B} \\
	\partial_t \bmsf{B} &= 0 \\
	\partial_t \bmsf{P} &= \bmsf{J} \\
	\partial_t \bmsf{J} &= 0 \\
	\partial_t \bmsf{Q} &= \bmsf{\upsigma} \\
	\partial_t \msf{\upsigma}_i &= 0,
    \end{cases}
    \quad \text{and} \quad
    \begin{cases}
        \partial_t \tilde{\bmsf{D}} &= 0 \\
	\partial_t \bmsf{B} &= - \mathbbm{d}_1 \bmsf{E} \\
	\partial_t \bmsf{P} &= 0  \\
	\partial_t \bmsf{J} &= \omega_p^2 \bmsf{E} - \omega_0^2 \bmsf{P} \\
	\partial_t \bmsf{Q} &= 0 \\
	\partial_t \msf{\upsigma}_i &= \omega_v^2 \left[ (\mathbb{M}_3^{-1})_{ij} \left( \Lambda_{3,j}, | \bm{E}_h |^2 \right)_{L^2(\Omega)}  - \msf{Q}_i \right].
    \end{cases}
\end{equation}
These two partial flows may be composed together to obtain approximations to the full flow to any reasonable order in time: e.g.\ $2^{nd}$ \cite{doi:10.1137/0705041}, $4^{th}$ or $6^{th}$ \cite{YOSHIDA1990262} order splitting methods. This paper only uses $2^{nd}$-order Strang splitting which is equivalent to the leapfrog method. 

Assuming stability of the time-stepper, which will be shown subsequently, one may use backwards error analysis to show that there exists a discrete energy which the time-stepper conserves \cite{geometric_numerical_integration}. That is, the $\Delta t$-time flow of the second order Strang splitting method corresponds exactly with the $\Delta t$-time flow of a Hamiltonian system with a modified Hamiltonian $\tilde{\mathsf{H}}$. This modified Hamiltonian is an $O(\Delta t^2)$ perturbation of the Hamiltonian of the continuous system, although its exact form is generally quite complicated to write down explicitly. Moreover, backwards error analysis further implies
\begin{equation}
    \mathsf{H}(\bmsf{E}^n, \bmsf{B}^n, \bmsf{P}^n, \tilde{\bmsf{J}}^n, \bmsf{Q}^n, \tilde{\bmsf{\upsigma}}^n) - \mathsf{H}(\bmsf{E}^0, \bmsf{B}^0, \bmsf{P}^0, \tilde{\bmsf{J}}^0, \bmsf{Q}^0, \tilde{\bmsf{\upsigma}}^0) = O \left( \Delta t^2 \right) \,.
\end{equation}
That is, the energy computed at each time-step oscillates around a mean value in a band with $O(\Delta t^2)$ width. 

\subsection{Numerical stability of time-stepper}
While it is possible to construct splitting methods with high temporal order, only second order Strang splitting \cite{doi:10.1137/0705041} is considered as the proof of numerical stability for this method is analytically tractable. Let $\varphi_{\bm{E}}$ be the flow generated by the Hamiltonian $\mathsf{H}_{\bm{E}}$ and $\varphi_{\bm{B}}$ be the flow generated by the Hamiltonian $\mathsf{H}_{\bm{B}}$. These partial flows are composed together to get a second order $\Delta t$-time flow map:
\begin{equation}
    \varphi_S(\Delta t) = \varphi_{\bm{E}}(\Delta t/2) \circ \varphi_{\bm{B}}(\Delta t) \circ \varphi_{\bm{E}}(\Delta t/2) \,.
\end{equation}
While this is typically written as a three stage method, it may be equivalently written as a two stage method with staggered time-steps: 
\begin{equation}
    \begin{cases}
        \bm{B}_h^{n+1/2} = \bm{B}_h^{n-1/2} - \Delta t \text{curl}_h \bm{E}_h^n \\
        \bm{J}_h^{n+1/2} = \bm{J}_h^{n-1/2} 
            + \omega_p^2 \Delta t \bm{E}_h^n - \omega_0^2 \Delta t \bm{P}_h^n \\
        \sigma_h^{n+1/2} = \sigma_h^{n-1/2} + \omega_v^2 \Delta t \left[ 
            \pi_h^3 | \bm{E}_h^n |^2  - Q_h^n \right] \,,
    \end{cases}
    \text{and} \quad
    \begin{cases}
        \bm{D}_h^{n+1} =\bm{D}_h^{n} 
            + \Delta t \text{curl}_h^* \bm{B}_h^{n+1/2} \\
        \bm{P}_h^{n+1} = \bm{P}_h^{n} + \Delta t \bm{J}_h^{n+1/2} \\
        Q_h^{n+1} = Q_h^{n} + \Delta t \sigma_h^{n+1/2} \,,
    \end{cases}
\end{equation}
where $\pi_h^3$ is the $L^2$ projection into $V^3_h$. 

\begin{theorem}
Time-stepping with Strang splitting is stable if
\begin{equation}
    \Delta t \max \left\{ \frac{\| \text{curl}_h \|_{V_h^1}}{C}, \frac{\omega_p^2}{C}, \frac{\omega_0^2}{C}, \frac{\omega_v^2}{C}, \frac{2 \| \pi_h^3 \|_{V_h^3} \| \bm{E}_h^n \| \omega_v^2}{C}, \| \text{curl}_h \|_{V_h^1} \right\} < 1 \,,
\end{equation}
where $\| \text{curl}_h \|_{V_h^1}$ is the norm of the discrete curl in $V_h^1$, $\| \pi_h^3 \|_{V_h^3}$ is the norm of the $L^2$ projection into $V_h^3$, and $C = c_L - c_N(\bm{E}, Q)$, where
\begin{equation} \label{eq:lin_bound}
    c_L = \sqrt{\frac{2 + \epsilon_\infty^2 + \sqrt{4 + \epsilon_\infty^4} }{2}} \,,
\end{equation}
and
\begin{equation} \label{eq:nonlin_bound}
    c_N(\bm{E}, Q) = \max \left\{ 1, a \left( \| 3 (1 - \theta) | \bm{E} |^2 + \theta ( |Q| + | \bm{E} | ) \|_{L^\infty(\Omega)} + 1 \right) \right\} \,.
\end{equation}
\end{theorem}
\noindent \textit{Proof:} Denote a second perturbed solution of this method with a tilde, e.g.\ $\bm{D}_h^n$ and $\tilde{\bm{D}}_h^n$. One wishes to bound the difference between these two solutions at the $n^{th}$ time-step by the difference at the previous time-step. Taking the difference and denoting $\epsilon_{\bm{A}}^n = \bm{A}^n_h - \tilde{\bm{A}}^n_h$ for each field $\bm{A} \in \{\bm{D}, \bm{B}, \bm{E}, \bm{P}, \bm{J}, Q, \sigma \}$, one finds
\begin{equation}
    \begin{cases}
        \epsilon_{\bm{B}}^{n+1/2} - \epsilon_{\bm{B}}^{n-1/2} = - \Delta t \text{curl}_h \epsilon_{\bm{E}}^n \\
        \epsilon_{\bm{J}}^{n+1/2} - \epsilon_{\bm{J}}^{n-1/2} =
            \omega_p^2 \Delta t \epsilon_{\bm{E}}^n - \omega_0^2 \Delta t \epsilon_{\bm{P}}^n \\
        \epsilon_{\sigma}^{n+1/2} - \epsilon_{\sigma}^{n-1/2} =
            \omega_v^2 \Delta t \left[ 
            \pi_h^3 ( | \bm{E}_h^n |^2 - | \tilde{\bm{E}}_h^n |^2 )  
            - \epsilon_{Q}^n \right] \,,
    \end{cases}
    \text{and} \quad
    \begin{cases}
        \epsilon_{\bm{D}}^{n+1} - \epsilon_{\bm{D}}^{n} 
            = \Delta t \text{curl}_h^* \epsilon_{\bm{B}}^{n+1/2} \\
        \epsilon_{\bm{P}}^{n+1} - \epsilon_{\bm{P}}^{n} 
            = \Delta t \epsilon_{\bm{J}}^{n+1/2} \\
        \epsilon_{Q}^{n+1} - \epsilon_{Q}^{n} 
            = \Delta t \epsilon_{\sigma}^{n+1/2} \,,
    \end{cases}
\end{equation}
where 
\begin{equation}
    \epsilon_{\bm{E}}^n 
    \coloneq
        \bm{E}(\bm{D}_h^n, \bm{P}_h^n, Q_h^n) - 
        \bm{E}(\tilde{\bm{D}}_h^n, \tilde{\bm{P}}_h^n, \tilde{Q}_h^n) \,.
\end{equation}
One can bound the error in difference of squares of $\bm{E}_h^n$ as follows:
\begin{equation}
    | \bm{E}_h^n |^2 - | \tilde{\bm{E}}_h^n |^2
    = ( \bm{E}_h^n + \tilde{\bm{E}}_h^n - \bm{E}_h^n + \bm{E}_h^n ) \cdot ( \bm{E}_h^n - \tilde{\bm{E}}_h^n )
    = 2 \bm{E}_h^n \cdot \epsilon_{\bm{E}}^n + | \epsilon_{\bm{E}}^n |^2 \,.
\end{equation}
Hence, it follows that
\begin{equation}
    \left\| \pi_h^3 \left( | \bm{E}_h^n |^2 - | \tilde{\bm{E}}_h^n |^2 \right) \right\| 
    \leq 
    \| \pi_h^3 \|_{V_h^3} \left( 2 \| \bm{E}_h^n \| \| \epsilon_{\bm{E}}^n \| 
    + \| \epsilon_{\bm{E}}^n \|^2 \right)\,.
\end{equation}
One may neglect the higher order error term. Note, it has been implicitly assumed that $ | \bm{E}_h^n |^2 \in L^2(\Omega)$. If $| \Omega | < \infty$, then one has the inclusion $L^p(\Omega) \subseteq L^q(\Omega)$ for $1 \leq q < p < \infty$. Hence, assuming $\bm{E}_h^n \in L^2(\Omega)$ for each $n$, this requirement is satisfied. Therefore, it is necessary to verify that $\bm{E}_h^n \in L^2(\Omega)$ at each time-step. It is natural that the stability of the nonlinear term in the time advance for $\sigma_h$ should depend on the magnitude of the field as the stability of time-steppers for nonlinear systems often depends on the state of the system. 

The half integer time-step error increments are bounded as
\begin{multline}
    \left( 
    \left\| \epsilon_{\bm{B}}^{n+1/2} - \epsilon_{\bm{B}}^{n-1/2} \right\|^2
    + 
    \left\| \epsilon_{\bm{J}}^{n+1/2} - \epsilon_{\bm{J}}^{n-1/2} \right\|^2
    + 
    \left\| \epsilon_{\sigma}^{n+1/2} - \epsilon_{\sigma}^{n-1/2} \right\|^2
    \right)^{1/2} 
    \leq \\
    \Delta t \max \left\{ \| \text{curl}_h \|_{V_h^1}, \omega_p^2, \omega_0^2, \omega_v^2, 2 \| \pi_h^3 \|_{V_h^3} \| \bm{E}_h^n \| \omega_v^2 \right\}
    \sqrt{ \| \epsilon_{\bm{E}}^n \|^2 + \| \epsilon_{\bm{P}}^n \|^2 + \| \epsilon_{Q}^n \|^2 } \,,
\end{multline}
and the integer time-step error increments are bounded as
\begin{multline}
    \left( 
    \left\| \epsilon_{\bm{D}}^{n+1} - \epsilon_{\bm{D}}^{n} \right\|^2
    +
    \left\| \epsilon_{\bm{P}}^{n+1} - \epsilon_{\bm{P}}^{n} \right\|^2
    + 
    \left\| \epsilon_{Q}^{n+1} - \epsilon_{Q}^{n} \right\|^2 
    \right)^{1/2}
    \leq \\
    \Delta t \| \text{curl}_h \|_{V_h^1} 
    \sqrt{ \| \epsilon_{\bm{B}}^{n+1/2} \|^2 + \| \epsilon_{\bm{J}}^{n+1/2} \|^2 + \| \epsilon_{\sigma}^{n+1/2} \|^2 } \,.
\end{multline}
For convenience, define the Hilbert space $X = (L^2(\Omega))^3 \times (L^2(\Omega))^3 \times L^2(\Omega)$ with inner product
\begin{equation}
    \left( (\bm{E}, \bm{P}, Q), (\tilde{\bm{E}}, \tilde{\bm{P}}, \tilde{Q}) \right)_X
    =
    \int_\Omega ( \bm{E} \cdot \tilde{\bm{E}} + \bm{P} \cdot \tilde{\bm{P}} + Q \tilde{Q} ) \mathsf{d}^3 \bm{x} \,.
\end{equation}
Its norm is given by
\begin{equation} 
    \| (\bm{E}, \bm{P}, Q) \|_X = \sqrt{ \int_\Omega ( | \bm{E} |^2 + | \bm{P} |^2 + Q^2 ) \mathsf{d}^3 \bm{x} } \,.
\end{equation}
This is the appropriate norm for bounding the error increments in each of the two stages of the time-stepper. 

One needs to bound the error in the implicitly defined $\bm{E}_h^n$ field by the errors in the dynamically evolving fields. To do so, it is necessary to take a closer look at the nonlinear coordinate change $(\bm{D}_h, \bm{B}_h, \bm{P}_h, \bm{J}_h, Q_h, \sigma_h) \mapsto (\bm{E}_h, \bm{B}_h, \bm{P}_h, \bm{J}_h, Q_h, \sigma_h)$. The principal difficulty is that $\bm{E}_h$ is implicitly defined by solving
\begin{equation}
    \bm{D}_h = \pi_h^1 \left( \epsilon_\infty \bm{E}_h + \bm{P}_h
        + a(1 - \theta) | \bm{E}_h |^2 \bm{E}_h + a \theta Q_h \bm{E}_h \right) \,,
\end{equation}
where $\pi_h^1: V^1 \to V^1_h$ is the $L^2$ projection. In appendix \ref{appendix:bound_e-field}, a Lipschitz bound for the electric field is obtained such that
\begin{equation}
    \| (\epsilon_{\bm{E}}^n, \epsilon_{\bm{P}}^n, \epsilon_Q^n) \|_X 
    \leq
    C^{-1} \| (\epsilon_{\bm{D}}^n, \epsilon_{\bm{P}}^n, \epsilon_Q^n) \|_X \,,
\end{equation}
where $C = c_L - c_N$ with $c_L$ given in equation \eqref{eq:lin_bound} and $c_N$ in equation \eqref{eq:nonlin_bound}. The result follows. \qed \\[1em]

In this work, only problems in which the fields remains $O(1)$ throughout the simulation are considered. Hence, one may select the time-step according to what the linear theory would prescribe. For this reason, it is generally sufficient to set the time-step to be such that
\begin{equation}
    \Delta t \| \text{curl}_h \|_{V_h^1} \leq 1 \,. 
\end{equation}
Therefore, it is generally sufficient to let $\Delta t = 0.8/\| \text{curl}_h \|_{V_h^1}$, where the operator norm of the curl is obtained by computing the spectral radius of its associated matrix:
\begin{equation}
    \| \text{curl}_h \|_{V_h^1}^2 = \max_{ \bm{f} \in V_h^1} \frac{ \left( \text{curl}_h \bm{f}, \text{curl}_h \bm{f} \right)_{L^2(\Omega)} }{ \| \bm{f} \|_{L^2(\Omega}^2 }
    =
    \rho \left( \mathbb{M}_1^{-1} \mathbbm{d}_2^T \mathbb{M}_2 \mathbbm{d}_1 \right) \,.
\end{equation}

\subsection{Picard iteration nonlinear solver}

Each time the partial time-step associated with the Hamiltonian $\mathsf{H}_{\bmsf{B}}$ is performed, one obtains a new value of $(\bmsf{D}, \bmsf{P}, \bmsf{Q})$. Before one can take another partial time-step with Hamiltonian the $\mathsf{H}_{\bmsf{E}}$, one needs to compute $\bmsf{E}$. This is accomplished by solving the nonlinear system:
\begin{equation}
    \tilde{\msf{D}}_i = \left( \Lambda_{1,i}, \left( \epsilon_\infty \bm{E}_h + \bm{P}_h + a(1 - \theta) | \bm{E}_h |^2 \bm{E}_h + a \theta Q_h \bm{E}_h \right) \right)_{L^2(\Omega)} \,.
\end{equation}
This is done with the fixed point iteration
\begin{equation}
    \mathsf{E}_i^{(k+1)} = \epsilon_\infty^{-1} \left( \mathsf{D}_i - \mathsf{P}_i 
    - a \sum_{ij} (\mathbb{M}_{1}^{-1})_{ij} \left( \Lambda_{1,j}, \left( \left[ (1 - \theta) \left| \bm{E}_h^{(k)} \right|^2 
        + \theta Q_h \right] \bm{E}_h^{(k)} \right) \right)_{L^2(\Omega)} \right) \,.
\end{equation}
Because the nonlinear parameter $a$ is small, this fixed point iteration generally converges in only a few steps if $\bmsf{E}$ is initialized on its value from the previous time-step. 

\subsection{Adding dissipative effects} \label{sec:cubic_maxwell_dissipation}
The gradient flow dissipative dynamics may be discretized in an analogous manner to the Hamiltonian dynamics to obtain the following discrete symmetric bracket:
\begin{equation}
    ( \mathsf{F}, \mathsf{G} ) = - \frac{1}{\epsilon_0} \left( 
        \omega_p^2 \lambda_0 \left( \frac{\partial \msf{F}}{\partial \tilde{\bmsf{J}}} \right)^T \mathbb{M}_1 \frac{\partial \msf{G}}{\partial \tilde{\bmsf{J}}} 
        + \frac{2 \omega_v^2 \lambda_v}{a \theta} \left( \frac{\partial \msf{F}}{\partial \tilde{\bmsf{\upsigma}}} \right)^T \mathbb{M}_3 \frac{\partial \msf{G}}{\partial \tilde{\bmsf{\upsigma}}} \right) \,.
\end{equation}
For a general functional $\msf{F}$, one then finds that its evolution is given by $\dot{\msf{F}} = \{ \msf{F}, \bar{\msf{H}} \} + (\msf{F}, \bar{\msf{H}})$. Therefore, one finds that the spatially semi-discretized flow has the energy dissipation rate
\begin{equation} \label{eq:discrete_dissipation_rate}
    \dot{\bar{\msf{H}}} = (\bar{\msf{H}}, \bar{\msf{H}}) = - \epsilon_0 \left( \frac{\lambda_0}{\omega_p^2} \bmsf{J}^T \mathbb{M}_1 \bmsf{J} + \frac{a \theta \lambda_v}{2 \omega_v^2} \bmsf{\upsigma}^T \mathbb{M}_3 \bmsf{\upsigma} \right) \,.
\end{equation}
Because the coupled Hamiltonian and gradient flow structure was discretized in a manner which obtained an explicit bracket structure for the spatially discrete dynamics, the system recovers the appropriate energy dissipation rate.

Applying the same Hamiltonian splitting procedure as before but now including the symmetric bracket, one again obtains two exactly integrable partial flows:
\begin{equation}
    \begin{cases}
        \partial_t \tilde{\bmsf{D}} &= \mathbbm{d}_1^T \mathbb{M}_2 \bmsf{B} \\
	\partial_t \bmsf{B} &= 0 \\
	\partial_t \bmsf{P} &= \bmsf{J} \\
	\partial_t \bmsf{J} &= 0 \\
	\partial_t \bmsf{Q} &= \bmsf{\upsigma} \\
	\partial_t \msf{\upsigma}_i &= 0,
    \end{cases}
    \quad \text{and} \quad
    \begin{cases}
        \partial_t \tilde{\bmsf{D}} &= 0 \\
	\partial_t \bmsf{B} &= - \mathbbm{d}_1 \bmsf{E} \\
	\partial_t \bmsf{P} &= 0  \\
	\partial_t \bmsf{J} + \lambda_0 \bmsf{J} &= \omega_p^2 \bmsf{E} - \omega_0^2 \bmsf{P} \\
	\partial_t \bmsf{Q} &= 0 \\
	\partial_t \msf{\upsigma}_i + \lambda_v \msf{\upsigma}_i &= \omega_v^2 \left[ (\mathbb{M}_3^{-1})_{ij} \left( \Lambda_{3,j}, | \bm{E}_h |^2 \right)_{L^2(\Omega)}  - \msf{Q}_i \right].
    \end{cases}
\end{equation}
Only the update rules for $\bmsf{J}$ and $\bmsf{\upsigma}$ change from the dissipation-free case. The update rule for $\bmsf{J}$ is now
\begin{equation}
    \bmsf{J}^{n+1} = \exp \left( - \lambda_0 \Delta t \right) \bmsf{J}^n 
        + \left( \frac{1 - \exp \left( - \lambda_0 \Delta t \right)}{\lambda_0} \right) \left( \omega_p^2 \bmsf{E} - \omega_0^2 \bmsf{P} \right) \,,
\end{equation}
and for $\bmsf{\upsigma}$ is now
\begin{equation}
    \msf{\upsigma}^{n+1}_i = \exp \left( - \lambda_v \Delta t \right) \msf{\upsigma}^n_i 
        + \left( \frac{1 - \exp \left( - \lambda_v \Delta t \right)}{\lambda_v} \right) \left( \omega_v^2 \left[ (\mathbb{M}_3^{-1})_{ij} \left( \Lambda_{3,j}, | \bm{E}_h |^2 \right)_{L^2(\Omega)}  - \msf{Q}_i \right] \right) \,.
\end{equation}
As before, one approximates the full $\mathsf{H} = \mathsf{H}_{\bmsf{E}} + \mathsf{H}_{\bmsf{B}}$ flow at $O(\Delta t^2)$ using Strang splitting \cite{doi:10.1137/0705041}. Finally, one can verify that in the limit as $(\lambda_0, \lambda_v) \to (0,0)$, one recovers the dissipation-free update rules. For extremely small parameter values, one should use a series expansion for $(1 - \exp(-\lambda \Delta t))/\lambda$ to avoid round off error.  

\subsection{Adding a time-dependent free-current source} \label{sec:current_source}

Suppose one adds a time-dependent free-current source to Amp{\`e}re's law:
\begin{equation}
    \partial_t \bm{D} = \nabla \times \bm{B} - \bm{J}_f \,.
\end{equation}
This is spatially semi-discretized via an $L^2$ projection:
\begin{equation}
    \partial_t \tilde{\bmsf{D}}
    =
    \mathbbm{d}_1^T \mathbb{M}_2 \bmsf{B}
    -
    \tilde{\bmsf{J}}_f \,,
\end{equation}
where $\msf{J}_i(t) = ( \Lambda_{1,i}, \bm{J}_f)_{L^2(\Omega)}$. Temporal integration proceeds as usual via Strang splitting with the free-current integrated in the Amp{\`e}re sub-step. 

%% file: numerical_results.tex
\section{Numerical results} \label{sec:numerical_results}

The numerical examples in this section utilize the B-spline FEEC implementation provided by the PSYDAC library \cite{Gl2022PSYDACAH}. Of particular interest in these examples is the conservation of the Casimir invariants. Additionally, in the non-dissipative case, energy should be conserved on average, while, in the dissipative case, the expected dissipation rate is known. Unless otherwise specified, the physical parameters in the following tests are chosen to correspond with those used in \cite{lyu_et_al_2022,BOKIL2017420, bokil_et_al_2018}:
\begin{multline} \label{eq:params_nonlinear_maxwell}
    \epsilon_{\infty} = 2.25 \,,
    \quad
    a = 0.07 \,,
    \quad
    \theta = 0.3 \,,
    \quad
    \omega_0 = 5.84 \,,
    \quad 
    \omega_p = 10.11 \,, \\
    \quad 
    \omega_v = 1.28 \,,
    \quad 
    \lambda_0 = 1.168 \times 10^{-5} \,,
    \quad
    \lambda_v = \frac{29.2}{32} \,.
\end{multline}
In all tests, the nonlinear solver tolerance is set to $\mathtt{tol} = 10^{-10}$. 

\subsection{Convergence studies with manufactured solutions}

An exact solution to equation \eqref{eq:cubic_nl_maxwell_sys} with a prescribed free current distribution, $\bm{J}_f$, on the right-hand side of Amp{\`e}re's law is sought using the method of manufactured solutions. This facilitates convergence studies for the numerical method. 

Suppose one posits a functional form
\begin{equation}
    \bm{E}(\bm{x},t) = \bm{A}(\bm{x}) \sin(2 \pi t) 
    \quad \implies \quad
    \bm{B}(\bm{x},t) = \frac{1}{2 \pi} \nabla \times \bm{A}(\bm{x}) \cos(2 \pi t) \,.
\end{equation}
The auxiliary fields mediating the temporally-dispersive effects are
\begin{multline}
    \bm{P}(\bm{x},t)
    =
    \frac{[ \omega_0 \sin(2 \pi t) - 2 \pi \sin(\omega_0 t) ] \omega_p^2}{\omega_0^3 - 4 \pi^2 \omega_0} \bm{A}(\bm{x}) \,, \\
    \quad \text{and} \quad
    Q(\bm{x},t)
    =
    \frac{\omega_v^2( \cos(4 \pi t) - 1) - 16 \pi^2 ( \cos(\omega_v t) - 1 )}{32 \pi^2 - 2 \omega_v^2} | \bm{A}(\bm{x}) |^2 \,.
\end{multline}
This implies that
\begin{multline}
    \bm{D}(\bm{x},t)
    =
    \left(
    \epsilon_\infty \sin(2 \pi t)
    +
    \frac{[ \omega_0 \sin(2 \pi t) - 2 \pi \sin(\omega_0 t) ] \omega_p^2}{\omega_0^3 - 4 \pi^2 \omega_0}
    \right)
    \bm{A}(\bm{x}) \\
    +
    \left(
    a(1 - \theta) \sin^3(2 \pi t)
    +
    a \theta
    \frac{[\omega_v^2( \cos(4 \pi t) - 1) - 16 \pi^2 ( \cos(\omega_v t) - 1 )] \sin(2 \pi t)}{32 \pi^2 - 2 \omega_v^2}
    \right)
    | \bm{A}(\bm{x}) |^2 \bm{A}(\bm{x}) \,.
\end{multline}
Hence, one finds that
\begin{equation}
    \bm{J}_f(\bm{x}, t)
    =
    \partial_t \bm{G} \,,
    \quad \text{where} \quad
    \bm{G}(\bm{x},t)
    =
    \bm{D}(\bm{x}, t)
    - \frac{1}{4 \pi^2} \nabla \times \nabla \times \bm{A}(\bm{x}) \sin(2 \pi t) \,.
\end{equation}
The update rule for Amp{\`e}re's law requires that one compute
\begin{equation}
    \int_{t_0}^{t_0 + \Delta t} \bm{J}_f(\bm{x},t) \mathsf{d} t
    =
    \left. \bm{G}(\bm{x},t) \right|_{t=t_0}^{t={t_0 + \Delta t}} 
    \,.
\end{equation}
Therefore, it is more convenient to work with the $\bm{G}$ rather than $\bm{J}_f$ itself. 

In order to isolate the spatial convergence rate, the time-step is set equal to $\Delta t = O(K^{-N/2})$ to ensure that errors from the second-order temporal integrator do not pollute the results. The solutions are computed on the interval $t \in [0,1]$, and the relative errors in the solution are computed in the $L^p([0,1]^d \times [0,1])$ norms for $p=2, \infty$. Homogeneous boundary conditions are used in both the one- and two-dimensional tests. 

\begin{figure}
    \centering
    \begin{subfigure}[t]{\textwidth}
    \centering
    \includegraphics[width=\linewidth]{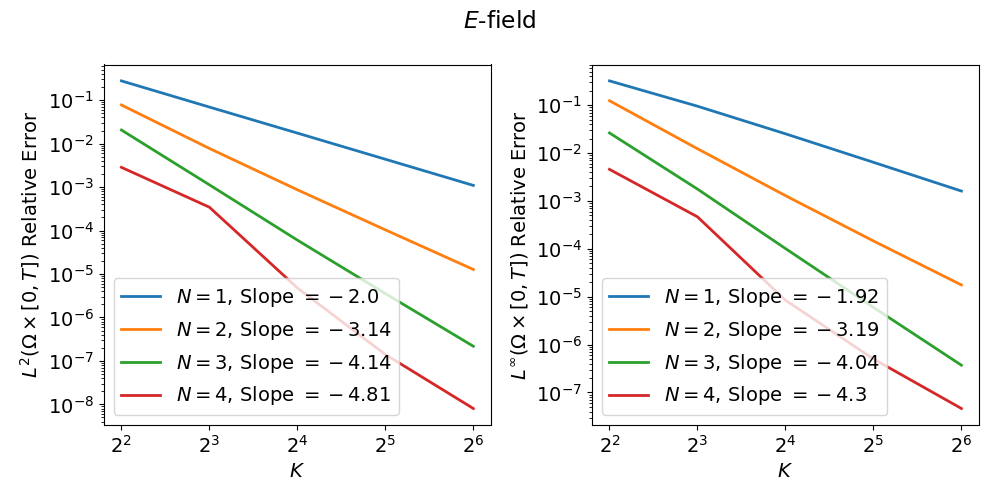}
    \end{subfigure}
    \begin{subfigure}[t]{\textwidth}
    \centering
    \includegraphics[width=\linewidth]{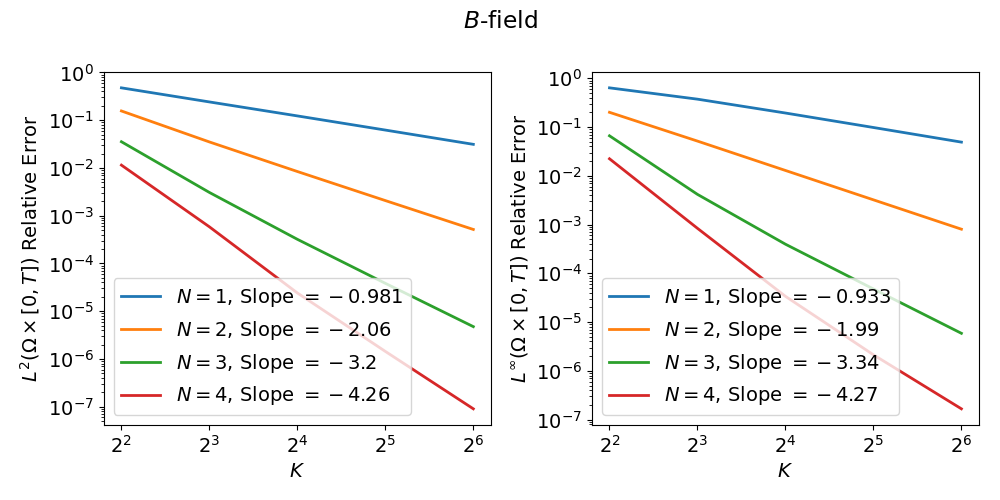}
    \end{subfigure}
    \caption{Convergence rates of the relative error in the electric and magnetic fields in the $L^2(\Omega \times [0,T])$ and $L^\infty(\Omega \times [0,T])$ norms for the one-dimensional manufactured solution test. Optimal convergence is achieved except for the $E$-field when $4^{th}$-order polynomial interpolation is used. }
    \label{fig:1d_conv}
\end{figure}

\begin{figure}
    \centering
    \begin{subfigure}[t]{\textwidth}
    \centering
    \includegraphics[width=\linewidth]{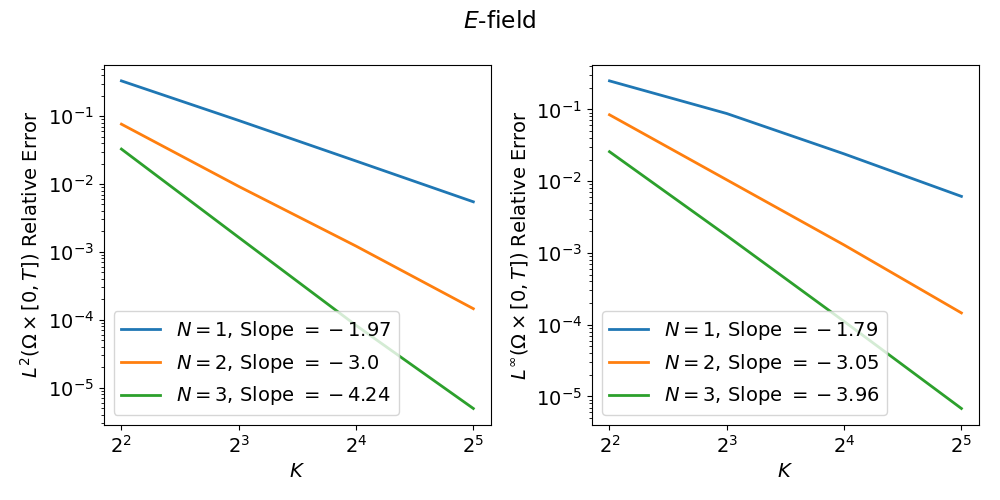}
    \end{subfigure}
    \begin{subfigure}[t]{\textwidth}
    \centering
    \includegraphics[width=\linewidth]{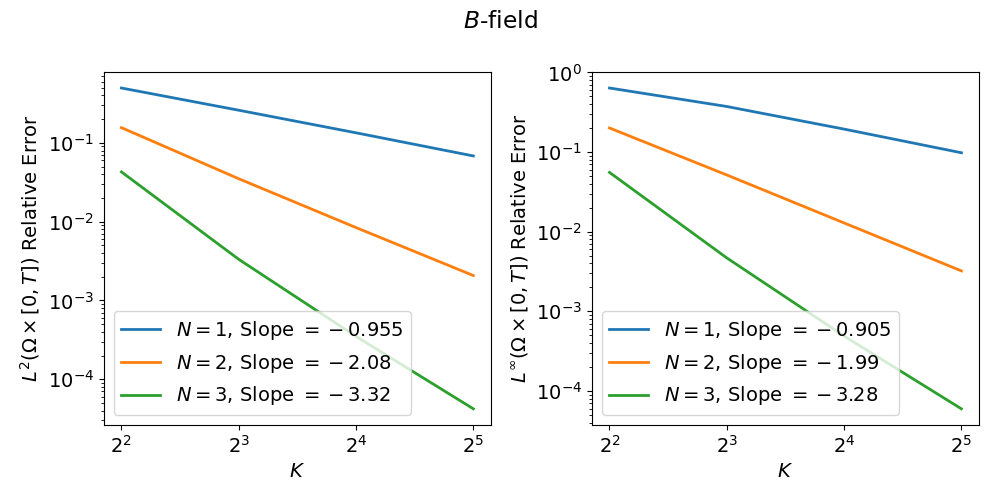}
    \end{subfigure}
    \caption{Convergence rates of the relative error in the electric and magnetic fields in the $L^2(\Omega \times [0,T])$ and $L^\infty(\Omega \times [0,T])$ norms for the two-dimensional manufactured solution test.}
    \label{fig:2d_conv}
\end{figure}

For the one-dimensional convergence study, the spatial component of the electric field was taken to be $A(x) = \sin(2 \pi x)$. See figure \ref{fig:1d_conv} for a visualization of the convergence as a function of polynomial degree and grid resolution. For the two-dimensional convergence study, the spatial component of the electric field was taken to be $\bm{A}(\bm{x}) = [\sin(2 \pi y), - \sin(2 \pi x) ]^T$. See figure \ref{fig:2d_conv} for a visualization of the convergence as a function of polynomial degree and grid resolution. 

\subsection{Numerical tests in one spatial dimension}
In a single spatial dimension, the only Casimir invariants are given in equation \eqref{eq:1d_disc_casimir}. In all experiments, they are conserved to machine precision. 

\subsubsection{Higher harmonic generation}
The nonlinear polarization causes different Fourier harmonics to mix in the temporal evolution of the system. In this example, the evolution of the sum of two Fourier modes is considered. Higher harmonics are generated as the system evolves. 

The domain is taken to be $\Omega = [0, 1]$ with periodic boundary conditions, and solutions are computed over the time-interval $t \in [0, 100]$. The initial conditions are taken to be $B_0(x) = \cos(2 \pi x) + \cos(4 \pi x)$ and $E_0(x) = 0$. For the numerical parameters, $N = 100$ cells are used and the polynomial degree is $p = 2$ for the $0$-forms and $p=1$ for the $1$-forms. This is more than sufficient for the initial data, but one should expect the time-evolution of the system to generate finer scales, hence the relatively high precision of the grid. With these parameters, the time-step is
\begin{equation}
    \Delta t = \frac{3}{4} \| \mathbb{M}_0^{-1} \mathbbm{d}_0^{T} \mathbb{M}_1 \mathbbm{d}_0 \|^{-1} \approx 0.002 \,.
\end{equation}
The simulation is performed both with the standard parameter set in equation \eqref{eq:params_nonlinear_maxwell} except that the nonlinear parameter is increased to $a=0.3$ so that the higher harmonics are activated over a shorter time-horizon. 

\begin{figure}
    \centering
    \includegraphics[width=\linewidth]{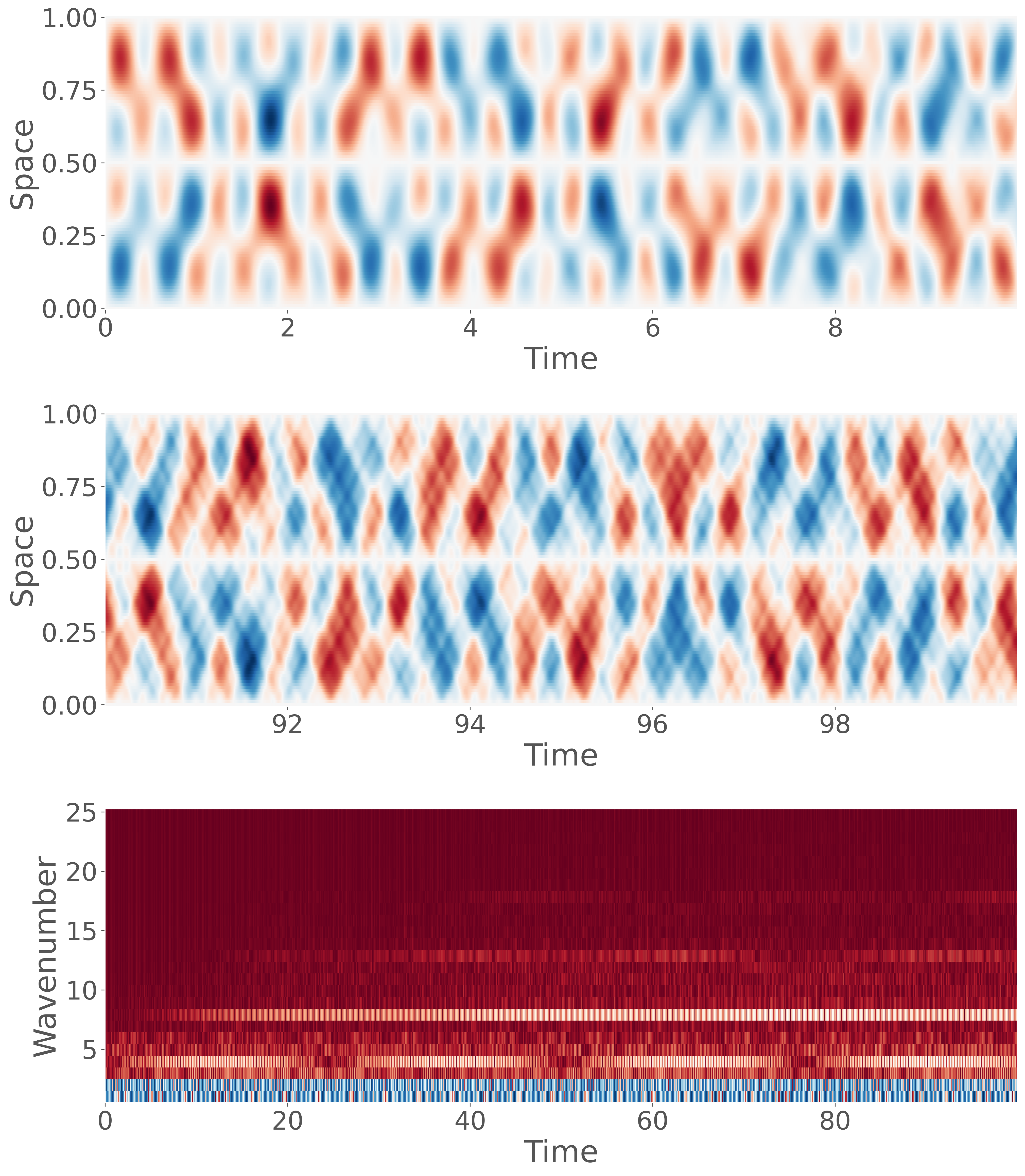}  
    \caption{Visualization of the evolution of the simple Fourier modes with dissipation. The top plot shows the solution in $t \in (0,10)$, the middle shows the solution in $t \in (90,100)$, and the bottom shows the solution in Fourier space.}
    \label{fig:1d_simple_soln}
\end{figure}

\begin{figure}
    \centering
    \includegraphics[width=\linewidth, height=7.4in]{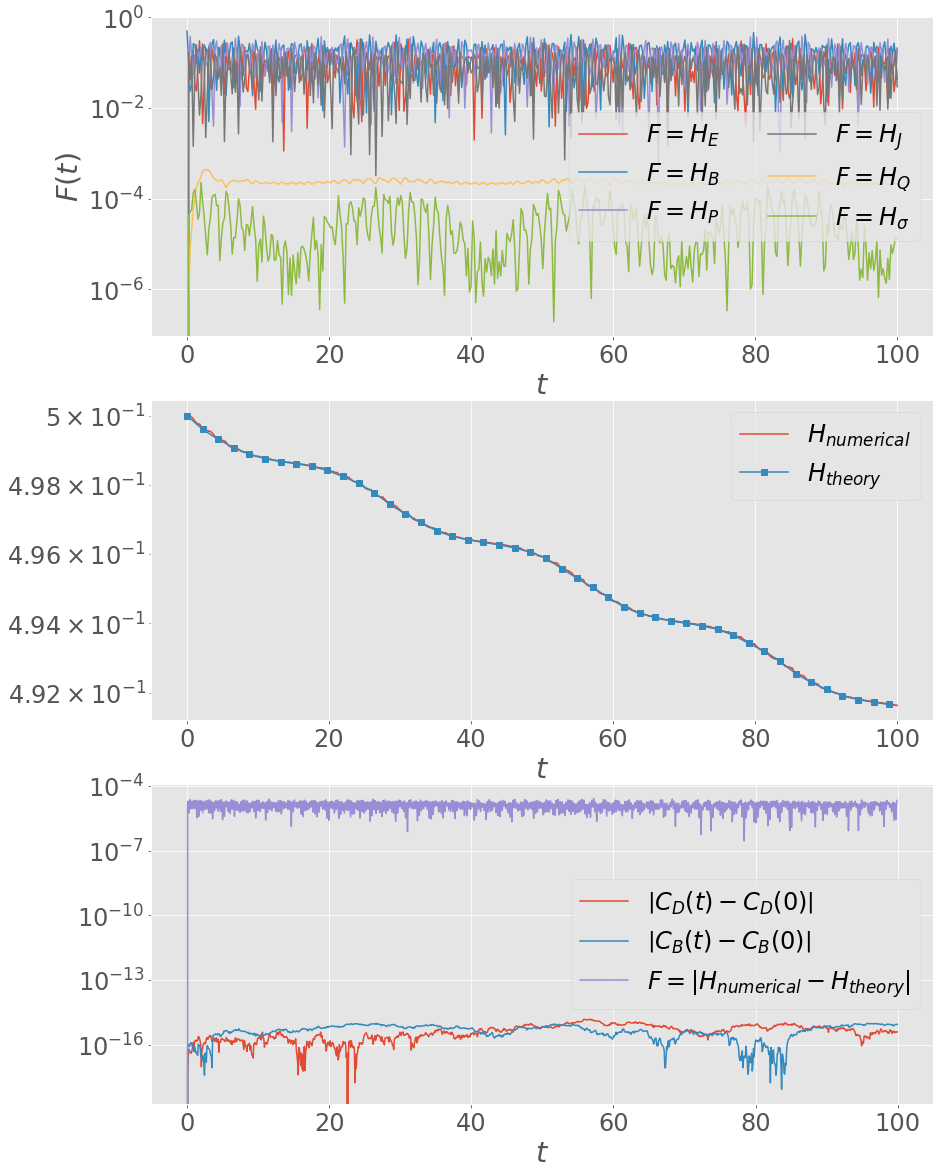}  
    \caption{Energy and Casimir conservation results for the simple Fourier mode test case with dissipation. The 1D discrete Casimir invariants are defined in equation \ref{eq:1d_disc_casimir}.}
    \label{fig:1d_simple_diss_diag}
\end{figure}

\begin{figure}
    \centering
    \includegraphics[width=\linewidth]{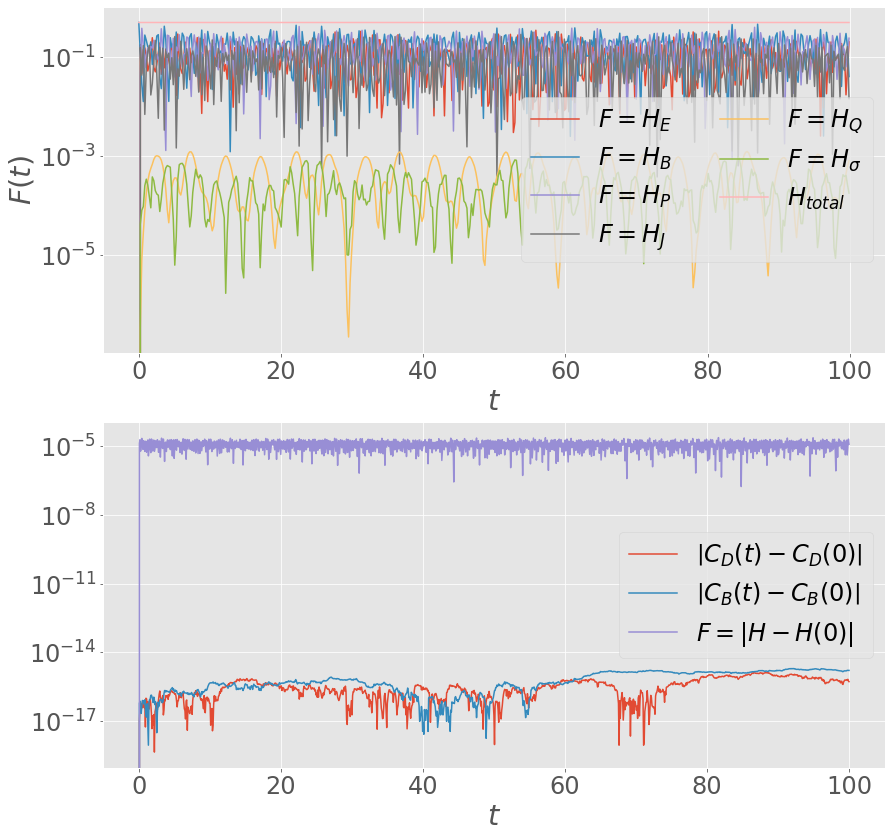}  
    \caption{Energy and Casimir conservation results for the simple Fourier mode test case without dissipation. The 1D discrete Casimir invariants are defined in equation \ref{eq:1d_disc_casimir}. One can see that the energy is conserved on average in the dissipation free case.}
    \label{fig:1d_simple_diag}
\end{figure}

See figure \ref{fig:1d_simple_soln} for a visualization of the solution in both standard and Fourier space. The plot looks the same in both dissipative and dissipation-free cases, so only one plot is given. See figure \ref{fig:1d_simple_diss_diag} for the energy dissipation and Casimir conservation results. One can see that the expected results are obtained. In the dissipation-free test, the dissipative parameters set to zero: $\lambda_0 = \lambda_v = 0$. This allows one to verify that energy is indeed conserved (within an $O(\Delta t^2)$ oscillating band) in the dissipation-free case, see figure \ref{fig:1d_simple_diag}. 

\subsubsection{Traveling wave problem}

\begin{figure}
    \centering
    \includegraphics[width=\linewidth]{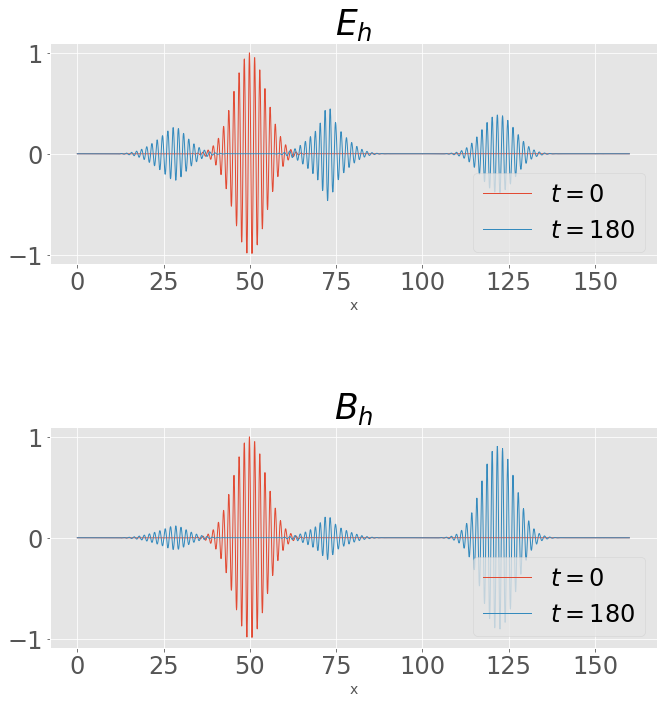}  
    \caption{Initial and final conditions for the Gaussian waveform test case.}
    \label{fig:1d_gaussian_soln}
\end{figure}

\begin{figure}
    \centering
    \includegraphics[width=\linewidth]{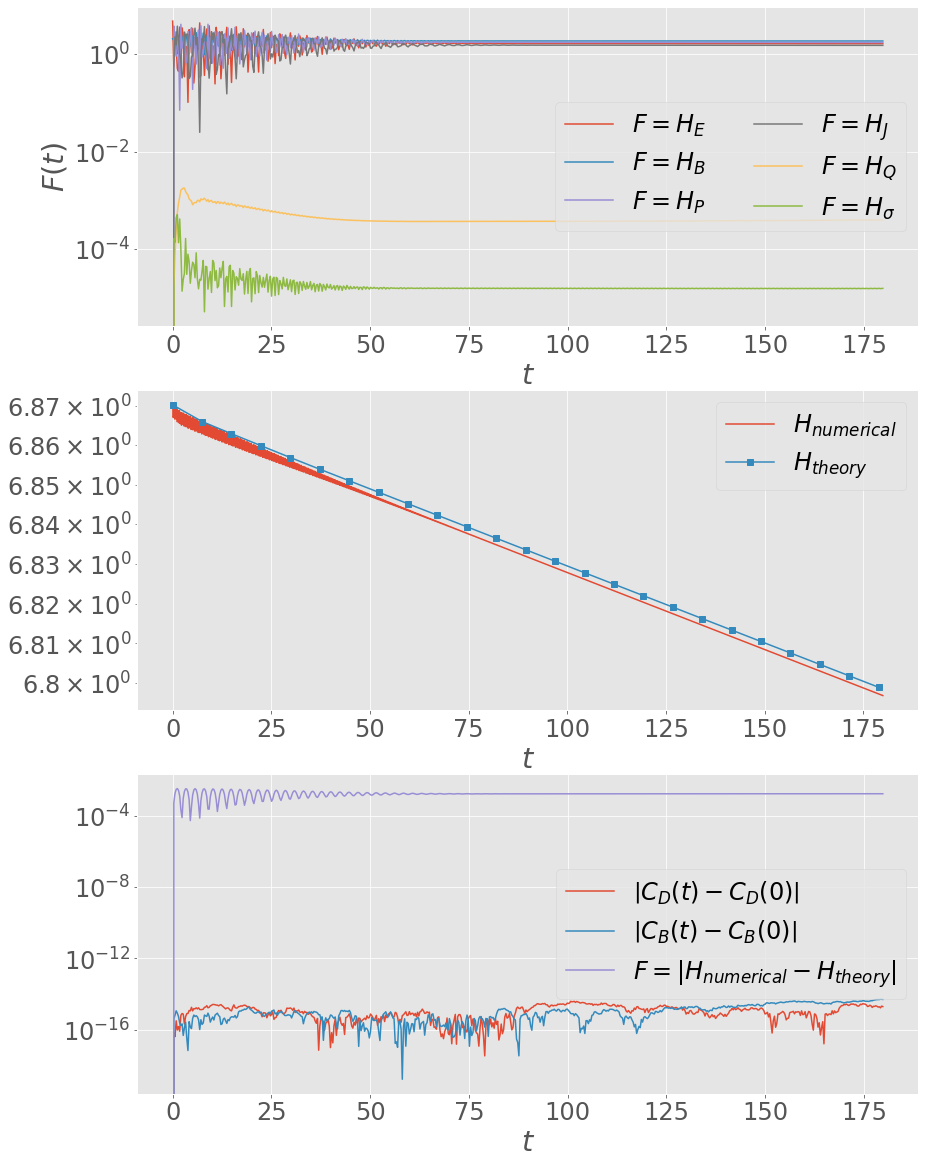}  
    \caption{Energy and Casimir conservation results for the Gaussian waveform test case. The 1D discrete Casimir invariants are defined in equation \ref{eq:1d_disc_casimir}.}
    \label{fig:1d_gaussian_diag}
\end{figure}

In this section, the propagation of Gaussian wave-packets is considered. The domain is taken to be $\Omega = [0, L]$ with periodic boundary conditions, and solutions are computed over the time-interval $t \in [0, 0.9L]$. The initial conditions are taken to be $E_0(x), B_0(x) = f_0(x)$, where
\begin{equation}
    f_0(x) = \cos \left( \frac{2 \pi k_0 x}{L} \right) g \left( \frac{x/L - \alpha}{s} \right) \,,
\end{equation}
where $L = 200$, $k_0 = 400/3$, $s = 1/30$, $\alpha = 0.25$, and $g(x) = \exp ( - x^2)$. The solution is only plotted on the subdomain $x \in [0, 180]$ because, in a medium with linear refractive index $\epsilon_{\infty} = 2.25$, the pulses begins centered $x=50$ and is centered at approximately $x=140$ at $t_{\text{fin}} = 270$. This ensures that the small portion of the wave which is deflected backwards does not interact with the forward propagating wave during the simulation. For the numerical parameters, $N = 6,000 = 30L$ cells are used and the polynomial degree is $p = 2$ for the $0$-forms and $p=1$ for the $1$-forms. A large number of cells are used because of the highly oscillatory initial data: this ensures that there are approximately $30$ grid-points per wave. With these numerical parameters, it was found that
\begin{equation}
    \Delta t = \frac{3}{4} \| \mathbb{M}_0^{-1} \mathbbm{d}_0^{T} \mathbb{M}_1 \mathbbm{d}_0 \|^{-1} \approx 0.0073 \,.
\end{equation}
See figure \ref{fig:1d_gaussian_soln} for a visualization of the simulation at $t=0$ and $t=t_{\text{fin}}$. Figure \ref{fig:1d_gaussian_diag} plots the energy and Casimir invariant conservation laws. The top figure shows the breakdown of each contribution to the energy, the middle figure shows the true energy decay versus the predicted theoretical energy decay from equation \eqref{eq:discrete_dissipation_rate}, and the bottom figure shows the errors in the Casimir invariants and the difference between the predicted and computed energy of the solution. 

\subsubsection{Exact kink-shaped wave solutions} \label{sec:kink_tw_test}

\begin{figure}
    \centering
    \includegraphics[width=0.9\linewidth]{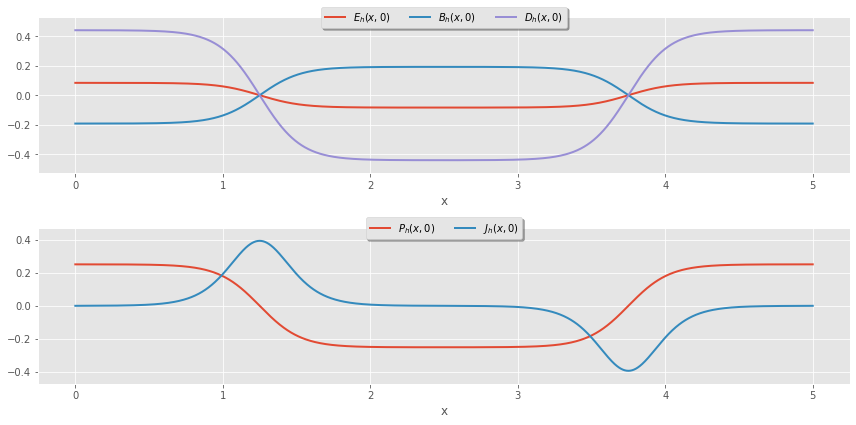}
    \caption{Initial conditions for the kink-shaped traveling wave numerical test.}
    \label{fig:kink_init}
\end{figure}

It was found in \cite{PhysRevE.71.036602} that the one-dimensional, cubicly nonlinear Maxwell equation possesses exact kink-shaped traveling-wave solutions. That is, given the model
\begin{equation} 
    \begin{aligned}
        &\partial_t B = \partial_x E \,, 
        \quad
        \partial_t D = \partial_x B \\
        &\partial_t P = J \,, 
        \quad
        \partial_t J = - \omega_0^2 P + \omega_p^2 E \,, \\
        &D = \epsilon_\infty E + a E^3 + P \,,
    \end{aligned}
\end{equation}
if one makes the traveling wave Ansatz, i.e. $E(x,t) = E(x - vt)$ (and similarly for the remaining fields), a system of ordinary differential equations describing traveling wave solutions to the full partial differential equation is obtained. Note, this model is the one-dimensional variant of equation \eqref{eq:nondim_nonlinear_maxwell} with $\theta = 0$. The traveling wave solutions are obtained by solving the following differential equation:
\begin{equation} 
\begin{aligned}
    \frac{d E}{d \xi} &= \Phi \\
    \frac{d \Phi}{d \xi} &= \frac{6 a v^2 E \Phi^2 + ( \epsilon_\infty \omega_0^2 + \omega_p^2 - \omega_0^2 / v^2) E + a \omega_0^2 E^3}{1 - \epsilon_\infty v^2 - 3 a v^2 E^2} \,.
\end{aligned}
\end{equation}
A method of generating accurate approximations to this traveling-wave differential equation is described in appendix \ref{appendix:kink_shaped_waves}. The physical parameters used in all tests in this section are as follows:
\begin{multline}
    \epsilon_\infty = 2.25 \,,
    \quad
    \epsilon_s = 5.25 \,,
    \quad\beta = \epsilon_s - \epsilon_\infty \,,
    \quad 
    \omega_0 = 52.943161119580196 \,, \\
    \quad
    \omega_p = \omega_0 \sqrt{\beta} \,,
    \quad
    a = \epsilon_\infty/3 \,,
    \quad 
    v = \frac{0.9995}{\sqrt{\epsilon_s}} \approx 0.43621756258174876 \,.
\end{multline}
A visualization of the initial conditions is given in figure \ref{fig:kink_init}. 

This numerical study using traveling wave solutions proceeds much in the same manner as in \cite{BOKIL2017420}, albeit with slightly different parameters and initial conditions. The solutions are computed on a periodic domain $\Omega = [0, 5]/\sim$ (where $\sim$ identifies the endpoints) on the time interval $T = [0, 5/v]$ so that the wave has traveled one full period. Second order Strang splitting is used in all tests. 

\begin{figure}
    \centering
    \includegraphics[width=\linewidth]{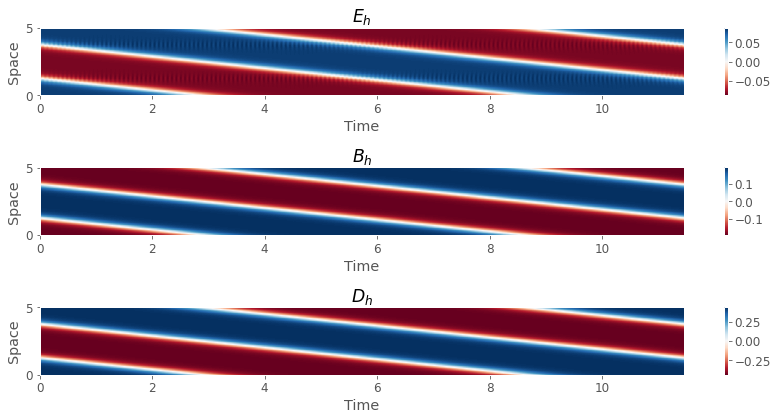}
    \caption{Computed fields $(E_h, B_h, D_h)$ for the kink-shaped traveling wave test computed using $K=1600$ elements and degree-$4$ polynomials for $V_h^0$ and degree-$3$ polynomials for $V_h^1$.}
    \label{fig:kink_soln}
\end{figure}

\begin{figure}
    \centering
    \includegraphics[width=\linewidth]{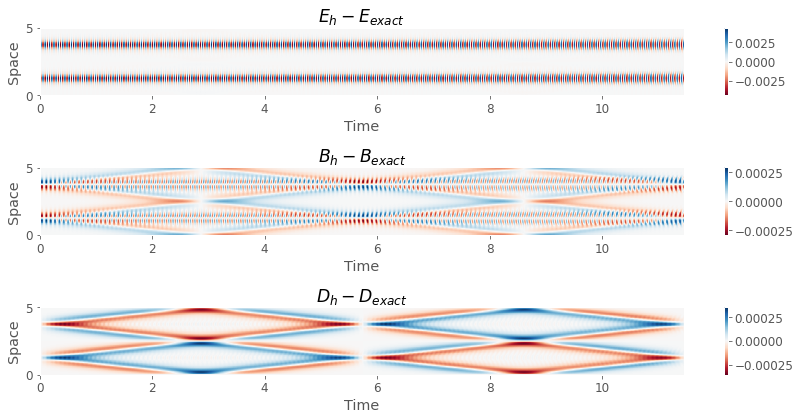}
    \caption{The error between the exact traveling wave solution and the reference numerical solution given in figure \ref{fig:kink_soln}.}
    \label{fig:error_fields}
\end{figure}

\begin{figure}
    \centering
    \includegraphics[width=\linewidth]{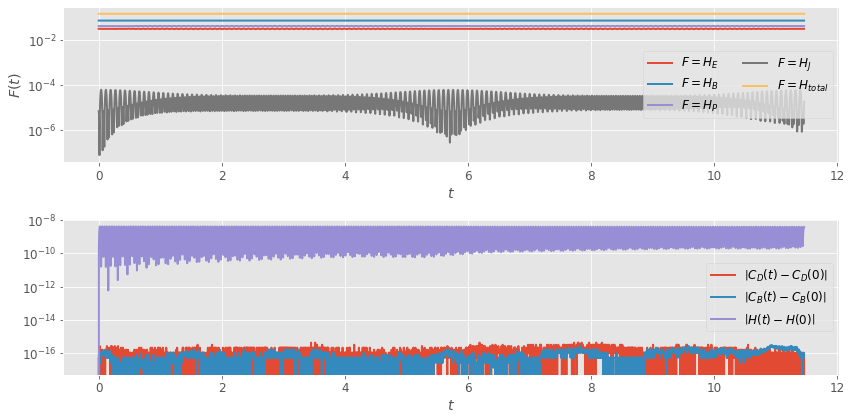}
    \caption{Conservation laws for the kink-shaped traveling wave test computed using $K=1600$ elements and degree-$4$ polynomials for $V_h^0$ and degree-$3$ polynomials for $V_h^1$.}
    \label{fig:kink_cons_laws}
\end{figure}

A solution computed using $K=1600$ elements with $4^{th}$-order interpolation for $V_h^0$ and $3^{th}$-order interpolation for $V_h^1$ and a time-step size of $\Delta t = 2 \| \text{curl}_h \|^{-1} / (5v)$ is visualized in figure \ref{fig:kink_soln} and the energy and Casimir conservation laws are shown in figure \ref{fig:kink_cons_laws}. One can see that the qualitative behavior is as expected with the wave translating across the domain unchanged, but there are small oscillations in the electric field as it evolves in time. It is possible to isolate these oscillations by subtracting the exact traveling wave solution from the computed solution, see figure \ref{fig:error_fields}. It was found that the amplitude of these oscillations relatively insensitive to the spatio-temporal resolution used. It is therefore plausible to conclude that the method is indeed converging, but not to the traveling wave solution predicted by the continuous dynamics; rather, it seems to converge to a solution which is a small perturbation from an exact traveling wave solution. This suggests a defect in the initial data which was obtained from solving a stiff ordinary differential equation. No amount of optimization of the initial data successfully eliminated these oscillations from the simulation. Because of these persistent small errors in the initial condition, the traveling wave exact solution could not be used in a convergence study as was done in \cite{BOKIL2017420}. 

\subsection{Numerical tests in two spatial dimensions}

In two spatial dimensions, one has Casimir invariants given by equation \eqref{eq:disc_casimir_2d}. Of particular interest is the conservation of Gauss's law and near-conservation of energy.

\subsubsection{Higher harmonic generation} \label{section:higher_harmonic_2d}

\begin{figure}
    \centering
    \begin{subfigure}[t]{0.5\textwidth}
    \includegraphics[width=\linewidth]{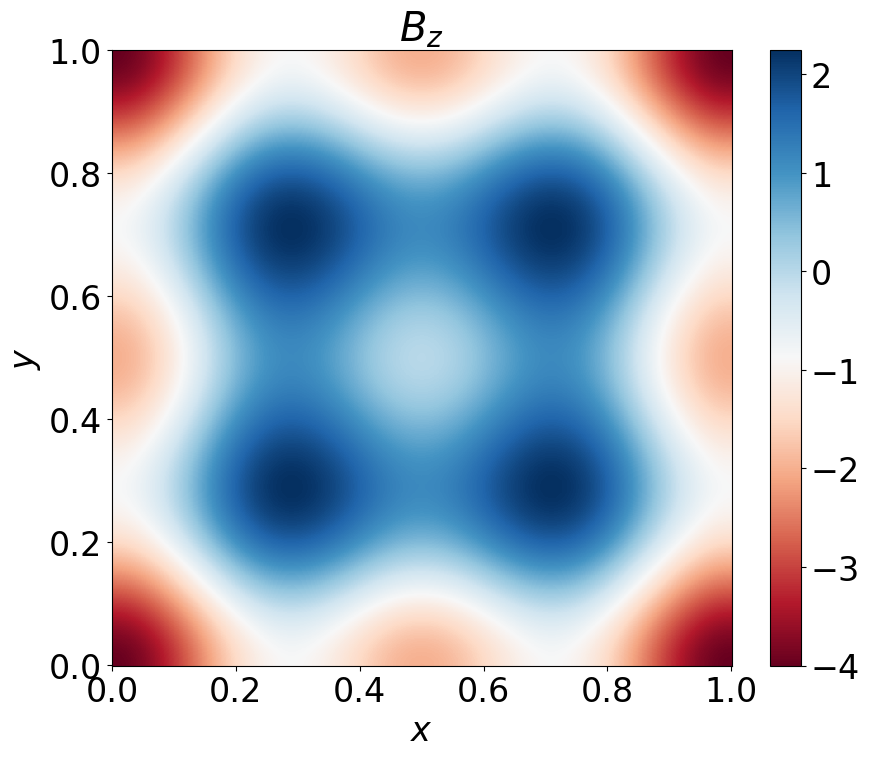}
    \caption{Magnetic field at $t=0.0$.}
    \end{subfigure}%
    \begin{subfigure}[t]{0.5\textwidth}
    \includegraphics[width=\linewidth]{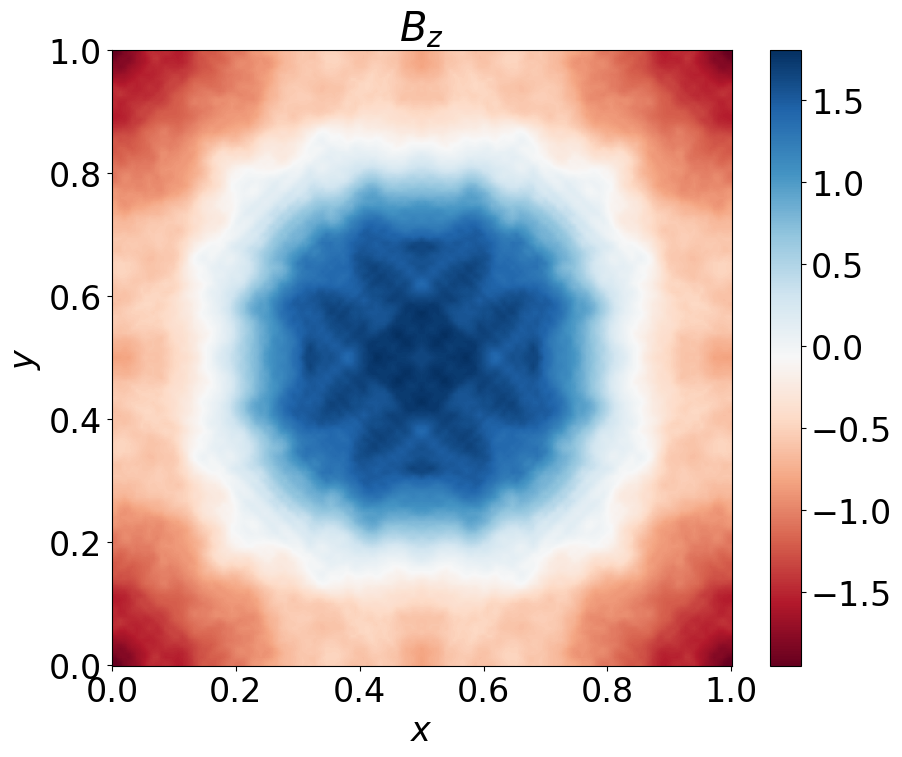}
    \caption{Magnetic field at $t=100.0$.}
    \end{subfigure}
    \begin{subfigure}[t]{0.5\textwidth}
    \includegraphics[width=\linewidth]{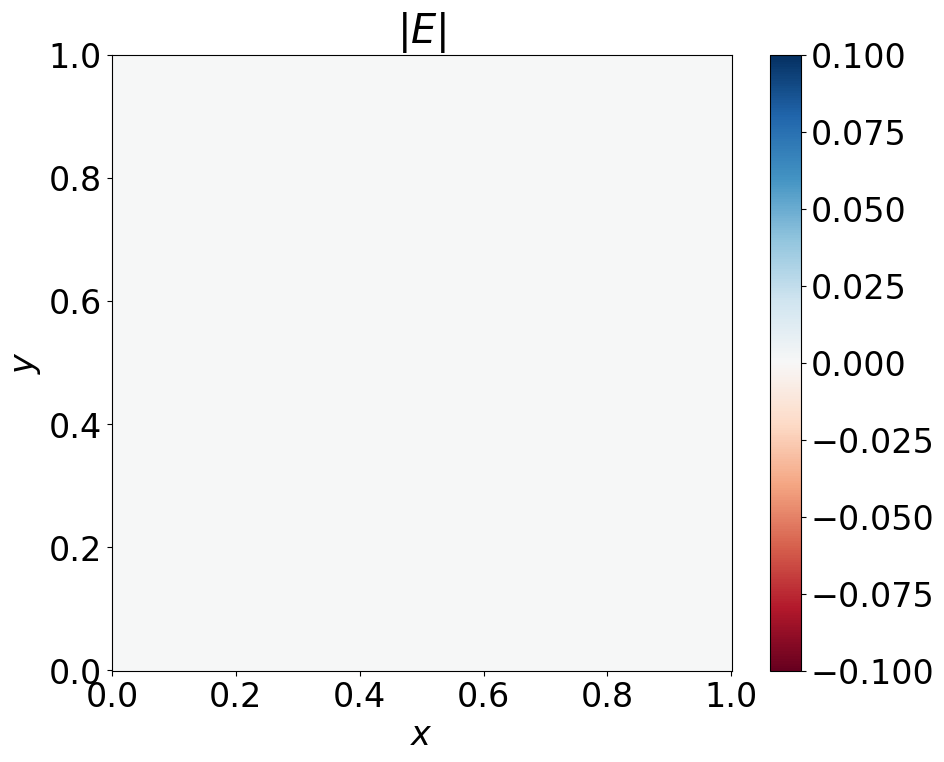}
    \caption{Electric field magnitude at $t=0.0$.}
    \end{subfigure}%
    \begin{subfigure}[t]{0.5\textwidth}
    \includegraphics[width=\linewidth]{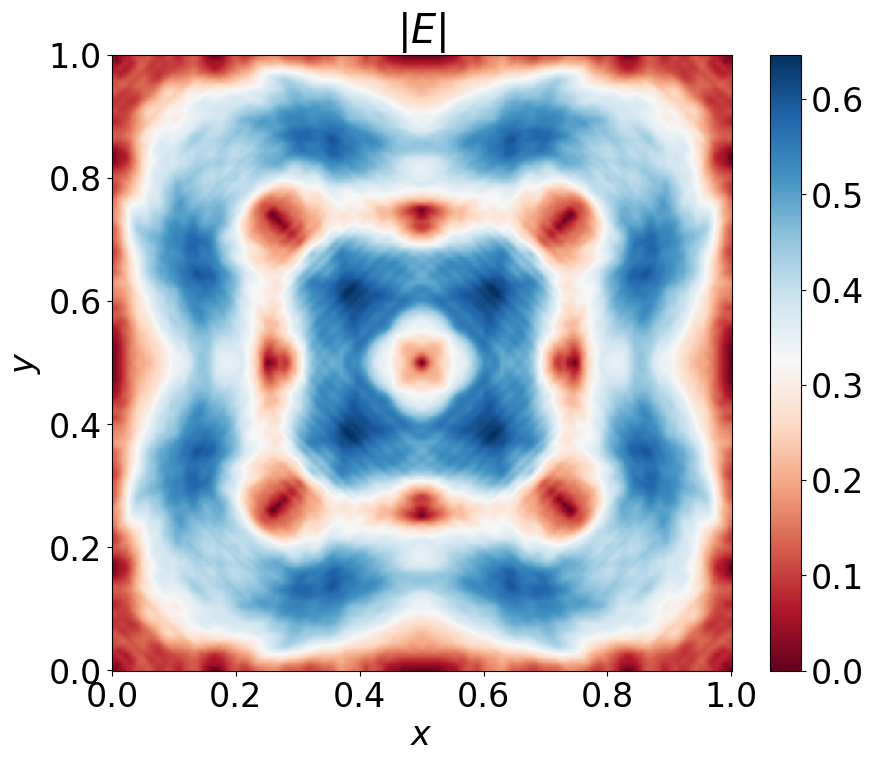}
    \caption{Electric field magnitude at $t=100.0$.}
    \end{subfigure}
    \caption{Visualization of initial and final states for the two-dimensional harmonic generation test.}
    \label{fig:harmonic_gen_vis}
\end{figure}

\begin{figure}
    \centering
    \includegraphics[width=0.75\linewidth]{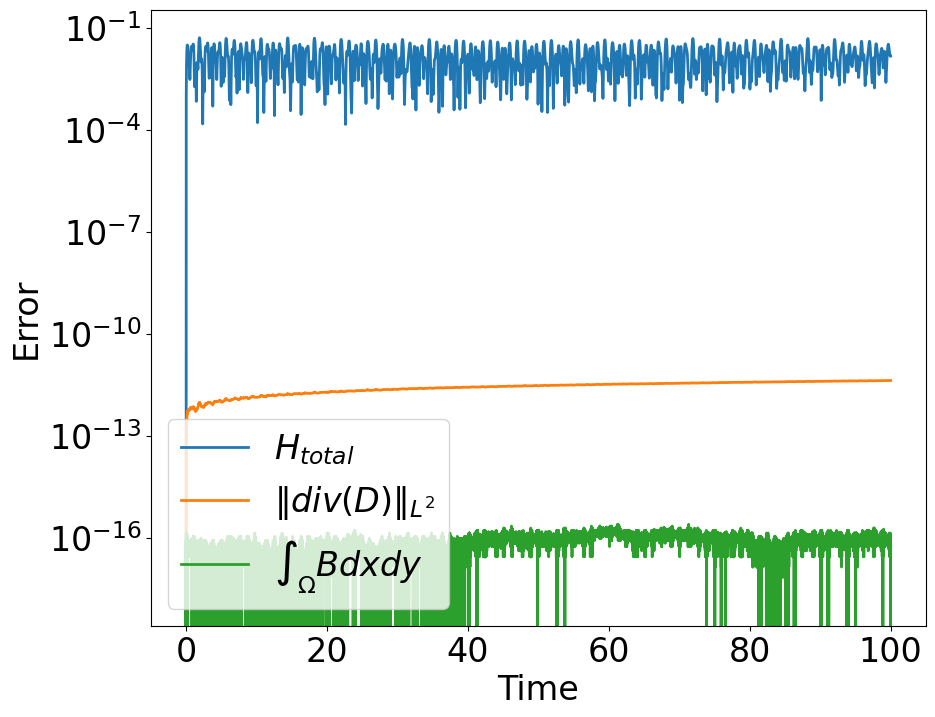}
    \caption{Error in the conservation laws in the two-dimensional harmonic generation test.}
    \label{fig:harmonic_err}
\end{figure}

As in the one-dimensional case, one wishes to investigate the generation of higher harmonics from a sum of simple Fourier modes. The domain is taken to be $\Omega = [0, 1]^2$ with homogeneous boundary conditions, and solutions are computed over the time-interval $t \in [0, 15]$. The initial conditions are
\begin{equation}
    B_0(x,y)
    =
    \cos(2 \pi x) + \cos(2 \pi y) + \cos(4 \pi x) + \cos(4 \pi y) \,,
\end{equation}
and $\bm{E}_0(x,y) = (0,0)$. The simulation is performed both with the standard parameter set in equation \eqref{eq:params_nonlinear_maxwell} but with zero dissipation and the nonlinear parameter increased to $a=0.5$ so that the higher harmonics are activated faster. For the numerical parameters, in both the $x$ and $y$ directions, $K=150$ cells are used and the polynomial degree is $p = 2$ for the $0$-forms and $p=1$ for the $1$-forms. Hence, the temporal resolution is
\begin{equation}
    \Delta t =  \frac{4 \| \text{curl}_h \|_{V_h^1}^{-1} }{5} \approx 0.001051 \,.
\end{equation}
The simulation is performed over the time interval $T=[0.0,100.0]$. See figure \ref{fig:harmonic_gen_vis} for a visualization of the initial and final state. One can see the the appearance of fine structures in the solution at $t=100.0$ as a result of the activation of higher harmonics as the solution evolves. See figure \ref{fig:harmonic_err} for a visualization of the conservation of energy and Casimir invariants. 

\subsubsection{Interface problem}
As a final test, this section considers the interaction of a wave with an interface between a linear and nonlinear medium. The domain is taken to be $\Omega = [0,2] \times [0,1]$, with homogeneous boundary conditions. The domain is partitioned into a linear and a nonlinear portion by introducing spatial dependence to the Hamiltonian:
\begin{equation}
	\begin{aligned}
		H[\bm{E}, \bm{B}, \bm{P}, \bm{J}, Q, \sigma] 
			= 
			\frac{1}{2} \int_\Omega \bigg[ \epsilon_\infty | \bm{E} |^2 
			+ w(\bm{x}) \bigg( \frac{3 a ( 1 - \theta) | \bm{E} |^4}{2} + a \theta Q | \bm{E} |^2 \\
			+ \frac{\omega_0^2}{\omega_p^2} | \bm{P} |^2 + \frac{1}{\omega_p^2}  | \bm{J} |^2
			+ \frac{a \theta}{2} Q^2 + \frac{a \theta}{2 \omega_v^2} \sigma^2 \bigg) 
			+ B^2 \bigg] \mathsf{d}^2 \bm{x}, 
	\end{aligned}
\end{equation}
where $w: \Omega \to [0,1]$ allows one to spatially vary the material properties. To model a vacuum electromagnetic wave entering a nonlinear medium, a diffuse interface model is used:
\begin{equation}
    w(\bm{x}) = \frac{1}{\pi} \arctan (\alpha(x - 1) ) + \frac{1}{2} \,,
\end{equation}
where $\alpha = 100$. Similarly, in \cite{lyu_et_al_2021}, a spatially dependent nonlinear constitutive relation was used to consider a scattering problem in which a nonlinear medium has small air holes specified by the subdomain $D$. The standard material parameters are used but with dissipation set to zero.

\begin{figure}
    \centering
    \begin{subfigure}[t]{0.5\textwidth}
    \includegraphics[width=\linewidth]{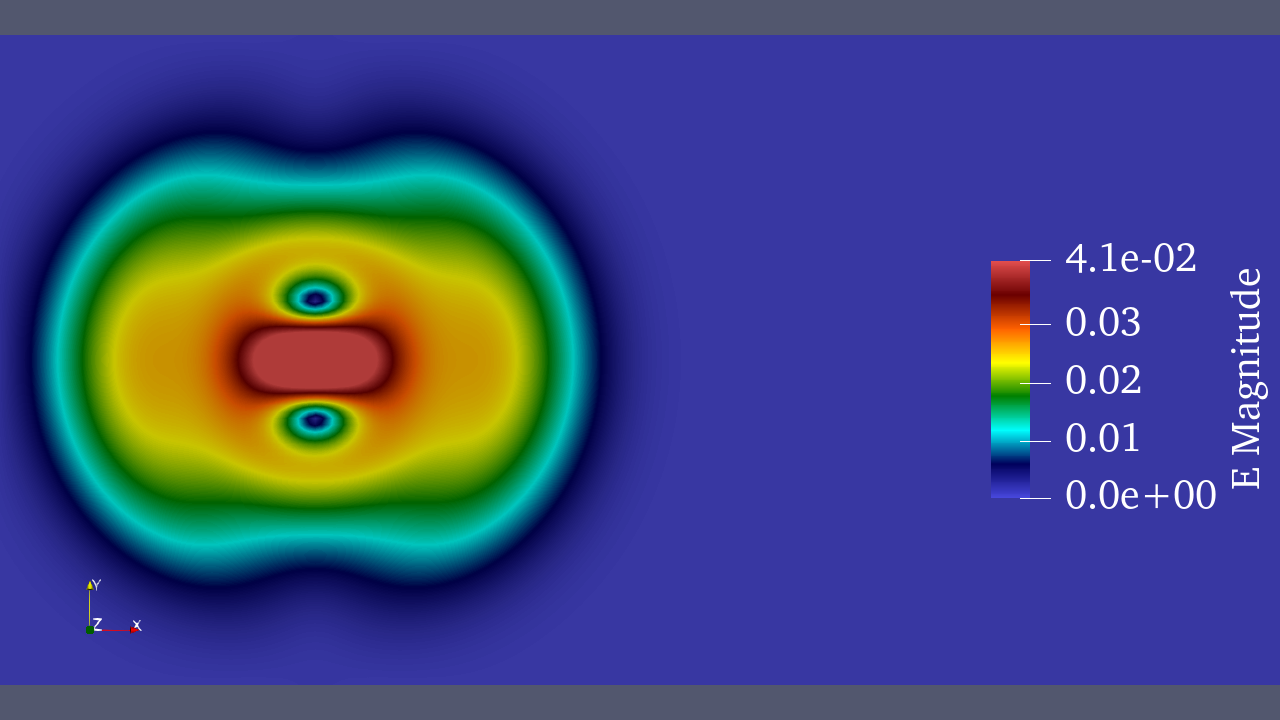}
    \caption{$t=0.5$}
    \end{subfigure}%
    \begin{subfigure}[t]{0.5\textwidth}
    \includegraphics[width=\linewidth]{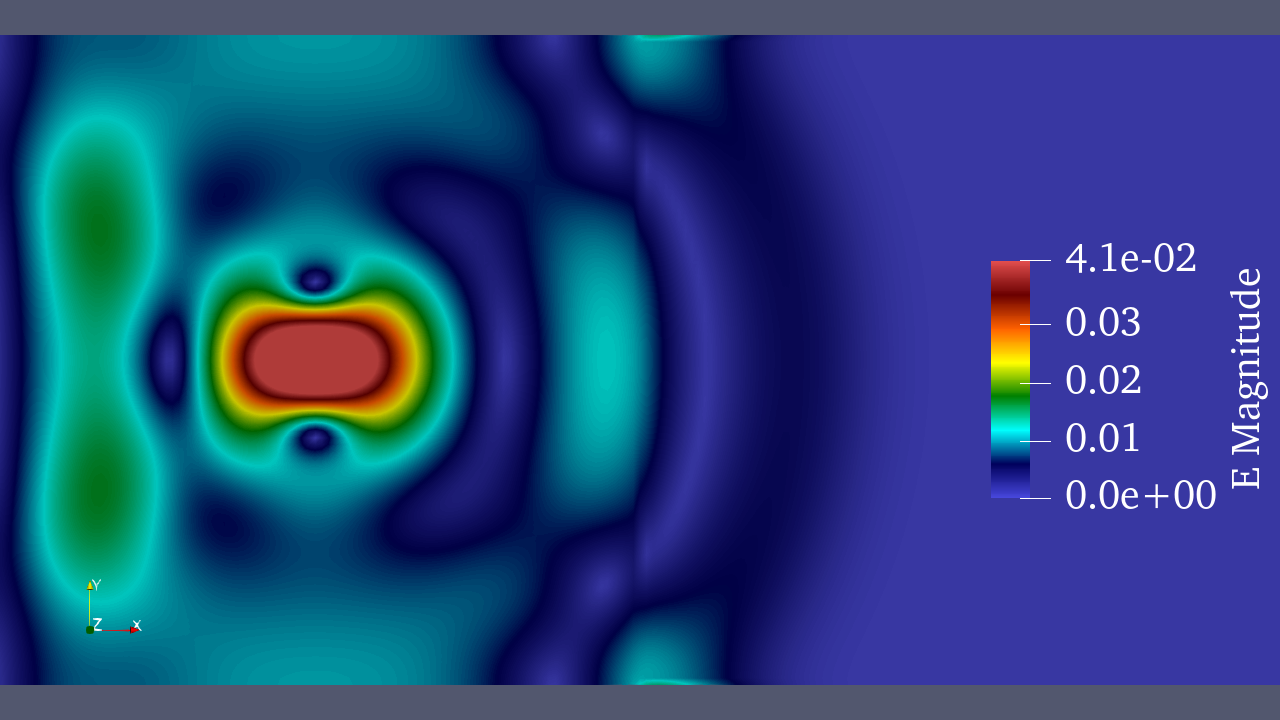}
    \caption{$t=1.0$}
    \end{subfigure}
    \begin{subfigure}[t]{0.5\textwidth}
    \includegraphics[width=\linewidth]{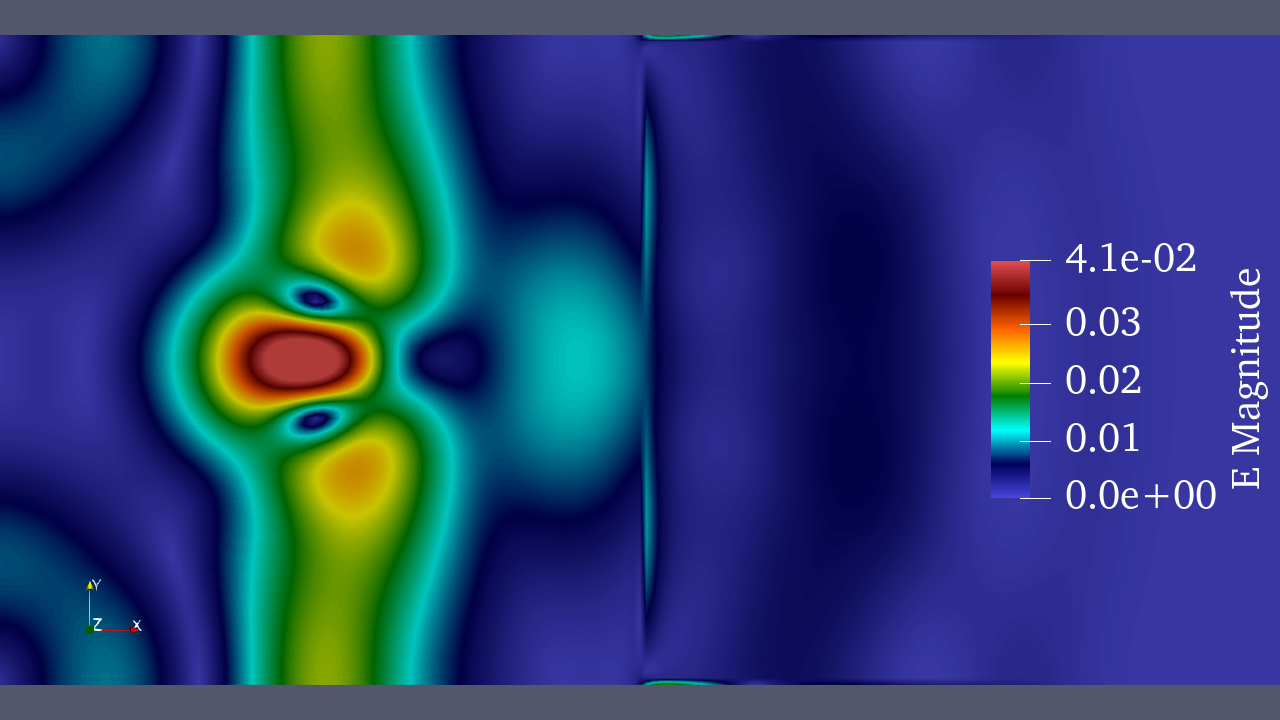}
    \caption{$t=1.5$}
    \end{subfigure}%
    \begin{subfigure}[t]{0.5\textwidth}
    \includegraphics[width=\linewidth]{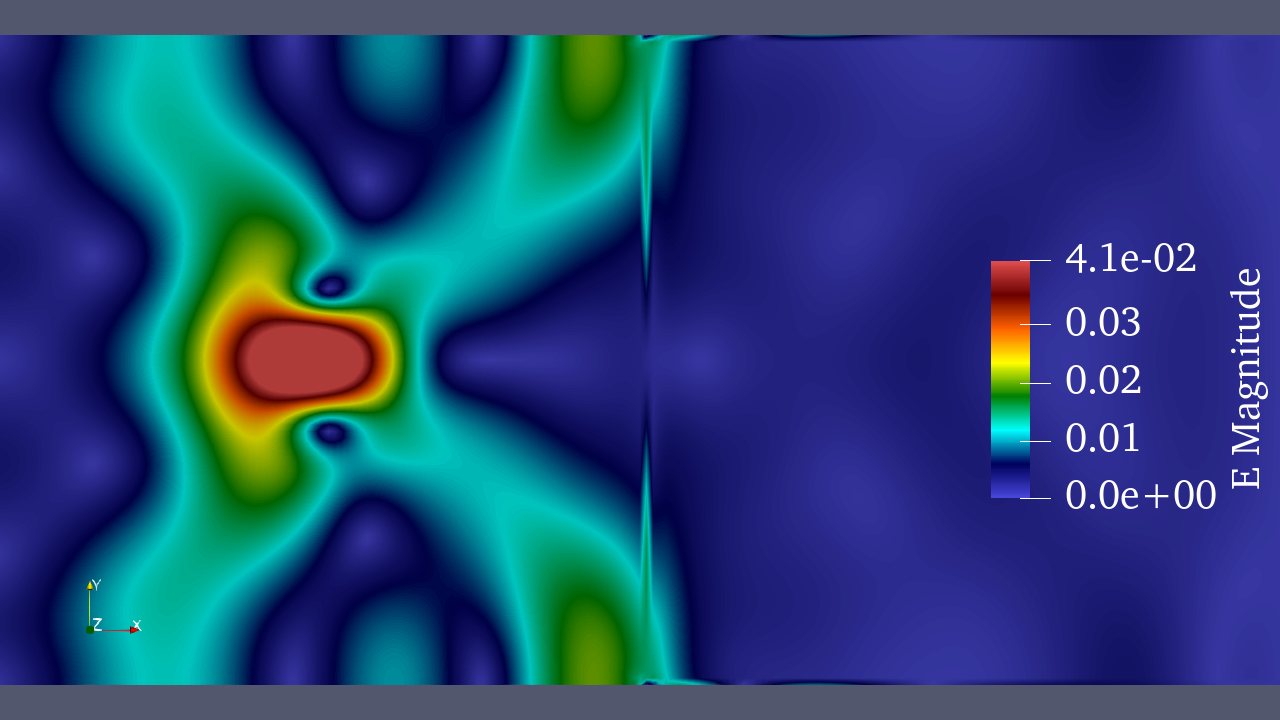}
    \caption{$t=2.0$}
    \end{subfigure}
    \begin{subfigure}[t]{0.5\textwidth}
    \includegraphics[width=\linewidth]{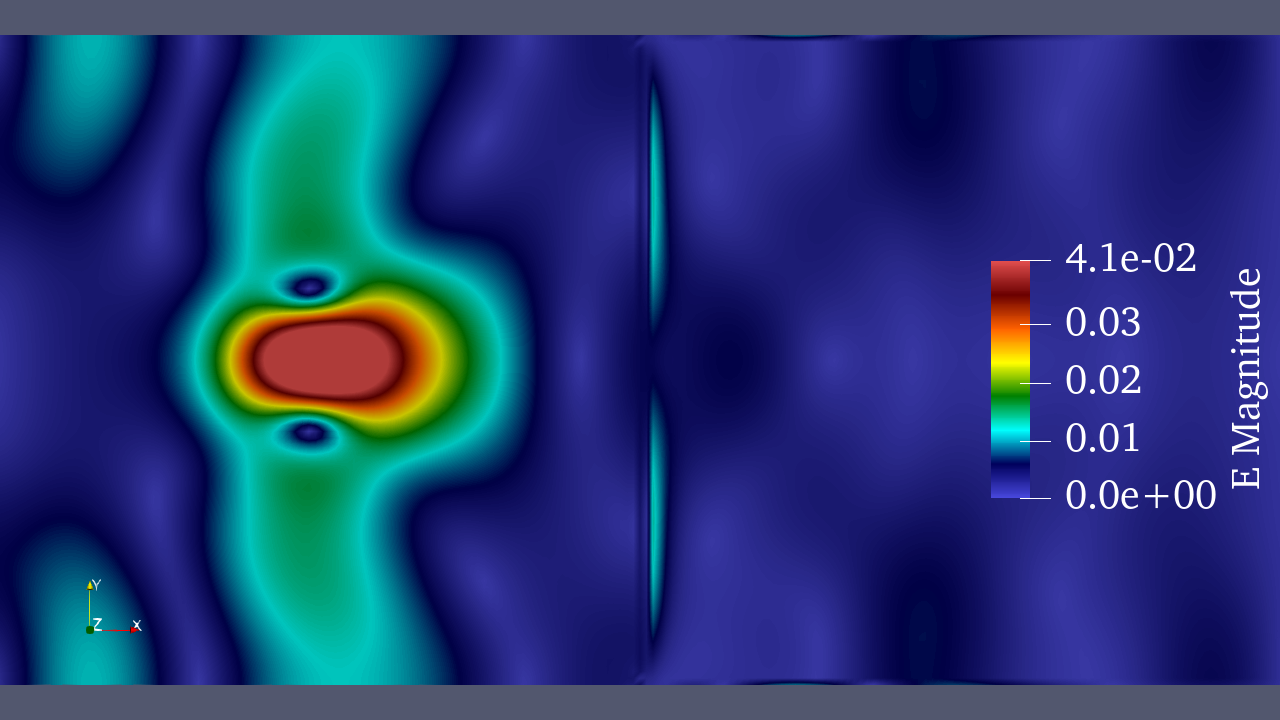}
    \caption{$t=2.5$}
    \end{subfigure}%
    \begin{subfigure}[t]{0.5\textwidth}
    \includegraphics[width=\linewidth]{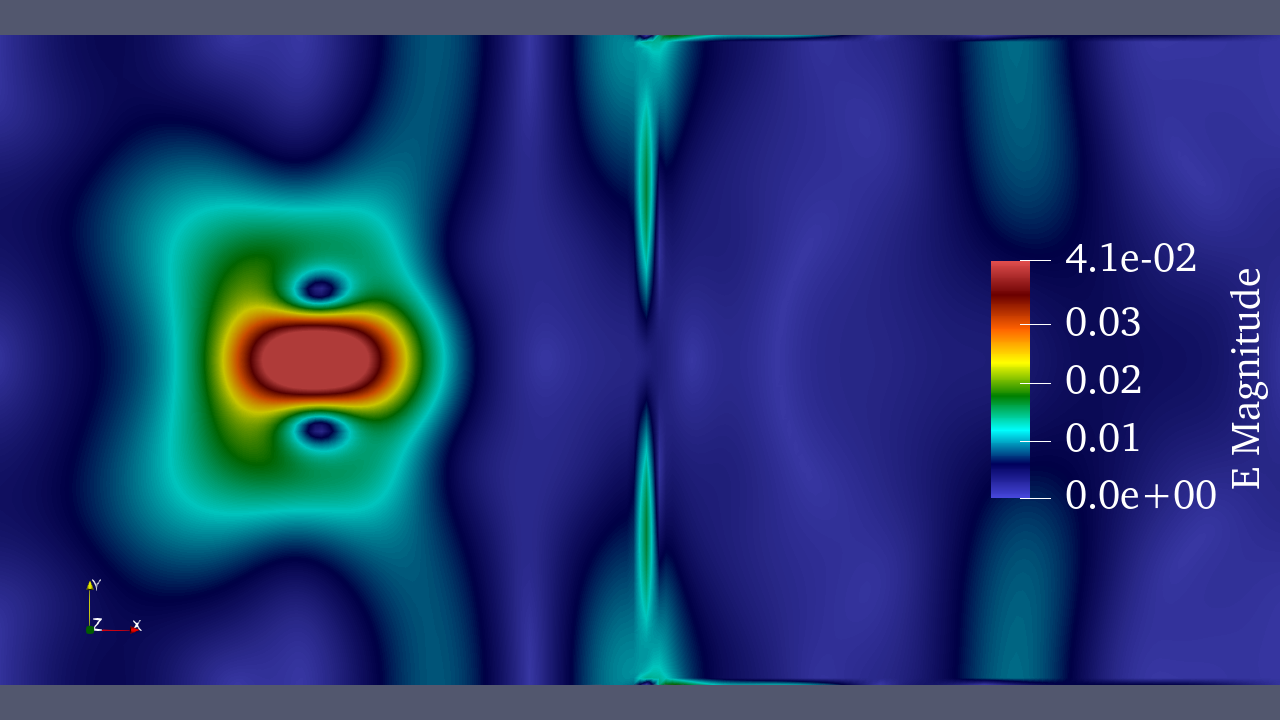}
    \caption{$t=3.0$}
    \end{subfigure}
    \begin{subfigure}[t]{0.5\textwidth}
    \includegraphics[width=\linewidth]{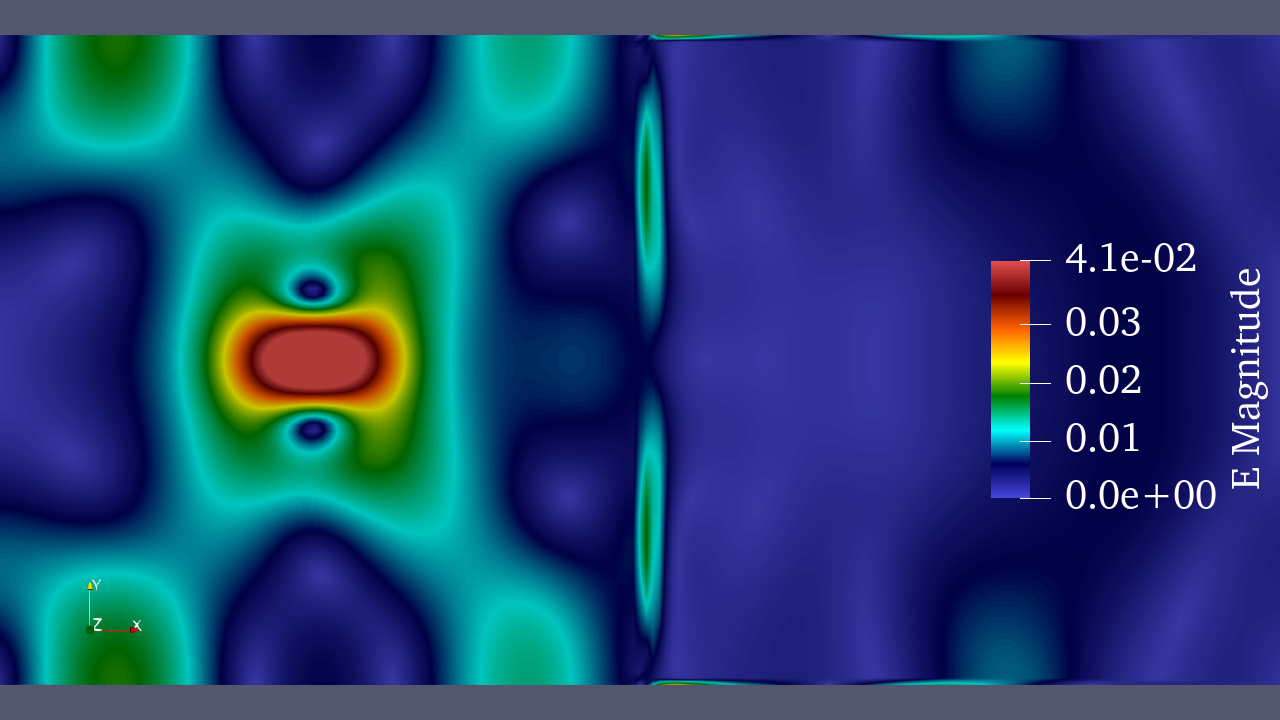}
    \caption{$t=3.5$}
    \end{subfigure}%
    \begin{subfigure}[t]{0.5\textwidth}
    \includegraphics[width=\linewidth]{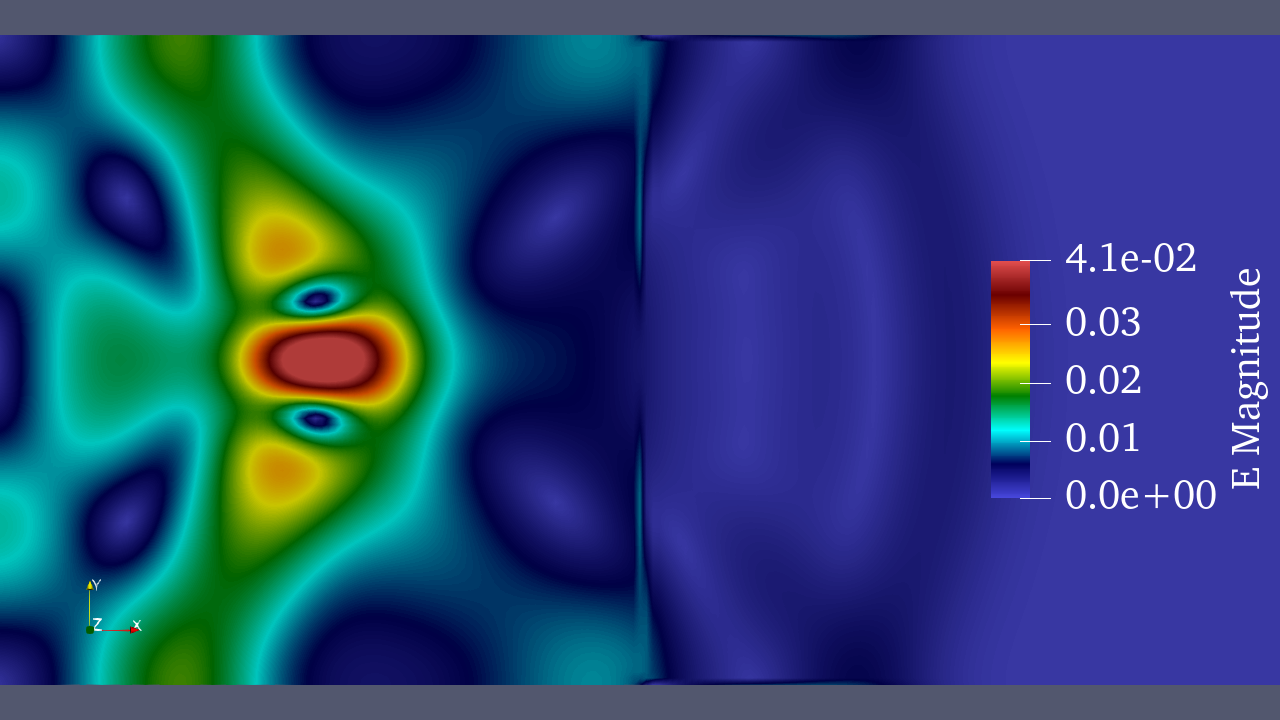}
    \caption{$t=4.0$}
    \end{subfigure}
    \caption{Magnitude of $\bm{E}$ for the interface problem.}
    \label{fig:interface}
\end{figure}

\begin{figure}
    \centering
    \includegraphics[width=0.75\linewidth]{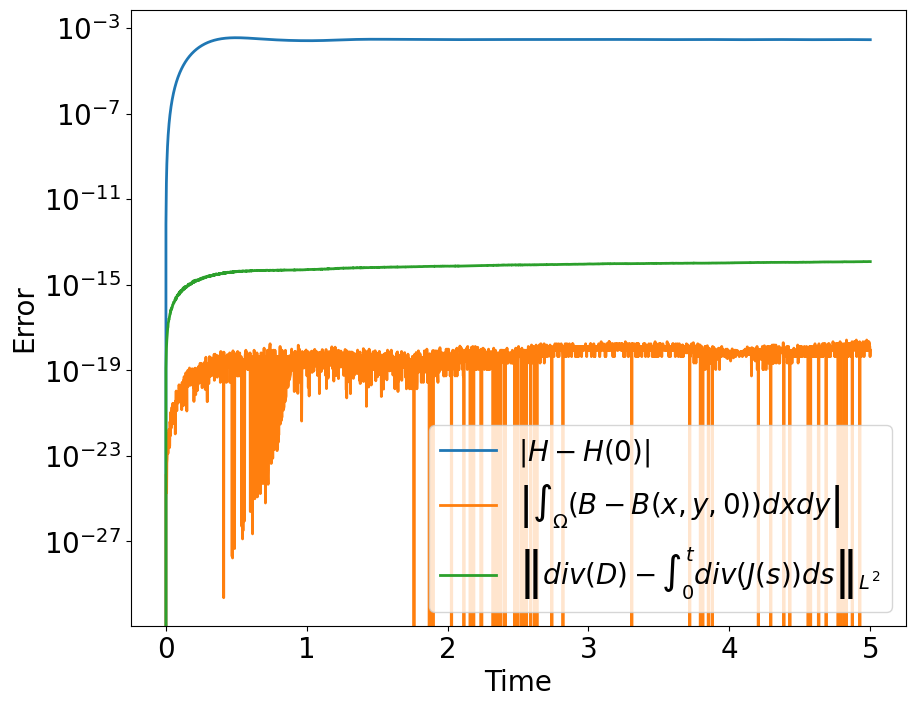}
    \caption{Error in the conservation laws for the interface problem.}
    \label{fig:interface_err}
\end{figure}

The simulation uses $K = (K_x, K_y) = (200, 100)$ elements with polynomial order $N = 2$, and the time-step is taken to be
\begin{equation}
    \Delta t = \frac{4 \| \text{curl}_h \|_{V_h^1}^{-1} }{5} 
    \approx 0.001576 \,.
\end{equation}
The time interval for the simulation is $T = [0, 5.0]$. The initial data is set equal to zero, and time-dependent right-hand side forcing is imposed via a free current distribution $\bm{J}_f = (J_{f,x}, J_{f,y})$ with $J_{f,x} = 0$ and
\begin{equation}
    J_{f,y}
    =
    A \exp(-t/\tau) \sin(2 \pi t) \exp \left( - \frac{(x - x_0)^2 + (y-y_0)^2}{\sigma^2} \right) \,,
\end{equation}
where $\tau = 0.5$, $(x_0, y_0) = (0.5,0.5)$, and $A = 2$. This current is not divergence free, and therefore causes $\msf{div}(\bm{D}) = \rho_f$ to evolve as
\begin{equation}
    \partial_t \msf{div}(\bm{D}) = \partial_t \rho_f = \msf{div} \bm{J}_f \,.
\end{equation}
See figures \ref{fig:interface} for a visualization of the magnitude of the electric field as the simulation progresses. See \ref{fig:interface_err} for the error in the conservation laws as a function of time. 

%% file: appendix.tex
\appendix

\input{appendices/variational_nl_maxwell}

\input{appendices/scalar_integrals}

\input{appendices/lipschitz}

\input{appendices/kink_solns}

%% file: appendices/variational_nl_maxwell.tex
\section{A variational formulation of nonlinear Maxwell models} \label{appendix:variational-NL-maxwell}

This appendix briefly considers the reason for the particular form of the nonlinear constitutive relations in the Hamiltonian formulation of the nonlinear Maxwell equations. In SI units, $\bm{D}$ and $\bm{A}$ are are canonically conjugate. One may define the electromagnetic ``kinetic energy'' and ``potential energy'' as
\begin{equation}
	T = \frac{\epsilon_0}{2} \int | \dot{\bm{A}} |^2 \mathsf{d}^3 \bm{x}
	\quad \text{and} \quad
	V = \frac{\mu_0^{-1}}{2} \int | \nabla \times \bm{A} |^2 \mathsf{d}^3 \bm{x}.
\end{equation}
respectively. The Lagrangian is then given by
\begin{equation}
	L = T - V. 
\end{equation}
One then defines the canonical momentum via a Legendre transform: $\bm{D} = \delta L/\delta \dot{\bm{A}}$, and the Hamiltonian is given by
\begin{equation}
    H = \int \bm{D} \cdot \dot{\bm{A}} \mathsf{d}^3 \bm{x} 
    - L 
    = \int \bm{D} \cdot \dot{\bm{A}} \mathsf{d}^3 \bm{x} 
    - \frac{1}{2} \int \left(  \epsilon_0 | \dot{\bm{A}} |^2 
    - \mu_0^{-1} | \nabla \times \bm{A} |^2  \right) \mathsf{d}^3 \bm{x}
\end{equation}
Letting $\bm{E} = \dot{\bm{A}}$ and $\bm{B} = \nabla \times \bm{A}$, one finds
\begin{equation}
    H = \int \bm{D} \cdot \bm{E} \mathsf{d}^3 \bm{x} 
    - \frac{1}{2} \int \left( \epsilon_0 | \bm{E} |^2 
    - \mu_0^{-1} | \bm{B} |^2 \right) \mathsf{d}^3 \bm{x}
\end{equation}
which is the usual electromagnetic energy in a vacuum since $\bm{D} = \epsilon_0 \bm{E}$ for this choice of $T$. Hence, one recovers the vacuum Maxwell equations. 

It is possible to generalize this construction to nonlinear media in which the refractive index is a perturbation from the identity. Letting
\begin{equation}
	T = \frac{\epsilon_0}{2} \int | \dot{\bm{A}} |^2 \mathsf{d}^3 \bm{x} - K,
\end{equation}
where $K$ is an arbitrary functional of the fields, one finds
\begin{equation}
    H = \int \bm{D} \cdot \dot{\bm{A}} \mathsf{d}^3 \bm{x} 
    + K \\
    - \frac{1}{2} \int \left( \epsilon_0 | \dot{\bm{A}} |^2 
    - \mu_0^{-1} | \nabla \times \bm{A} |^2  \right) \mathsf{d}^3 \bm{x}
\end{equation}
where, since $\bm{E} = \dot{\bm{A}}$, 
\begin{equation}
    \bm{D} = \frac{\delta L}{\delta \bm{\dot{A}}} 
    = \epsilon_0 \dot{\bm{A}} - \frac{\delta K}{\delta \dot{\bm{A}}}
    = \epsilon_0 \bm{E} - \frac{\delta K}{\delta \bm{E}} \,,
\end{equation}
and
\begin{multline}
    H = \int \bm{D} \cdot \bm{E} \mathsf{d}^3 \bm{x} 
    + K - \frac{1}{2} \int \left( \epsilon_0 | \bm{E} |^2 
    - \mu_0^{-1} | \bm{B} |^2  \right) \mathsf{d}^3 \bm{x} \\
    = K - \int \frac{\delta K}{\delta \bm{E}} \cdot \bm{E} \mathsf{d}^3 \bm{x} 
    + \frac{1}{2} \int \left( \epsilon_0 | \bm{E} |^2 
    + \mu_0^{-1} | \bm{B} |^2 \right) \mathsf{d}^3 \bm{x} \,,
\end{multline}
where $H = H[\bm{D}, \bm{B}]$. Hence, each occurrence of $\bm{E}$ is an implicit function of the dynamical fields $(\bm{D}, \bm{B})$. This explains the particular form of the electromagnetic energy in a perturbed nonlinear medium. 

The canonical bracket is simply
\begin{equation}
    \{F, G\} = \int \left( \frac{\delta F}{\delta \bm{D}} \cdot \frac{\delta G}{\delta \bm{A}} 
    - \frac{\delta G}{\delta \bm{D}} \cdot \frac{\delta F}{\delta \bm{A}} \right) \mathsf{d}^3 \bm{x}. 
\end{equation}
Hence, in the variables $(\bm{D}, \bm{B})$, the bracket is 
\begin{equation}
    \{F, G\} = \int \left( \frac{\delta F}{\delta \bm{D}} \cdot \nabla \times \frac{\delta G}{\delta \bm{B}} 
    - \frac{\delta G}{\delta \bm{D}} \cdot \nabla \times \frac{\delta F}{\delta \bm{B}} \right) \mathsf{d}^3 \bm{x}. 
\end{equation}
Assuming the Lagrangian is hyperregular, because the Legendre transform which maps $\dot{\bm{A}}$ to $\bm{D}$ is involutive, one finds that $\delta H/\delta \bm{D} = \dot{\bm{A}} = \bm{E}$. Finally, the auxiliary magnetic field is simply defined via $\bm{H} = \delta H / \delta \bm{B} = \mu_0^{-1} \bm{B} + \delta K/\delta \bm{B}$. One therefore recovers the macroscopic Maxwell equations. 

This has several consequences when considering the Lagrangian and Hamiltonian formulation of Maxwell's equations in general media. Any asymptotic procedure which introduces changes to the electromagnetic energy must likewise induce the appropriate compensating polarization and magnetization prescribed by the formalism discussed herein. Moreover, even in linear media, $(\bm{D}, \bm{A})$ are the naturally canonically conjugate variables with $\bm{D}$ and $\bm{E}$ being related to each other in the same manner in which the canonical momentum and velocity are related in classical mechanics. This framework facilitates building theories of nonlinear electrodynamics which are not perturbations of the linear theory. Finally, straightforward modifications of this procedure facilitate a Lagrangian and Hamiltonian formulation of nonlinear waves in general. 

%% file: appendices/scalar_integrals.tex
\section{Computing integrals of functions in the scalar FEEC spaces} \label{appendix:integral_formulas}
This appendix demonstrates that integrals of functions in the scalar function spaces in tensor product or mapped tensor product de Rham sequences with a partition of unity basis for the $0$-forms (e.g.\ B-spline or spectral element FEEC methods) are simple to compute with an explicit formula. This result holds in arbitrary dimensions since the $0$-form basis is simply the $n$-fold product of the $0$-form basis in $1$D, and the $n$-form basis is the $n$-fold product of the $1$-form basis in $1$D. Hence, the result is shown in general although it is only used in $1$D and $2$D in this work. 

\begin{lemma} \label{eq:integral_formulas}
Let $\text{dim}(\Omega) = n$. Suppose $v_h \in V^0_h$ and $u_h \in V^n_h$. Then denoting $v_h = \sum_{\bm{\alpha}} \Lambda_{0,\bm{\alpha}} \msf{v}_{\bm{\alpha}}$ and $u_h = \sum_{\bm{\alpha}} \Lambda_{n,\bm{\alpha}} \msf{u}_{\bm{\alpha}}$ where $\bm{\alpha}$ is a multiindex representing the degrees of freedom, it follows that 
\begin{equation}
    \int_\Omega v_h(\bm{x}) \mathsf{d}^n \bm{x} = \sum_{ \bm{\alpha} , \bm{\beta} } (\mathbb{M}_0)_{\bm{\alpha} \bm{\beta}} \msf{v}_{\bm{\beta}} = \mathbbm{1}^T \mathbb{M}_0 \bmsf{v}
    \quad \text{and} \quad
    \int_\Omega u_h \mathsf{d}^n \bm{x} = \sum_{\bm{\alpha}} \msf{u}_{\bm{\alpha}} = \mathbbm{1}^T \bmsf{u}
\end{equation}
where $\mathbbm{1}_{\bm{\alpha}} = 1 \ \forall \bm{\alpha}$. 
\end{lemma}

\noindent \textit{Proof:} The proof proceeds by direct calculation and application of the definitions. First, the result is proven for the $0$-forms. Suppose the domain is made up of mapped elements such that $\Omega = \bigcup_k \Omega_k$ and suppose each element is related to a reference element by $F_k: \hat{\Omega} \to \Omega_k$. Let the multiindex $\alpha = (\bm{I}, k)$ where $k$ denotes the $k^{th}$ element and $\bm{I}$ is a multiindex for the degrees of freedom within each element. Hence,
\begin{equation}
	\begin{aligned}
		\int_\Omega v_h \mathsf{d}^n \bm{x} 
			&= \sum_{\bm{I}, k} \int_{\Omega_k} \Lambda_{0,(\bm{I},k)}( \bm{x} ) \msf{v}_{\bm{I},k} \mathsf{d}^n \bm{x} \\
			&= \sum_{\bm{I}, k} \left( \int_{\hat{\Omega}} \hat{\Lambda}_{0,\bm{I}}( \bm{\xi} ) (\det(F_k))^{-1}(\bm{\xi}) \mathsf{d}^n \bm{\xi} \right) \msf{v}_{\bm{I},k}.
	\end{aligned}
\end{equation}
Recall, the $0$-form shape functions are a partition of unity in a B-spline FEEC method: $\sum_{\bm{I}} \hat{\Lambda}_{0,\bm{I}}(\bm{\xi}) = 1$ $\forall \bm{\xi} \in \hat{\Omega}$. Hence,
\begin{multline}
	\sum_{\bm{I}, k} \left( \int_{\hat{\Omega}} \hat{\Lambda}_{0,\bm{I}}( \bm{\xi} ) (\det(F_k))^{-1}(\bm{\xi}) \mathsf{d}^n \bm{\xi} \right) \msf{v}_{\bm{I},k} \\
		= \sum_{\bm{I}, \bm{J}, k} \left( \int_{\hat{\Omega}} \hat{\Lambda}_{0,\bm{I}}( \bm{\xi} ) 
			\hat{\Lambda}_{0,\bm{J}}( \bm{\xi} ) (\det(F_k))^{-1}(\bm{\xi}) \mathsf{d}^n \bm{\xi} \right) \msf{v}_{\bm{I},k} \\
		= \sum_{\bm{I}, \bm{J}, k} ( \mathbb{M}_0^{(k)} )_{\bm{I}, \bm{J}} \msf{v}_{\bm{I}, k} = \mathbbm{1}^T \mathbb{M}_0 \bmsf{v}.
\end{multline}
where $\mathbb{M}_0^{(k)}$ represents the $0$-form mass matrix for the $k^{th}$ element. 

Now, the result is proven for the $n$-forms. Recall, the reference element is subdivided into sub-rectangles such that $\hat{\Omega} = \bigcup_{\bm{J}} \hat{\Omega}_{\bm{J}}$ so that the $n$-form degrees of freedom operator (on the reference domain) might be defined as integrals over these sub-rectangles:
\begin{equation}
	\hat{\sigma}_{\bm{J}}( \hat{v} ) = \int_{\hat{\Omega}_{\bm{J}}} \hat{v}(\bm{\xi}) \mathsf{d}^n \bm{\xi} 
\end{equation}
Hence, one finds
\begin{equation}
	\begin{aligned}
		\int_\Omega u_h(\bm{x}) \mathsf{d}^n \bm{x} 
			&= \sum_{\bm{I}, k} \int_{\Omega_k} \Lambda_{1,(\bm{I},k)}(\bm{x}) \msf{u}_{\bm{I},k} \mathsf{d}^n \bm{x} 
				= \sum_{\bm{I}, k} \left( \int_{\hat{\Omega}} \hat{\Lambda}_{1,\bm{I}}( \bm{\xi} ) \mathsf{d}^n \bm{\xi} \right) \msf{u}_{\bm{I},k} \\
			&= \sum_{\bm{I}, \bm{J}, k} \left( \int_{\hat{\Omega}_{\bm{J}}} \hat{\Lambda}_{1,\bm{I}}(\bm{\xi}) \mathsf{d}^n \bm{\xi} \right) \msf{u}_{\bm{I},k} 
				= \sum_{\bm{I}, \bm{J}, k} \delta_{\bm{I} \bm{J}} \msf{u}_{\bm{I},k} 
                    		= \sum_{\bm{I}, k} \msf{u}_{\bm{I},k} 
				= \mathbbm{1}^T \bmsf{u}
	\end{aligned}
\end{equation}
where the fact that $F_k^*(\Lambda_{1,\bm{I},k}(\bm{x}) \mathsf{d}^n \bm{x}) = \hat{\Lambda}_{1, \bm{I}}(\bm{\xi}) \mathsf{d}^n \bm{\xi}$ and
\begin{equation}
	\int_{\hat{\Omega}_{\bm{J}}} \hat{\Lambda}_{1, \bm{I}}(\bm{\xi}) \mathsf{d}^n \bm{\xi} 
		= \hat{\sigma}_{\bm{J}}( \hat{\Lambda}_{1, \bm{I}}(\bm{\xi}) ) = \delta_{\bm{I}\bm{J}} 
\end{equation} 
were used.

\qed

%% file: appendices/lipschitz.tex
\section{A Local Lipschitz bound for the inverse of the nonlinear perturbation of the identity} \label{appendix:lipschitz-bound}

In this appendix a Lipschitz bound is sought for the implicit map defined as the inverse of the perturbation of the identity operator by a nonlinear map in some Banach space. This provides a technical lemma needed to prove the CFL condition for the time-stepping method of the nonlinear Maxwell solver. 

\subsection{Result in general Banach spaces}
As a first step, a general Lipschitz bound for implicitly defined weakly nonlinear maps in general Banach spaces is derived. 
\begin{theorem}
Suppose $(X, \| \cdot \|_X)$ is a Banach space, and let $g: X \to X$ be a Fr{\'e}chet differentiable map in a neighborhood of the origin. Moreover, suppose $Dg(0) = 0$ so that $Df(0) = I$. Define $f(x) \coloneq x + \epsilon g(x)$. Then $\exists r, L_-, L_+ > 0$ such that for $x, y \in B_r(0)$, 
\begin{equation}
    L_- \| x - y \|_X \leq \| f(x) - f(y) \|_X \leq L_+ \| x - y \|_X \,.
\end{equation}
In particular, this ensures that, where $f^{-1}$ is defined,
\begin{equation}
    \| f^{-1}(x) - f^{-1}(y) \| \leq L_-^{-1} \| x - y \|_X \,,
\end{equation}
so that the implicitly defined inverse operator is bounded. 
\end{theorem}

\noindent \textit{Proof:}
First note that the hypotheses of the inverse function theorem are satisfied in some neighborhood of the origin. Hence, $f^{-1}: B_r(0) \to X$ is well defined for some $r>0$. Evidently, because the map, $g$, is Fr{\'e}chet differentiable,
\begin{equation}
    \frac{\| f(x) - f(y) \|_X}{ \| x - y \|_X } \leq  \sup_{u \in B_r(0)} \| I + \epsilon D g(u) \|_X \,.
\end{equation}
Moreover, one has
\begin{equation}
\begin{aligned}
    \| x - y \|_X &= \| x - y - (f(x) - f(y)) + (f(x) - f(y)) \|_X \\
    &\leq \epsilon \sup_{u \in B_r(0)} \| D g(u) \|_X \| x - y \|_X + \| f(x) - f(y) \|_X \\
    &\quad \implies \quad 
    \left( 1 - \epsilon \sup_{u \in B_r(0)} \| D g(u) \|_X \right) \| x - y \|_X
        \leq \| f(x) - f(y) \|_X \,.
\end{aligned}
\end{equation}
Assuming one has chosen $r$ such that $\sup_{u \in B_r(0)} \| D g(u) \|_X < \epsilon^{-1}$, the result follows. \qed \\[0.1em]

\noindent Hence, one has a Lipschtz bound on the implicitly defined inverse map:
\begin{equation}
    \| f^{-1} (x) - f^{-1}(y) \|_X \leq \left( 1 - \epsilon \sup_{u \in B_r(0)} \| D g(u) \|_X \right)^{-1} \| x - y \|_X \,.
\end{equation}
The result still holds if $g$ is only defined on a complete subspace of $X$, that is, if one redefines $g: D(X) \supset X \to X$ where $D(X)$ is itself Banach. 

More broadly, suppose one not only has a perturbation by a nonlinear map, but also a linear change of coordinates, that is, a map of the form $f(x) \coloneq Ax + \epsilon g(x)$, where $A:X \to X$ is an invertible bounded linear map. 

\begin{theorem} \label{thm:bounded_weakly_nl_map}
Suppose $(X, \| \cdot \|_X)$ is a Banach space, and let $g: X \to X$ be a Fr{\'e}chet differentiable map in a neighborhood of the origin. Moreover, suppose $Dg(0) = 0$ so that $Df(0) = I$. Let $f(x) \coloneq A x + \epsilon g(x)$ where $A: X \to X$ is an invertible bounded linear map. Then $\exists r, L_-, L_+ > 0$ such that for $x, y \in B_r(0)$, 
\begin{equation}
    L_- \| x - y \|_X \leq \| f(x) - f(y) \|_X \leq L_+ \| x - y \|_X \,.
\end{equation}
In particular, this ensures that, where $f^{-1}$ is defined,
\begin{equation}
    \| f^{-1}(x) - f^{-1}(y) \| \leq L_-^{-1} \| x - y \|_X \,,
\end{equation}
so that the implicitly defined inverse operator is bounded. 
\end{theorem}

\noindent \textit{Proof:}
The proof proceeds in an identical manner to the previous case for the upper Lipschitz constant:
\begin{equation}
    \frac{\| f(x) - f(y) \|_X}{ \| x - y \|_X } \leq  \sup_{u \in B_r(0)} \| A + \epsilon D g(u) \|_X \,.
\end{equation}
For the lower Lipschitz constant, one begins by stating the result of the previous theorem if one instead perturbs by $A^{-1} g$:
\begin{equation}
\begin{aligned}
    \| x - y \|_X &= \| x - y - (A^{-1} f(x) - A^{-1} f(y)) + (A^{-1} f(x) - A^{-1} f(y)) \|_X \\
    &\leq \epsilon \sup_{u \in B_r(0)} \| A^{-1} \|_X \| D g(u) \|_X \| x - y \|_X + \| A^{-1} \|_X \| f(x) - f(y) \|_X \\
    &\quad \implies \quad 
    \left( \| A^{-1} \|_X - \epsilon \sup_{u \in B_r(0)} \| D g(u) \|_X \right) \| x - y \|_X
        \leq \| f(x) - f(y) \|_X \,.
\end{aligned}
\end{equation}
Assuming one has chosen $r$ such that $\sup_{u \in B_r(0)} \| D g(u) \|_X < \epsilon^{-1} \| A^{-1} \|_X $, the result follows. \qed \\[0.1em]

\noindent Hence, one has a Lipschtz bound on the implicitly defined inverse map:
\begin{equation}
    \| f^{-1} (x) - f^{-1}(y) \|_X \leq \left( \| A^{-1} \|_X - \epsilon \sup_{u \in B_r(0)} \| D g(u) \|_X \right)^{-1} \| x - y \|_X \,.
\end{equation}
This provides a suitable local bound for the inverse map of a change of coordinates which has a linear part which is perturbed by a nonlinear part. 

\subsection{A local Lipschitz bound for the electric field in cubicly nonlinear media} \label{appendix:bound_e-field}

The result from the previous section is now used to bound the electric field as a function of the dynamical fields $(\bm{D}, \bm{P}, Q)$ as defined in equation \eqref{eq:cubic_nl_maxwell_sys}. Note, one may ignore the remaining fields in this coordinate transformation, $(\bm{B}, \bm{J}, \sigma)$, as they do not contribute to the nonlinear coordinate change. Recall, 
\begin{equation}
    \bm{D} = \bm{P} + \left[ \epsilon_\infty + a \left( (1 - \theta) | \bm{E} |^2 + \theta Q \right) \right] \bm{E} \,,
\end{equation}
so that one may split the transformation into the sum of a linear map and a nonlinear map:
\begin{equation}
    A[\bm{E}, \bm{P}] = \epsilon_\infty \bm{E} + \bm{P} \,,
    \quad \text{and} \quad
    g(\bm{E}, Q) = \left( 
    (1 - \theta) | \bm{E} |^2 + \theta Q
    \right) \bm{E} \,.
\end{equation}
The mapping is then defined to be
\begin{equation}
    \Phi(\bm{E}, \bm{P}, Q) = 
    ( A[\bm{E}, \bm{P}] + a g(\bm{E}, Q), \bm{P}, Q) 
    = (\bm{D}, \bm{P}, Q) \,.
\end{equation}
Several assumptions are made regarding the range of values the parameters may take: $\epsilon_\infty > 0$, $1 \gg a > 0$, and $\theta \in [0, 1]$. While it is not required that $\epsilon_\infty$ be small, it is generally $O(1)$ in physically meaningful scenarios. On the other hand, for the mapping $\Phi$ to be invertible, one must assume that the nonlinearity is relatively weak: $0 < a \ll 1$. This is likewise physically realistic. 

In order to derive a CFL condition for the solver derived in this work, one seeks a local Lipschitz bound on the implicitly defined inverse mapping, $\Phi^{-1}(\bm{D}, \bm{P}, Q) = (\bm{E}, \bm{P}, Q)$. To do so, one first bounds $\| A \|$ and $\| D \Phi(u) \|$, where $u = (\bm{E}, \bm{P}, Q)$. These bounds will be sought in the space $X = (L^2(\Omega))^3 \times (L^2(\Omega))^3 \times L^2(\Omega)$ with norm
\begin{equation} \label{eq:X_norm}
    \| (\bm{E}, \bm{P}, Q) \| = \int_\Omega ( | \bm{E} |^2 + | \bm{P} |^2 + Q^2 ) \mathsf{d}^3 \bm{x} \,.
\end{equation}
This is a Hilbert space with inner product
\begin{equation}
    \left( (\bm{E}, \bm{P}, Q), (\tilde{\bm{E}}, \tilde{\bm{P}}, \tilde{Q}) \right)
    =
    \int_\Omega ( \bm{E} \cdot \tilde{\bm{E}} + \bm{P} \cdot \tilde{\bm{P}} + Q \tilde{Q} ) \mathsf{d}^3 \bm{x} \,.
\end{equation}
It is easier to compute the operator norm of a linear map in this space. 

The linear part of the full transformation may be written
\begin{equation}
    \Phi_L[\bm{E}, \bm{P}, Q] =
    \begin{pmatrix}
        \epsilon_\infty & 1 & 0 \\
        0 & 1 & 0 \\
        0 & 0 & 1
    \end{pmatrix}
    \begin{pmatrix}
        \bm{E} \\
        \bm{P} \\
        Q
    \end{pmatrix} \,.
\end{equation}
One begins by finding the operator norm of the linear part of the transformation.
\begin{lemma} \label{lemma:linear_map}
Assuming $\epsilon_\infty > 0$, in the Hilbert space $X$ defined above, one has
\begin{equation} \label{eq:linear_norm}
    \| \Phi_L \|
    =
    \sqrt{\frac{2 + \epsilon_\infty^2 + \sqrt{4 + \epsilon_\infty^4} }{2}} \,.
\end{equation}
\end{lemma}
\noindent \textit{Proof:} The operator norm in the Hilbert space $X$ is the $L^2(\Omega)$ norm of the finite dimensional $2$-norm of the fields applied pointwise. Hence, the problem reduces to finding the matrix $2$-norm of
\begin{equation}
    B = 
    \begin{pmatrix}
        \epsilon_\infty & 1 & 0 \\
        0 & 1 & 0 \\
        0 & 0 & 1
    \end{pmatrix} \,.
\end{equation}
This is simply its maximum singular value of $B$ which is the righthand side of equation \eqref{eq:linear_norm}. \qed \\[0.1em]

Now, consider the nonlinear part of the transformation which is denoted $\Phi_N$. The linear polarization $\bm{P}$ is unchanged by the nonlinear portion of the transformation. That is, $D \Phi_N(\bm{E}, Q, \bm{P})$ will take the form
\begin{equation}
    D \Phi_N(\bm{E}, Q, \bm{P})
    =
    \begin{pmatrix}
        a D \bar{\Phi}_N(\bm{E},Q) & 0 \\
        0 & I
    \end{pmatrix} \,,
\end{equation}
where $\bar{\Phi}_N(\bm{E}, Q) = (a g(\bm{E}, Q), Q)$. It may be shown that the matrix $p$-norm of a block diagonal matrix is simply the maximum of the $p$-norms of each block. Therefore, $\| D \Phi_N \| = \max \{ a \| D \bar{\Phi}_N \|, 1 \}$. 

The derivative of the nonlinear part of the transformation rule for $\bm{D}$ is given by
\begin{equation}
    Dg(\bm{E}, Q) [\delta \bm{E}, \delta Q]
    =
    \left[ 
    ( 1 - \theta) | \bm{E} |^2 \delta \bm{E} 
    + 2 ( 1 - \theta) (\bm{E} \cdot \delta \bm{E}) \bm{E} 
    + \theta Q \delta \bm{E} + \theta \bm{E} \delta Q 
    \right] \,.
\end{equation}
Hence, one finds that the derivative of the full nonlinear transformation is
\begin{equation}
    D \bar{\Phi}_N(\bm{E}, Q)[\delta \bm{E}, \delta Q]
    =
    \begin{pmatrix}
        \left[ (1 - \theta)( | \bm{E} |^2 I + 2 \bm{E} \otimes \bm{E}) + \theta Q I \right] & \theta \bm{E} \\
        0 & 1 
    \end{pmatrix}
    \begin{pmatrix}
        \delta \bm{E} \\
        \delta Q
    \end{pmatrix} \,.
\end{equation}
One likewise bounds the operator norm of the derivative of the nonlinear part of the transformation. 
\begin{lemma}
In the Hilbert space $(L^2(\Omega))^3 \times L^2(\Omega)$, one has
\begin{equation}
    \| D \bar{\Phi}_N(\bm{E}, Q) \| \leq
    \| 3 (1 - \theta) | \bm{E} |^2 + \theta |Q|
    + \theta | \bm{E} | \|_{L^\infty(\Omega)} + 1 \,,
\end{equation}
where $| \cdot |$ is the vector $2$-norm when applied to vectors and the absolute value when applied to scalars. 
\end{lemma}
\noindent \textit{Proof:} One may rewrite the problem of finding the operator norm of $D\Phi_N$ as an optimization problem of the form
\begin{equation}
    \sigma_{max}(u) = \max_{ \delta u, \delta \tilde{u} \in B_1(0) } \left(\delta u, M(u) \delta \tilde{u} \right) \,,
\end{equation}
where $u = (\bm{E}, Q)$ (and similarly for $\delta u$ and $\delta \tilde{u}$), and
\begin{equation}
    M(u) =
    \begin{pmatrix}
        \left[ (1 - \theta)( | \bm{E} |^2 I + 2 \bm{E} \otimes \bm{E}) + \theta Q I \right] & \theta \bm{E} \\
        0 & 1 
    \end{pmatrix} \,.
\end{equation}
Denote the maximum singular value of $M(u)$ by $\sigma_*(u)$. Then it follows
\begin{multline}
    \sigma_{max}(u) = \max_{ \delta u, \delta \tilde{u} \in B_1(0) } \left(\delta u, \sigma_*(u) \delta \tilde{u} \right) 
    \leq
    \| \sigma_*(u) \|_{L^\infty(\Omega)} \\
    \eqcolon \inf \{ C \geq 0 : | \sigma_*(u) | \leq C \ a.e. \ \text{in} \ \Omega \} \,.
\end{multline}
Hence, if one can bound the maximum singular value of $M(u)$, then the proof is complete. But this value is the $2$-norm of a block upper triangular matrix. In general, one has that if
\begin{equation}
    M =
    \begin{pmatrix}
        A & B \\
        0 & C
    \end{pmatrix} \,,
\end{equation}
then $\|M \|_2 \leq \|A \|_2 + \|B \|_2 + \| C \|_2$. Moreover, if a matrix is an outer product, $M = u v^T$, then $\|M \|_2 = \|u\|_2 \|v \|_2$. Hence, using these facts and the Cauchy-Schwartz inequality, one finds
\begin{multline}
    \left\|
    \begin{pmatrix}
        \left[ (1 - \theta)( | \bm{E} |^2 I + 2 \bm{E} \otimes \bm{E}) + \theta Q I \right] & \theta \bm{E} \\
        0 & 1 
    \end{pmatrix}
    \right\| \\
    \leq
    \| (1 - \theta)( | \bm{E} |^2 I + 2 \bm{E} \otimes \bm{E}) + \theta Q I \|
    + \theta | \bm{E} | + 1 \\
    \leq
    3 (1 - \theta) | \bm{E} |^2 + \theta | Q |
    + \theta | \bm{E} | + 1
\end{multline}
Hence, one may bound the operator norm as follows:
\begin{equation}
    \sigma_{max}(u) \leq \| 3 (1 - \theta) | \bm{E} |^2 + \theta |Q|
    + \theta | \bm{E} | \|_{L^\infty(\Omega)} + 1 \,.
\end{equation}
\qed \\[0.1em]

\begin{lemma} \label{lemma:nl_map}
In the Hilbert space $X$ defined above, one has
\begin{equation}
    \| D \Phi_N(\bm{E}, Q) \| \leq \max \left\{ 1, a \left(
    \| 3 (1 - \theta) | \bm{E} |^2 + \theta |Q|
    + \theta | \bm{E} | \|_{L^\infty(\Omega)} + 1 \right) \right\} \,,
\end{equation}
where $| \cdot |$ is the vector $2$-norm when applied to vectors and the absolute value when applied to scalars. 
\end{lemma}
\noindent \textit{Proof:} This is a corollary of the previous lemma and the fact that matrix $p$-norms of a block diagonal matrix is simply the maximum of the $p$-norms of each block. \qed

Finally, one may use the general result for finding local Lipschitz bounds of implicitly defined weakly nonlinear maps on Banach spaces to obtain a local Lipschitz bound on the electric field. 

\begin{theorem}
Let $X = (L^2(\Omega))^3 \times (L^2(\Omega))^3 \times L^2(\Omega)$ be the Hilbert space with norm given in equation \eqref{eq:X_norm}. Suppose $\bm{D} = \left( \epsilon_\infty \bm{E} + \bm{P} + a(1 - \theta) | \bm{E} |^2 \bm{E} + a \theta Q \bm{E} \right)$, and define the nonlinear change of coordinates $\Phi(\bm{E}, \bm{P}, Q) = (\bm{D}, \bm{P}, Q)$. Then $\exists r > 0$ such that for $(\bm{D}, \bm{P}, Q), (\tilde{\bm{D}}, \tilde{\bm{P}}, \tilde{Q}) \in B_r(0) \subset X$, the implicitly defined inverse map $\Phi^{-1}(\bm{D}, \bm{P}, Q) = (\bm{E}, \bm{P}, Q)$ is well defined, and one has the local Lipschitz bound
\begin{equation}
    \| \Phi^{-1}(\bm{D}, \bm{P}, Q) - \Phi^{-1}(\tilde{\bm{D}}, \tilde{\bm{P}}, \tilde{Q}) \|_X
    \leq
    C^{-1} \| (\bm{D}, \bm{P}, Q) - (\tilde{\bm{D}}, \tilde{\bm{P}}, \tilde{Q}) \|_X \,,
\end{equation}
where $C = c_L - c_N > 0$, where
\begin{equation}
    c_L = \sqrt{\frac{2 + \epsilon_\infty^2 + \sqrt{4 + \epsilon_\infty^4} }{2}} \,,
\end{equation}
and
\begin{equation}
    c_N = \max \left\{ 1, a \left( \| 3 (1 - \theta) | \bm{E} |^2 + \theta ( |Q|
    + | \bm{E} | ) \|_{L^\infty(\Omega)} + 1 \right)\right\} \,.
\end{equation}
\end{theorem}
\noindent \textit{Proof:} This follows from theorem \ref{thm:bounded_weakly_nl_map} and lemmas \ref{lemma:linear_map} and \ref{lemma:nl_map} if one selects $r > 0$ be such $c_N < c_L$. \qed \\[0.1em]

\noindent As $0 < a \ll 1$ for general nonlinear media, in many cases one will find that $c_N = 1$. Moreover, one has that $c_L \geq \sqrt{2} \approx 1.41$. Hence, the condition that $c_N < c_L$ will be met in all but the most extreme cases in which the magnitude of the fields is enormous. 

%% file: appendices/kink_solns.tex
\section{On computing traveling wave solutions to the one-dimensional nonlinear Maxwell system} \label{appendix:kink_shaped_waves}

Consider the one-dimensional Maxwell equations in cubicly nonlinear media with no damping and no time-dispersive nonlinear response:
\begin{equation} \label{eq:nonlinear_maxwell_1d}
    \begin{aligned}
        &\partial_t B = \partial_x E \,, 
        \quad
        \partial_t D = \partial_x B \\
        &\partial_t P = J \,, 
        \quad
        \partial_t J = - \omega_0^2 P + \omega_p^2 E \,, \\
        &D = \epsilon_\infty E + a E^3 + P \,.
    \end{aligned}
\end{equation}
This is the one-dimensional variant of equation \eqref{eq:nondim_nonlinear_maxwell} with $\theta = 0$. If one makes the traveling wave Ansatz, i.e. $E(x,t) = E(x - vt)$ (and similarly for the remaining fields), a system of ordinary differential equations is obtained. This differential equation was first derived in \cite{PhysRevE.71.036602}:
\begin{equation} \label{eq:tw_ode}
\begin{aligned}
    \frac{d E}{d \xi} &= \Phi \\
    \frac{d \Phi}{d \xi} &= \frac{6 a v^2 E \Phi^2 + ( \epsilon_\infty \omega_0^2 + \omega_p^2 - \omega_0^2 / v^2) E + a \omega_0^2 E^3}{1 - \epsilon_\infty v^2 - 3 a v^2 E^2} \,.
\end{aligned}
\end{equation}
The solution to this equation, which gives traveling wave solutions to the full partial differential equation, provides a convenient tool for a convergence study of solvers for the full partial differential equation as was done in \cite{BOKIL2017420}. Section \ref{sec:numerical_results} uses these traveling wave solutions as a numerical example. Therefore, a high-accuracy periodic solution of this differential equation is sought to provide an initial condition for the convergence study. That the solution be highly accurate is critical lest the convergence studies be polluted by propagating errors in the initial data. The solution to equation \eqref{eq:tw_ode} was obtained using shooting methods in \cite{BOKIL2017420}, but a different strategy is employed here. 

\subsection{The noncanonical Hamiltonian structure of the traveling wave differential equation}

Because this system given in equation \eqref{eq:tw_ode} arises as a particular solution to a Hamiltonian system of partial differential equations, see section \ref{sec:cts_cubic_maxwell}, one expects the traveling wave ordinary differential equation to possess a Hamiltonian formulation. With the Hamiltonian structure of this system, it is possible to derive a quadrature relating the energy of the system to the period of orbits on that energy level for a given selection of parameters. This then allows one to choose parameters and initial conditions which yield an exact periodic orbit. A Hamiltonian formulation of this system was in fact discussed in \cite{WEBB200449}, but in different coordinates than those considered here. 

The noncanonical Hamiltonian formulation of the traveling wave differential equation may be derived as follows. The Hamiltonian and Poisson bracket are postulated to take the form
\begin{equation}
    H[E, \Phi] = f(E) \Phi^2 + g(E) \,,
    \quad \text{and} \quad
    J(E) = 
    \begin{pmatrix}
        0 & (2 f(E) )^{-1} \\
        - (2 f(E) )^{-1} & 0 
    \end{pmatrix} \,.
\end{equation}
This yields the following differential equation for $z = (E, \Phi)$:
\begin{equation}
    \dot{z} = J DH(z)
    =
    \begin{pmatrix}
        \Phi \\
        - \frac{1}{2} ( \log f(E) )' \Phi^2
        - \frac{g'(E)}{2 f(E)}
    \end{pmatrix} \,.
\end{equation}
Hence, the problem of finding the Hamiltonian structure reduces to solving the differential equation
\begin{equation}
    ( \log f(E) )' = \frac{- 12 a v^2 E}{1 - \epsilon_\infty v^2 - 3 a v^2 E^2}
\end{equation}
for $f(E)$, and then integrating
\begin{equation}
    g'(E) = - 2 f(E) \left[ \frac{( \epsilon_\infty \omega_0^2 + \omega_p^2 - \omega_0^2 / v^2) E + a \omega_0^2 E^3}{1 - \epsilon_\infty v^2 - 3 a v^2 E^2} \right] \,.
\end{equation}
One finds that
\begin{equation}
    f(E) = (1 - v^2(\epsilon_\infty + 3 a E^2) )^2 \,,
\end{equation}
and
\begin{equation}
    g(E) = \frac{1}{2} \left[ \omega_p^2 \left( 2 - v^2 \left( 2 \epsilon_\infty + 3 a E^2 \right) \right) 
    - \frac{2 \omega_0^2}{v^2} \left( 1 - v^2(\epsilon_\infty + 3 a E^2) \right)^2
    \right] E^2 \,.
\end{equation}
Therefore, the Hamiltonian is simply a polynomial in $(E, \Phi)$. 

It is possible to describe the system with a smaller set of free parameters (see \cite{PhysRevE.71.036602}). Define $\epsilon_S > \epsilon_\infty > 0$ and let $\omega_p^2 = (\epsilon_S - \epsilon_\infty) \omega_0^2$. Further, let
\begin{multline} \label{eq:scaling1}
    x = k_x \tilde{x} \,,
    \quad 
    t = k_t \tilde{t} \,,
    \quad
    \xi = k_x ( \tilde{x} - \tilde{v} \tilde{t} ) \,,
    \quad
    v = \frac{k_x}{k_t} \tilde{v} \,, \\
    \quad
    E(\xi) = k_E \tilde{E}(\tilde{\xi} ) \,,
    \quad 
    \Phi(\xi) = k_{\Phi} \tilde{\Phi}(\tilde{\xi}) \,,
\end{multline} 
where
\begin{equation} \label{eq:scaling2}
    k_t = \frac{1}{\omega_0} \,,
    \quad
    k_x = \frac{1}{\omega_0 \sqrt{\epsilon_\infty}} \,,
    \quad 
    k_E = \sqrt{\frac{\epsilon_\infty}{3 a}} \,,
    \quad \text{and} \quad
    k_{\Phi} = 
    \frac{\epsilon_\infty \omega_0}{\sqrt{3 a}} \,.
\end{equation}
Note, this is not a non-dimensionalization as the system was already dimensionless. The dependent and independent variables have simply been rescaled in a convenient manner. In the following, the tildes are dropped for notational convenience. The equations of motion become
\begin{equation} \label{eq:tw_diff_eq_nondim}
    \frac{d E}{d \xi} = \Phi \,,
    \quad 
    \frac{d \Phi}{d \xi}
    =
    \frac{2 v^2 E \Phi^2 - (1/v^2 - \epsilon_S/\epsilon_\infty) E + E^3/3}{1 - v^2 ( 1 + E^2) } \,.
\end{equation}
In these units, the functions used to define the Hamiltonian structure are
\begin{equation}
    f(E) = (1 - v^2(1 + E^2))^2
    \,,
\end{equation}
and
\begin{equation}
    g(E) = 
    \frac{(1 - v^2)(1 - \epsilon_S/\epsilon_\infty v^2)}{v^2} E^2 
    + \frac{3 \epsilon_S / \epsilon_\infty v^2 + ( v^2 - 4)}{6} E^4 +
    \frac{v^2}{9} E^6
    \,.
\end{equation}
Only a two-dimensional parameter space needs to be searched to find the desired periodic solutions: $\mathcal{P} = (v, \epsilon_S/\epsilon_\infty)$. 

\subsection{Computing periodic orbits}

The phase-space of the traveling wave system has three fixed points:
\begin{equation}
    z_c = (0,0) \,,
    \quad \text{and} \quad
    z_\pm =
    \left( \pm \sqrt{3 \left( \frac{1}{v^2} - \frac{\epsilon_S}{\epsilon_\infty} \right) }, 0 \right) \,.
\end{equation}
If the traveling wave speed is in the interval
\begin{equation}
    \sqrt{\frac{2}{3 \epsilon_S / \epsilon_\infty - 1}} < | v | < \sqrt{ \frac{\epsilon_\infty}{\epsilon_S}} \,,
\end{equation}
then $z_c$ is a center while $z_\pm$ are hyperbolic fixed points.
Additionally, the Poisson bracket becomes singular at
\begin{equation}
    E = \pm \sqrt{\frac{1 - v^2}{v^2}} = \pm E_\infty \,.
\end{equation}
This is not problematic for the purposes of this study as the bounded trajectories of interest are contained in the interval $(-E_\infty, E_\infty)$. 

To aid in numerically computing periodic orbits, a quadrature formula may be derived which computes the period of a bound orbit for a given initial condition and parameter choice. One may write $\Phi$ as a function of $(E, H)$ which yields a quadratrue:
\begin{equation}
    \Phi(E, H) = \pm \sqrt{ \frac{H - g(E)}{f(E)} } \,,
    \quad \implies \quad
    \int d \xi =
    \pm \int \sqrt{ \frac{f(E)}{H - g(E)} } d E \,.
\end{equation}
Because the Hamiltonian only contains even powers in $\Phi$ and $E$, phase-space is up-down and left-right symmetric. Hence, one can get the period of a closed orbit by integrating a quarter turn along the orbit and multiplying by four. That is, letting $\Xi(H_0)$ denote the period of a closed trajectory with energy $H_0$,
\begin{equation}
    \Xi(H_0) = 4 \Delta \xi 
    =
    4 \int_{0}^{E_0} \sqrt{ \frac{f(E)}{H_0 - g(E)} } d E \,,
\end{equation}
where $z_0 = (E_0, 0)$ is chosen as a convenient initial point and $H_0 = H(E_0, 0)$. Because of the symmetry of phase-space, one has traveled a quarter of the way around the periodic orbit going from $E = E_0$ to $E = 0$. The parametric dependence has been notationally suppressed, but all of these quantities (as well as the existence of periodic orbits) depend on the choice of parameters. 

A periodic orbit will be sought with parameters
\begin{equation}
    \epsilon_S = 5.25 \,,
    \quad
    \epsilon_\infty = 2.25 \,,
    \quad \text{and} \quad
    v = 0.9995 \sqrt{\frac{\epsilon_\infty}{\epsilon_S}} \approx 0.6543263438726231 \,.
\end{equation}
The value of $v$ is chosen so as to make the traveling wave speed very nearly the maximum allowed value. A large number of decimal places is needed to ensure that the numerically computed periodic orbit is accurate nearly to machine precision. Even small errors in the initial data greatly impact the validity of numerical tests initialized on the traveling wave solution. In order to be as close to the separatrix as possible (i.e.\ the boundary between closed and unbounded orbits in the traveling wave phase-space), let
\begin{multline}
    E_0 = 0.9995 \sqrt{3 \left( \frac{1}{v^2} - \frac{\epsilon_S}{\epsilon_\infty} \right) }
    \approx
    0.08365554374935166 \\
    \implies
    H_0 = H(E_0,0) \approx 4.668996808871565 \times 10^{-6} \,.
\end{multline}
Numerical quadrature yields
\begin{equation}
    \Xi(H_0) =
    4 \int_{E_0}^0 \sqrt{ \frac{f(E)}{H_0 - g(E)} } d E 
    \approx
    397.0737083968515 \,.
\end{equation}
Recall, these quantities are in the rescaled units. In order to obtain periodic solutions in the original units, one must undo the rescaling done in equations \eqref{eq:scaling1} and \eqref{eq:scaling2}. This requires that one make choices for the remaining parameters which were removed by the rescaling. Let $a = \epsilon_\infty/3$, and finally choose $\omega_0$ to fix the domain length. In order to let the physical domain have length $L = 5$ and to allow for one full periodic orbit, one finds
\begin{equation}
    5 = L = \tilde{L} k_x =  \Xi(H_0) k_x = \frac{\Xi(H_0)}{\omega_0 \sqrt{\epsilon_\infty}} 
    \quad \implies \quad
    \omega_0 \approx 52.943161119580196 \,.
\end{equation}
Finally, recall that
\begin{equation}
    \omega_p = \omega_0 \sqrt{\epsilon_S - \epsilon_\infty} \approx 91.70024497241806 \,.
\end{equation}
With these parameters, it is possible to obtain an exact periodic solution to use as initial data in the numerical study performed in section \ref{sec:numerical_results}. 

The differential equation given in equation \eqref{eq:tw_diff_eq_nondim} with the parameters derived in the previous paragraph must be integrated with high accuracy in order to obtain initial conditions which recover the traveling wave solution. To this end, these equations are solved as an initial value problem with $(E_0, \Phi_0) = (E_0, 0)$ using an implicit Runge-Kutta method of the Radau IIA family of order 5 (see \cite{hairer1996solving}). The solution features a rapid change between its maximum and minimum values; hence the choice of a solver well suited to stiff differential equations. 

In order to obtain the parameters and initial conditions to input into the solver for the full nonlinear Maxwell system, one needs to undo the scaling performed in equations \eqref{eq:scaling1} and \eqref{eq:scaling2}. The parameters are
\begin{multline}
    \epsilon_\infty = 2.25 \,,
    \quad
    \epsilon_s = 5.25 \,,
    \quad\beta = \epsilon_s - \epsilon_\infty \,,
    \quad 
    \omega_0 = 52.943161119580196 \,, \\
    \quad
    \omega_p = \omega_0 \sqrt{\beta} \,,
    \quad
    a = \epsilon_\infty/3 \,,
    \quad 
    v = \frac{0.9995}{\sqrt{\epsilon_s}} \approx 0.43621756258174876 \,.
\end{multline}
Notice, only the traveling wave speed, $v$, was rescaled to return to the physical domain. The initial conditions are given by
\begin{equation}
    E(x,0) = \sqrt{\frac{\epsilon_\infty}{3 a}} \tilde{E_0}( \omega_0 \sqrt{\epsilon_\infty} x ) \,,
    \quad \text{and} \quad
    \Phi(x,0) = \frac{\epsilon_\infty \omega_0}{\sqrt{3 a}} \tilde{\Phi}_0( \omega_0 \sqrt{\epsilon_\infty} x )  \,,
\end{equation}
where $(\tilde{E}_0, \tilde{\Phi}_0)$ are the periodic solutions using the reduced parameter set obtained in the previous section. The remaining fields may be initialized by applying the traveling wave Ansatz to the rest of the variables in equation \eqref{eq:nonlinear_maxwell_1d}. One finds
\begin{multline}
    B(x,0) = - \frac{1}{v} E(x,0) \,,
    \quad
    D(x,0) = \frac{1}{v^2} E(x,0) \,, \\
    \quad 
    P(x,0) = \left( \frac{1}{v^2} - \epsilon_\infty \right) E(x,0) - a E(x,0)^3 \,, \\
    \quad
    J(x,0) = \left( \epsilon_\infty v - \frac{1}{v} \right) \Phi(x,0) + 3 a v E(x,0)^2 \Phi(x, 0) \,.
\end{multline}
This may be used to obtain the initial data to numerically compute traveling wave solutions to the full nonlinear Maxwell partial differential equation.